\documentclass[prx,twocolumn,preprintnumbers,superscriptaddress,amsmath,amssymb,longbibliography]{revtex4-1}
\usepackage{graphicx}
\usepackage{mathtools}
\usepackage{subfigure}
\usepackage{mathrsfs}
\usepackage{amsfonts}
\usepackage{times}
\usepackage{amsmath}
\usepackage{leftidx}
\usepackage{tikz}
\usepackage{tikz-network}
\usepackage{color}
\usepackage[colorlinks,linkcolor=blue,citecolor=blue]{hyperref}

\newcommand{\Tr}{\operatorname{Tr}}
\newcommand{\Pf}{\operatorname{Pf}}

\newtheorem{theorem}{Theorem}
\newtheorem{lemma}{Lemma}
\usepackage{bbold}
\usepackage{braket}
\usepackage{mathtools}

\begin{document}
\title{Lieb-Robinson Bounds on Entanglement Gaps %-Spectrum Splittings 
from Symmetry-Protected Topology} %Dynamical Stability of Symmetry-Protected Topological Order in One Dimension: Lieb-Robinson Bounds on Entanglement Gaps}
\author{Zongping Gong}
\affiliation{Department of Physics, University of Tokyo, 7-3-1 Hongo, Bunkyo-ku, Tokyo 113-0033, Japan}
\author{Naoto Kura}
\affiliation{Department of Physics, University of Tokyo, 7-3-1 Hongo, Bunkyo-ku, Tokyo 113-0033, Japan}
%\author{J. Ignacio Cirac}
%\affiliation{Max-Planck-Institut f\"ur Quantenoptik, Hans-Kopfermann-Stra{\ss}e 1, D-85748 Garching, Germany}
\author{Masatoshi Sato}
\affiliation{Yukawa Institute for Theoretical Physics, Kyoto University, Kyoto 606-8502, Japan}
\author{Masahito Ueda}
\affiliation{Department of Physics, University of Tokyo, 7-3-1 Hongo, Bunkyo-ku, Tokyo 113-0033, Japan}
\affiliation{Institute for Physics of Intelligence, University of Tokyo, 7-3-1 Hongo, Bunkyo-ku, Tokyo 113-0033, Japan}
\affiliation{RIKEN Center for Emergent Matter Science (CEMS), Wako, Saitama 351-0198, Japan}
\date{\today}

\begin{abstract}
A quantum quench is the simplest protocol to investigate nonequilibrium many-body quantum dynamics. Previous studies on the entanglement properties of quenched quantum many-body systems mainly focus on the growth of entanglement entropy. Several rigorous results and phenomenological guiding principles have been established, such as the no-faster-than-linear entanglement growth generated by generic local Hamiltonians and the peculiar logarithmic growth for many-body localized systems. However, little is known about the dynamical behavior of the full entanglement spectrum, which is a refined character closely related to the topological nature of the wave function. Here, we establish a \emph{rigorous} and universal result for the entanglement spectra of one-dimensional symmetry-protected topological (SPT) systems evolving out of equilibrium. Our result is derived both for free-fermion SPT systems and interacting ones. For free-fermion systems with Altland-Zirnbauer symmetries, we prove that the single-particle entanglement gap after quenches obeys essentially the same Lieb-Robinson bound as that on the equal-time correlation, provided that there is no dynamical symmetry breaking. As a notable byproduct, we obtain a new type of Lieb-Robinson velocity which is related to the band dispersion with a complex wave number and reaches the minimum as the maximal (relative) group velocity. Within the framework of tensor networks, i.e., for SPT matrix-product states evolved by symmetric and trivial matrix-product unitaries, we also identify a Lieb-Robinson bound on the many-body entanglement gap for general quenched interacting SPT systems. This result suggests high potential of tensor-network approaches for exploring rigorous results on long-time quantum dynamics. Influence of partial symmetry breaking, effects of disorder, and the relaxation property in the long-time limit are also discussed. Our work establishes a paradigm for exploring rigorous results of SPT systems out of equilibrium.
\end{abstract}
\maketitle

\section{Introduction} 
Recent years have witnessed remarkable experimental developments in atomic, molecular and optical physics, which have enabled us to engineer and control artificial quantum many-body systems at the level of individual atoms, ions and photons \cite{Bloch2012,Blatt2012,Koch2012}. Particular attention is focused on nonequilibrium quantum dynamics \cite{Choi2017,Bernien2017,Zhang2017,Zhang2017b}, of which the arguably simplest situation is \emph{quantum quenches} \cite{Calabrese2007,Fagotti2016,Mitra2018} --- the system is initialized as a wave function $|\Psi_0\rangle$ which then evolves unitarily by a Hamiltonian $H$, with respect to which $|\Psi_0\rangle$ is typically a highly excited superposition state.  The wave function at time $t$ is then formally given by $|\Psi_t\rangle=e^{-iHt}|\Psi_0\rangle$. To model realistic quantum simulators, especially ultracold atoms and superconducting circuits with short-range interactions, we usually assume $H$ to be \emph{local}, in the sense that it can be written as a sum of short-range operators. In light of the rapid development of topological material science \cite{Kane2010,Qi2011,Bernevig2013,Ryu2016}, there is growing interest in topological aspects of quench dynamics \cite{Foster2013,Rigol2015,Cooper2015,Vajna2015,Heyl2016,Budich2016,Cooper2016,Refael2016,Balatsky2016,Zhai2017,Weitenberg2019,Weitenberg2018,Yang2018,Chang2018,Ezawa2018,Liu2018,Pan2018,Gong2018b,Cooper2018b,Cooper2019,Cooper2019b,Yu2019}. A fundamental question in this context is: given $|\Psi_0\rangle$ as the ground state of a gapped Hamiltonian $H_0$, which may be trivial or topological, whether the topology of $|\Psi_t\rangle$ will change during time evolution, and, if yes, in what way. To make the topology well-defined,  we may have to impose certain symmetries. For the sake of concreteness, we assume that $H_0$ and $H$ share the same symmetries, if any. The answer to the above question has recently been given and is somewhat negative: for unitary symmetries or/and anti-unitary anti-symmetries \cite{antisym}, $|\Psi_t\rangle$ stays in the same \emph{symmetry-protected topological} (SPT) phase \cite{Rigol2015,Cooper2015,Cooper2018b,Cooper2019}. For anti-unitary symmetries or/and unitary anti-symmetries, the topological number of $|\Psi_t\rangle$ generally becomes ill-defined (or reduces) due to \emph{dynamical symmetry breaking} \cite{Cooper2018b,Cooper2019}. To understand this, we only have to note that $|\Psi_t\rangle$ is the ground state of \cite{Cooper2018b,Gong2018b}
\begin{equation}
H(t)\equiv e^{-iHt}H_0e^{iHt}, 
\label{Ht}
\end{equation}
which shares the same spectrum as $H_0$. The conservation of topological number follows from the fact that $H(t)$ is gapped and continuously deformed from $H_0$ in a symmetry-preserving manner. 

Since topological numbers are rather abstract quantities and take very different forms depending on the specific systems, we need a universal topological indicator to formalize the above qualitative analysis into a general, rigorous and, in principle, experimentally verifiable statement. Entanglement turns out to be an ideal candidate to demonstrate the persistence of topology. We can trace the time evolution of the \emph{entanglement spectrum} (ES) for a proper bipartition, which contains crucial information of the entanglement pattern and is arguably the most widely used universal topological indicator that is applicable to both noninteracting \cite{Vishwanath2010,Fidkowski2010,Hughes2011,Chang2014} and interacting systems \cite{Haldane2008,Pollmann2010,Thomale2010,Pollmann2011,Fidkowski2011,Cirac2011b}. The ES is expected to be accessible in near-future ultracold-atom and trapped-ion experiments \cite{Pichler2016,Blatt2017,Dalmonte2018}. The persistence of SPT order thus manifests itself in that the ES stays gapless or degenerate \cite{Cooper2018b,Cooper2019}.

However, a vital point is missing in the above argument of topology conservation --- defining SPT phases requires \emph{locality} in the Hamiltonian \cite{Wen2017}, while $H(t)$ in Eq.~(\ref{Ht}) may become highly nonlocal after a long time. That is to say, the dynamically generated non-locality could obscure the SPT order in experimentally relevant length scales, which are often limited. It is thus practically, and of course theoretically important to understand at which length scale SPT order survives. Intuitively, we expect from the \emph{bulk-edge correspondence} \cite{Hatsugai1993} that the SPT order should persist up to a time scale $t^*$ %$t^*=\frac{l}{2v_{\rm LR}}$, where $l$ is interpreted as the length scale of the system \cite{Cooper2019}. 
when topological edge modes at different boundaries start to interfere and diffuse into the bulk, though a more refined analysis is needed to reach a definite conclusion.

A key concept to make the intuitive argument rigorous is the \emph{Lieb-Robinson bound}. A natural assumption on locality has a striking consequence known as the \emph{light-cone effect} --- a local operator evolved by $H$ spreads no faster than an emergent ``light velocity" $v_{\rm LR}$. Precisely speaking, there can be a nonzero leakage outside the light cone, which nevertheless decays exponentially with respect to the distance from the light cone. This rigorous result was derived by Lieb and Robinson nearly half a century ago \cite{Lieb1972}, and is known as the Lieb-Robinson bound. An important implication of this bound is that, for two remote local operators separated by $l$, the equal-time correlation should stay exponentially small up to a time scale  $t^*=\frac{l}{2v_{\rm LR}}$, provided that the initial correlation decays exponentially \cite{Bravyi2006}. Such a light-cone spreading of correlation as well as its breakdown for long-range interactions has recently been examined in ultracold-atom and trapped-ion experiments \cite{Bloch2012b,Monroe2014,Roos2014}.

\begin{figure}
\begin{center}
 \includegraphics[width=8.5cm, clip]{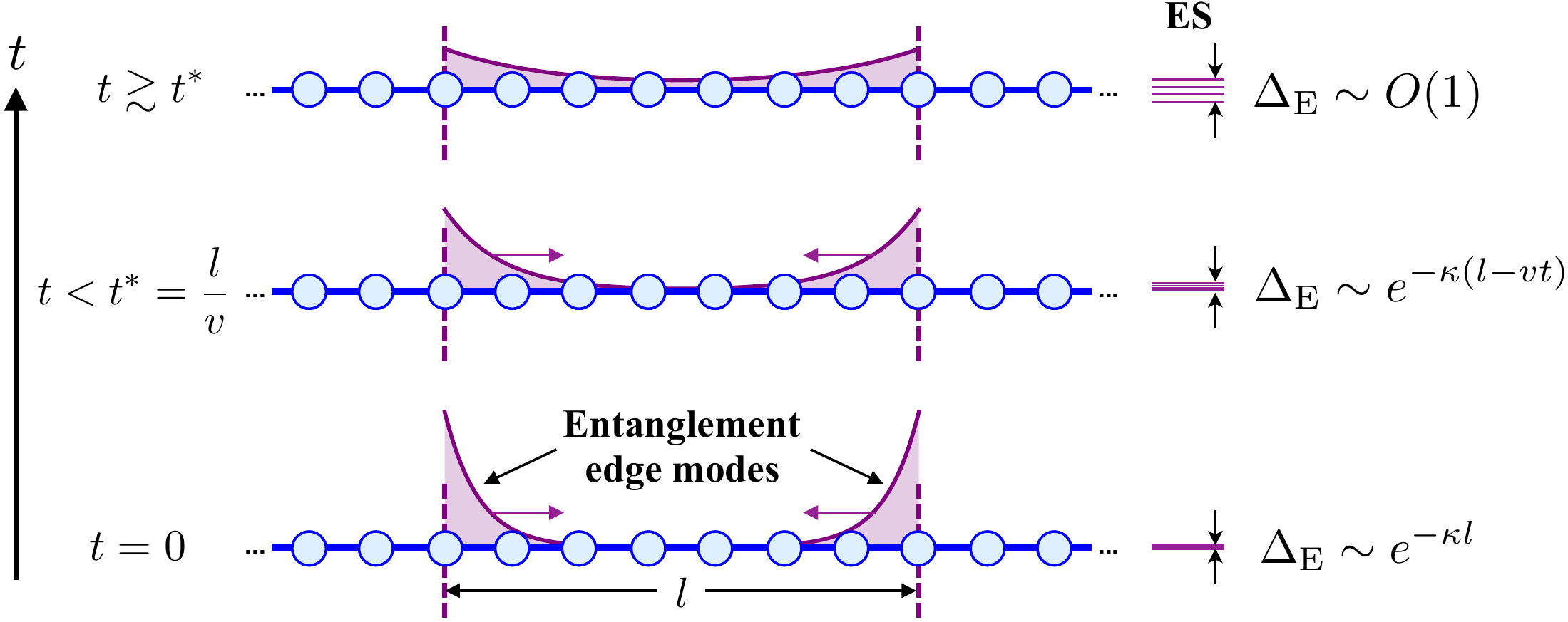}
       \end{center}
   \caption{Schematic illustration of the main result. (Bottom) Initially, the entanglement gap $\Delta_{\rm E}$ of a length-$l$ segment embedded in a 1D topological system is of the order of $e^{-\kappa l}$ due to an exponentially small overlap between the tails of the two entanglement edge modes. (Middle) After quench, these edge modes diffuse no faster than linearly, leading to an exponential increase in $\Delta_{\rm E}$. (Top) After a time proportional to $l$, these modes significantly diffuse into the bulk and the quasi-degeneracy in the ES is clearly lifted.}
   \label{fig1}
\end{figure}

The Lieb-Robinson bound has been employed to reveal universal behaviors of quantum many-body systems. In addition to correlations between observables, \emph{entanglement entropy} \cite{Vidal2003,Calabrese2004,Ryu2006}, which quantifies genuinely quantum correlation between a subsystem and its complement, has also been widely studied in quench dynamics \cite{Calabrese2005,Chiara2006,Fagotti2008,Daley2013,Hauke2013}. Again, locality of $H$ sets very fundamental limitations on the growth of entanglement entropy. From a quasi-particle viewpoint \cite{Altman2015}, we can infer from the Lieb-Robinson bound that the entanglement growth cannot be faster than linearly in time \cite{Bravyi2006}. In particular, with the Lieb-Robinson bound applied to quasiadiabatic continuation \cite{Hastings2005}, it has been proved that given $|\Psi_1\rangle$ and $|\Psi_2\rangle$ as the ground states of two gapped local Hamiltonians $H_1$ and $H_2$ that can be adiabatically connected to each other, $|\Psi_1\rangle$ obeys an entanglement \emph{area law} \cite{Hastings2007} if and only if $|\Psi_2\rangle$ also does \cite{Verstraete2013}. In one dimension (1D), where all the gapped local Hamiltonians are adiabatically connected \cite{Chen2011,Chen2011b,Schuch2011}, we always obtain an area law by choosing $|\Psi_1\rangle$ to be a product state. It is worth mentioning that, without referring to any Hamiltonian, all the 1D wave functions with exponentially decaying correlations have recently been proved to obey the area law \cite{Brandao2013,Brandao2015,Cho2018}. Moreover, the measurement of the entanglement (R\'enyi) entropy has been achieved in ultracold-atom and trapped-ion experiments \cite{Greiner2015,Greiner2016,Greiner2019,Roos2019}. 

According to the light-cone picture, we can anticipate that the time scale of the Leib-Robinson bound roughly estimates how long it takes for a topological edge mode to completely diffuse into the bulk, after which the SPT order becomes invisible. This argument may remain applicable if the physical edge is replaced by an artificial entanglement cut, which can always be done even if there are no physical boundaries. In this case, reinterpreting $l$ as a subsystem size, we expect that the ES should stay nearly gapless or degenerate with precision $e^{-O(l)}$ until $t^*$ \footnote{This phenomenon should be distinguished from the ES crossings emerging in quantum quenches from trivial to topological Hamiltonians \cite{Chang2018,Gong2018b,Yu2019}, which correspond to higher dimensional topology and are also observed in Floquet topological systems \cite{Potter2016,Yao2017b}.}. Such an expectation has been numerically verified in several free-fermion models \cite{Chung2013,Chung2016,Chung2017}. It is thus natural to conjecture the existence of a Lieb-Robinson bound on the \emph{entanglement gap} (to be exactly defined in a moment): 
\begin{equation}
\Delta_{\rm E}%\lesssim 
\le Ce^{-\kappa(l-vt)}.
\label{LRB}
\end{equation}
Here $v=2v_{\rm LR}$, $\kappa$ estimates an inverse correlation length and $C$ is a prefactor that is at most polynomial in terms of $l$ and $t$. See Fig.~\ref{fig1} for a schematic illustration for 1D systems.

\begin{figure}
\begin{center}
 \includegraphics[width=7.5cm, clip]{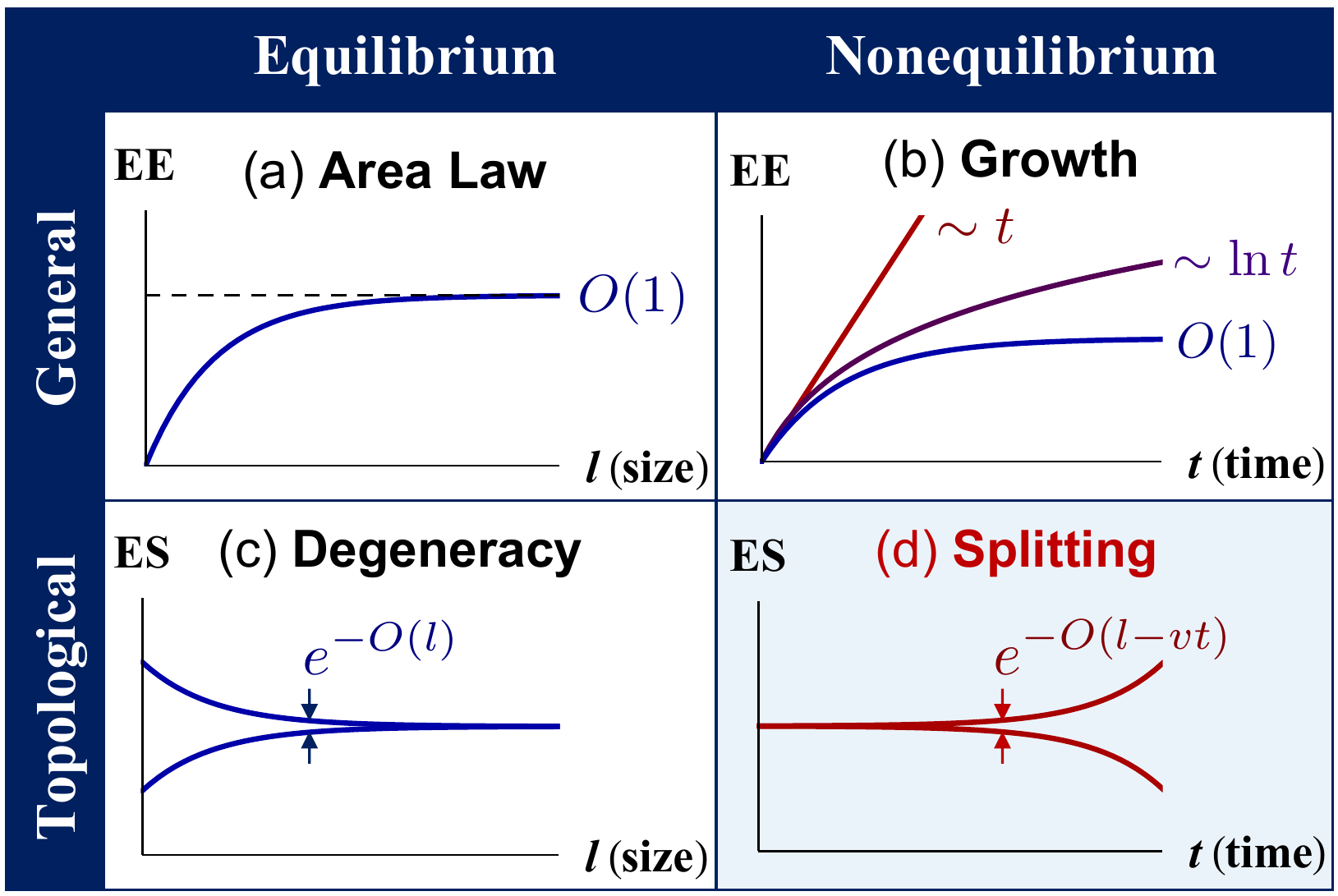}
       \end{center}
   \caption{Schematic illustrations of the most fundamental results on entanglement in 1D quantum many-body systems in and out of equilibrium, where the shaded part is rigorously established in this work. (a) The ground state of an arbitrary gapped local Hamiltonian obeys an entanglement-entropy (EE) area law. (b) After a quench, the entanglement-entropy growth exhibits universal features such as linear (red), logarithmic growth (purple) or saturation (blue), depending on whether the postquench Hamiltonian is thermal, many-body localized or Anderson localized, respectively. (c) If the system is in an SPT phase, the ES is degenerate up to an exponentially small correction. (d) Rigorous Lieb-Robinson bound on the lifting of ES degeneracy for SPT systems after quenches, which constitutes the primary result of this paper.}
   \label{fig2}
\end{figure}

In this paper, we \emph{prove} Eq.~(\ref{LRB}) for general 1D SPT systems undergoing nonequilibrium unitary evolution. In fact, a rigorous result has been obtained for intrinsic topological order from the perspective of local indistinguishability between degenerate ground states \cite{Bravyi2006}. However, unlike the derivation in Ref.~\cite{Bravyi2006} which is a rather straightforward application of the Lieb-Robinson bound on operator spreading \cite{Lieb1972} and is irrelevant to symmetries, the ES is not a conventional observable and symmetries play a crucial role in protecting 1D topological phases. Nevertheless, we find that the entanglement gap can be rigorously bounded by some correlation-related quantities, which obey the Lieb-Robinson bound. The central idea is based on the powerful \emph{Weyl's perturbation theorem} \cite{Bhatia1997}, which guarantees the spectral shift between two Hermitian operators to be rigorously bounded by the operator norm of their difference. In addition to the several well-known results such as the entanglement area law \cite{Plenio2005,Wolf2006,Wolf2008,Eisert2010,Nayak2013}, the bounds on entanglement growth \cite{Osborne2007,Verstraete2013} and the entanglement detection of topological phases \cite{Vishwanath2010,Fidkowski2010,Hughes2011,Chang2014,Haldane2008,Pollmann2010,Thomale2010,Pollmann2011,Fidkowski2011,Cirac2011b,Wen2006,Kitaev2006}, our work brings about yet another rigorous and fundamental result on the entanglement in quantum many-body systems \cite{Vedral2008,Latorre2009,Laflorencie2016}. As shown Fig.~\ref{fig2}, our work lays a corner stone for exploring \emph{exact results} on the entanglement properties of SPT systems out of equilibrium. Also, we hope that the ideas and methods developed in this work could stimulate further exploration of rigorous results on various spectra, which appear ubiquitously in physics. 

This paper is structured as follows. In Sec.~\ref{pre}, we review the basic properties of the ES and the notion of dynamical symmetry breaking. In Sec.~\ref{main}, we define the entanglement gap and present the main results, i.e., the explicit forms of Lieb-Robinson bounds. In Sec.~\ref{FF}, we focus on the single-particle entanglement gap in quenched free-fermion systems and derive the first main result (Theorem~\ref{ThmspEG}). In particular, we derive an almost optimal Lieb-Robinson bound for free-fermion systems and justify the quasi-particle picture. In Sec.~\ref{Interaktion}, we focus on the many-body entanglement gap in the tensor-network setting and derive the second main result (Theorem~\ref{MBEG}). We discuss the impact of partial symmetry breaking, the effects of disorder and long-time dynamics in Sec.~\ref{Diskussionen}. Finally, we summarize the main results of this paper and provide some outlook in Sec.~\ref{Zusammenfassung}. We relegate some technical details to appendices to avoid digressing from the main subject. %Appendix A ....

\section{Preliminaries}
\label{pre}
We begin by clarifying the definitions of single-particle and many-body ES for free-fermion and general systems. We also briefly review why 1D SPT order renders the many-body ES to be degenerate. Finally, we review the notion of dynamical symmetry breaking and point out the relevant symmetries that can protect SPT order in quench dynamics. %provide the concrete setup and the main theorems, which will be proved in the following sections.

\subsection{Definition of the ES}
\label{DefES}
We consider a 1D lattice with the total number of unit cells denoted as $L$. Each unit cell contains $d$ internal degrees of freedoms, including spins, orbitals, sublattices and so on. For spin systems, $d$ gives the Hilbert-space dimension of a single unit cell. Given a local basis $\{|j\rangle\}^d_{j=1}$, an arbitrary many-body wave function can be expressed as a superposition of Fock states $|j_1j_2...j_L\rangle\equiv|j_1\rangle\otimes|j_2\rangle\otimes...\otimes|j_L\rangle$ with $j_s=1,2,...,d$ for $\forall s=1,2,...,L$. For fermionic systems, the local Hilbert-space dimension is $2^d$ since each mode can be either occupied or unoccupied. A complete fermionic Fock basis is given by $\prod_{\{n_{ja}\}}(c^\dag_{j a})^{n_{ja}}|{\rm vac}\rangle$ with $n_{ja}=0,1$ for $\forall j=1,2,...,L$ and $a=1,2,...,d$, where $|{\rm vac}\rangle$ is the Fock vacuum and $c^\dag_{ja}$ creates a fermion with internal state $a$ in the $j$th unit cell and satisfies $\{c^\dag_{j'a'},c_{ja}\}=\delta_{j'j}\delta_{a'a}$ and $\{c_{j'a'},c_{ja}\}=0$.

\begin{figure}
\begin{center}
 \includegraphics[width=8.5cm, clip]{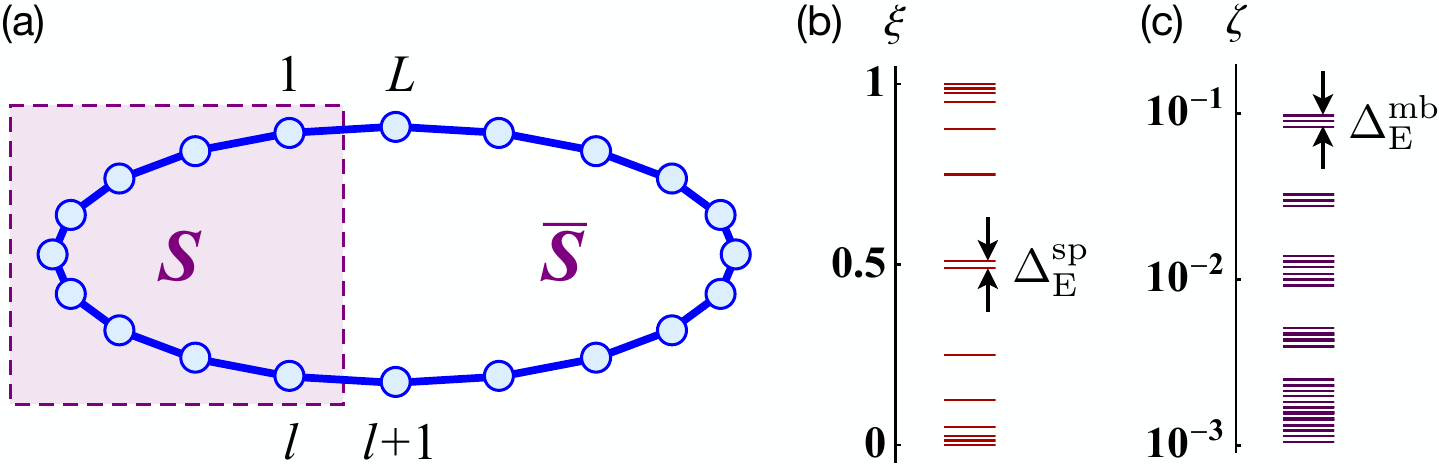}
       \end{center}
   \caption{(a) Entanglement bipartition in a 1D lattice subject to the periodic boundary condition. The reduced density operator $\rho_S$ is obtained by tracing out the degrees of freedom in $\bar S$. The total length and the length of subsystem $S$ are denoted as $L$ and $l$, respectively. (b) Typical single-particle ES of a topological particle-hole symmetric system. The single-particle entanglement gap $\Delta^{\rm sp}_{\rm E}$ (\ref{spEGdef}) is defined as the splitting between the two modes closest to one half in the ES. (c) Typical many-body ES of an SPT system. The many-body entanglement gap $\Delta^{\rm mb}_{\rm E}$ (\ref{mbEGdef}) is defined to be the width of those modes in the ground-state manifold of the many-body entanglement Hamiltonian that become generate in the thermodynamic limit.}
   \label{fig3}
\end{figure}

To study the bipartite entanglement properties, we divide the entire system into two subsystems $S$ and $\bar S$ (see Fig.~\ref{fig3}), which consists of $l$ and $L-l$ adjacent unit cells, respectively. For convenience, we label the unit cells in $S$ as $1,2,...,l$. Given a many-body wave function $|\Psi\rangle$, we can always define the \emph{many-body ES} %is simply defined 
as the eigenvalues of the reduced density operator
\begin{equation}
\rho_S=\Tr_{\bar S}|\Psi\rangle\langle\Psi|.
\end{equation}
%where $S$ is the subsystem of our interest and $\bar S$ is its complement (see Fig.~\ref{fig1}(a)). 
For simplicity, we require the ES to be positive since we can always truncate the Hilbert space of $S$ into a subspace, within which $\rho_S$ is positive definite. In particular, for free-fermion systems, $\rho_S$ is a \emph{Gaussian state} \cite{Bravyi2005} in the sense that there exists a quadratic entanglement Hamiltonian 
%\begin{equation}
$H_S=\sum^{l}_{j,j'=1}\sum^d_{a,a'=1} [H_S]_{j'a',ja}c^\dag_{j'a'} c_{ja}$ 
%(J_{j'a',ja}c^\dag_{j'a'} c_{ja}+F_{j'a',ja}c^\dag_{j'a'} c^\dag_{ja}+{\rm H.c.})
%\label{EH}
%\end{equation}
\footnote{Here we have tacitly assumed the particle-number conservation, as is the usually case for various quantum simulators. The generalization to the case with pairing terms is straightforward.} such that
%If $|\Psi\rangle$ is Gaussian, which is the case of free-fermion or harmonic-lattice systems, we can introduce a quadratic entanglement Hamiltonian $H_S=\sum_n\epsilon_na^\dag_na_n$ such that
\begin{equation}
\rho_S=\frac{e^{-H_S}}{\Tr e^{-H_S}}.
\end{equation}
Diagonalizing the entanglement Hamiltonian %(\ref{EH}) 
as $H_S=\sum^{ld}_{n=1} \epsilon_n f^\dag_n f_n$, where $f_n$'s are related to $c_{ja}$'s via a unitary transformation,  we can identify the \emph{single-particle ES} as \cite{Peschel2003} 
\begin{equation}
\xi_n=\frac{1}{e^{\epsilon_n}+1},\;\;\;\;n=1,2,...,ld,
\end{equation}
%where ``$+/-$" refers to fermions/bosons  
in terms of which the many-body ES reads
\begin{equation}
\zeta_{\{s_n\}}=\prod_{\{n:0<\xi_n<1\}}\left[\frac{1}{2}+\frac{1}{2}s_n(1-2\xi_n)\right], 
\label{ESspmb}
\end{equation}
where $s_n=\pm1$.

\subsection{SPT-enforced ES degeneracy} 
\label{ESdeg}
For free-fermion systems, it is known that $\{1-2\xi_n\}_n$ is \emph{exactly} the energy spectrum subject to the open boundary condition after band flattening \cite{Fidkowski2010}. To see this, we consider a gapped quadratic Hamiltonian in the diagonalized form $H=\sum_n E_n\psi^\dag_n\psi_n$, which may not correspond to a translation-invariant system. Assuming the Fermi energy to be zero without loss of generality, we can define the single-particle projector onto the Fermi sea as 
\begin{equation}
P_<\equiv\sum_{\{n:E_n<0\}}|\psi_n\rangle\langle\psi_n|,
\label{FSproj}
\end{equation}
where $|\psi_n\rangle\equiv\psi^\dag_n|{\rm vac}\rangle$ is a single-particle eigenstate. With $P_S$ denoted as the projector onto the single-particle Hilbert space of $S$, the single-particle ES $\{\xi_n\}_n$ coincides with the spectrum of $P_SP_<P_S$ \cite{Peschel2003}. The statement made in the beginning of this subsection follows from the fact that the involutory (i.e., being its own inverse) flattened Hamiltonian is given by 
\begin{equation}
H^{\rm flat}=\mathbb{I}^{\rm sp}-2P_<, 
\label{Hflat}
\end{equation}
where $\mathbb{I}^{\rm sp}$ is the identity operator in the single-particle sector. From this exact correspondence,  we know that there should be $\xi_n=\frac{1}{2}$ modes in the single particle ES for a topological insulator with boundary zero modes. According to Eq.~(\ref{ESspmb}), the corresponding many-body ES is necessarily degenerate.

To analyze noncritical interacting systems in 1D, we employ the \emph{matrix-product-state} (MPS) formalism \cite{Werner1992,Perez2007,Verstraete2008,Orus2014}. The validity of this formalism is rooted in the entanglement area law \cite{Hastings2007}. For simplicity, we focus on translation-invariant MPSs, which take the form of
\begin{equation}
|\Psi\rangle=\sum_{\{j_s\}^L_{s=1}}\Tr[A_{j_1}A_{j_2}...A_{j_L}]|j_1j_2...j_L\rangle,
\label{MPS}
\end{equation} 
where $j_s=1,2,..,d$ and $A_j$'s are $D\times D$ matrices with $D$ being the bond dimension. Each MPS is associated with a linear map on $\mathbb{C}^{D\times D}$:
\begin{equation}
\mathcal{E}(\cdot)\equiv\sum^d_{j=1} A_j(\cdot )A^\dag_j.
\label{unicha}
\end{equation}
Assuming the MPS to be \emph{normal} \cite{Cirac2017b}, which rules out the possibility of spontaneous symmetry breaking, we can always perform gauge transformations of $A_j$'s, i.e., $A_j\to XA_jX^{-1}$ that leaves $|\Psi\rangle$ in Eq.~(\ref{MPS}) invariant, such that $\mathcal{E}$ is a \emph{unital channel}: 
\begin{equation}
\mathcal{E}(\mathbb{1}_{\rm v})=\sum^d_{j=1}A_jA^\dag_j=\mathbb{1}_{\rm v} 
\end{equation}
where $\mathbb{1}_{\rm v}$ is the identity in $\mathbb{C}^{D\times D}$ acting on the \emph{virtual} Hilbert space.
%\begin{equation}
%\sum^d_{j=1}A_jA^\dag_j=\mathbb{1}_{\rm v}, 
%\label{Ajunital} 
%\end{equation}
%where $\mathbb{1}_{\rm v}$ denotes the identity in $\mathbb{C}^{D\times D}$. By defining a linear map on $\mathbb{C}^{D\times D}$ associated with $A_j$'s as
%\begin{equation}
%\mathcal{E}(\cdot)\equiv\sum^d_{j=1} A_j(\cdot )A^\dag_j,
%\end{equation}
%we can simplify Eq.~(\ref{Ajunital}) into $\mathcal{E}(\mathbb{1}_{\rm v})=\mathbb{1}_{\rm v}$, i.e., $\mathcal{E}$ is a unital channel. 
If we further impose an on-site unitary symmetry, i.e., $\rho^{\otimes L}_g|\Psi\rangle=|\Psi\rangle$ for $\forall g\in G$ with $G$ being a group and $\rho_g\in{\rm U}(d)$ %\mathbb{C}^{d\times d}$ 
being a unitary representation of $G$, we can find a \emph{projective representation} $V_g$ with $V_gV_h=\omega_{g,h}V_{gh},\;\omega_{g,h}\in{\rm U}(1)$ for $\forall g,h\in G$ such that \cite{Perez2008}
\begin{equation}
\sum^d_{j'=1}[\rho_g]_{jj'}A_{j'}=V^\dag_gA_jV_g,\;\;\;\;\forall j=1,2,...,d.
\end{equation}
In the thermodynamic limit of subsystem $S$ with length $l$, i.e., $\lim_{l\to\infty}\lim_{L\to\infty}$, the many-body ES is exactly given by $\{\lambda_\alpha\lambda_\beta\}^D_{\alpha,\beta=1}$, where $\{\lambda_\alpha\}^D_{\alpha=1}$ is the spectrum of $\Lambda$ ($>0$), that is uniquely determined from \cite{Perez2007}
\begin{equation}
\mathcal{E}^\dag(\Lambda)\equiv\sum^d_{j=1}A^\dag_j\Lambda A_j=\Lambda,\;\;\;\;
\Tr\Lambda=1. 
\label{Lam}
\end{equation}
If $|\Psi\rangle$ is in an SPT phase, $\omega_{g,h}$ must correspond to a nontrivial element in the second cohomology group $H^2(G,{\rm U}(1))$ \cite{Chen2011,Chen2011b,Schuch2011}, leading to degeneracy in $\lambda_\alpha$'s \cite{Pollmann2010}. This is because a nondegenerate $\lambda_\alpha$ implies that $\omega_{g,h}$ must be trivial, as will be proved in Sec.~\ref{EESD}.

\subsection{Dynamical symmetry breaking}
While there is a huge variety of symmetries, we can classify them into four groups depending on whether the symmetry operator $S$ commutes or anti-commutes with the Hamiltonian, and whether it is unitary or anti-unitary:
\begin{equation}
SHS^{-1}=a H,\;\;\;\; SiS^{-1}=b i,
\label{SHab}
\end{equation}
where $a,b=\pm$. Concretely, $S$ is said to be symmetric (an anti-symmetric) if $a=+$ ($a=-$), and $S$ is said to be unitary (anti-unitary) if $b=+$ ($b=-$). Note that $H$ in Eq.~(\ref{SHab}) can be either on the single-particle level (for free-fermion systems) or on the many-body level (for interacting systems).

\begin{table}[tbp]
\caption{Dynamical stability of unitary symmetries ($a=b=+$), anti-unitary symmetries ($a=-b=+$), unitary anti-symmetries ($a=-b=-$), and anti-unitary anti-symmetries ($a=b=-$), where $a$ and $b$ are given in Eq.~(\ref{SHab}).}
\begin{center}
\begin{tabular}{cccc}
\hline\hline
\;\;\;\;$a$\;\;\;\; & \;\;\;\;$b$\;\;\;\; & Dynamical stability & \;\;\;\;\;\;\;\;\;\;\;\;Example\;\;\;\;\;\;\;\;\;\;\;\;  \\
\hline
$+$ & $+$ & $\surd$ & Parity symmetry \\
$+$ & $-$ & $\times$ & Time-reversal symmetry \\
$-$ & $+$ & $\times$ & Chiral symmetry \\
$-$ & $-$ & $\surd$ & Particle-hole symmetry \\
\hline\hline
\end{tabular}
\end{center}
\label{table1}
\end{table}

Now let us impose Eq.~(\ref{SHab}) to both $H_0$ and $H$, i.e., the Hamiltonians before and after a quench. Regarding the symmetry action on the parent Hamiltonian (\ref{Ht}), we have
\begin{equation}
\begin{split}
SH(t)S^{-1}&=Se^{-iHt}H_0e^{iHt}S^{-1} \\
&=e^{-iabHt}aH_0e^{iabHt}=aH(ab t).
\end{split}
\end{equation}
Accordingly, when $ab=1$, which means that $S$ is either unitary and symmetric or anti-unitary and anti-symmetric, we have $[S,H(t)]=0$ for $\forall t$. Otherwise, $H(t)$ no longer respects $S$ in general due to the fact that $H(t)$ generally differs from $H(-t)$. This phenomenon is dubbed \emph{dynamical symmetry breaking}, which reduces the number of symmetries relevant to SPT orders in quench dynamics \cite{Cooper2018b}. See Table~\ref{table1} for a brief summary.

\section{Main results}
\label{main}
%Having in mind the definitions of ES and their relation to SPT order, we are ready to introduce 
We are now in a position to present the exact statement of our main results --- Lieb-Robinson bounds on entanglement gaps. 

\begin{figure}
\begin{center}
 \includegraphics[width=7cm, clip]{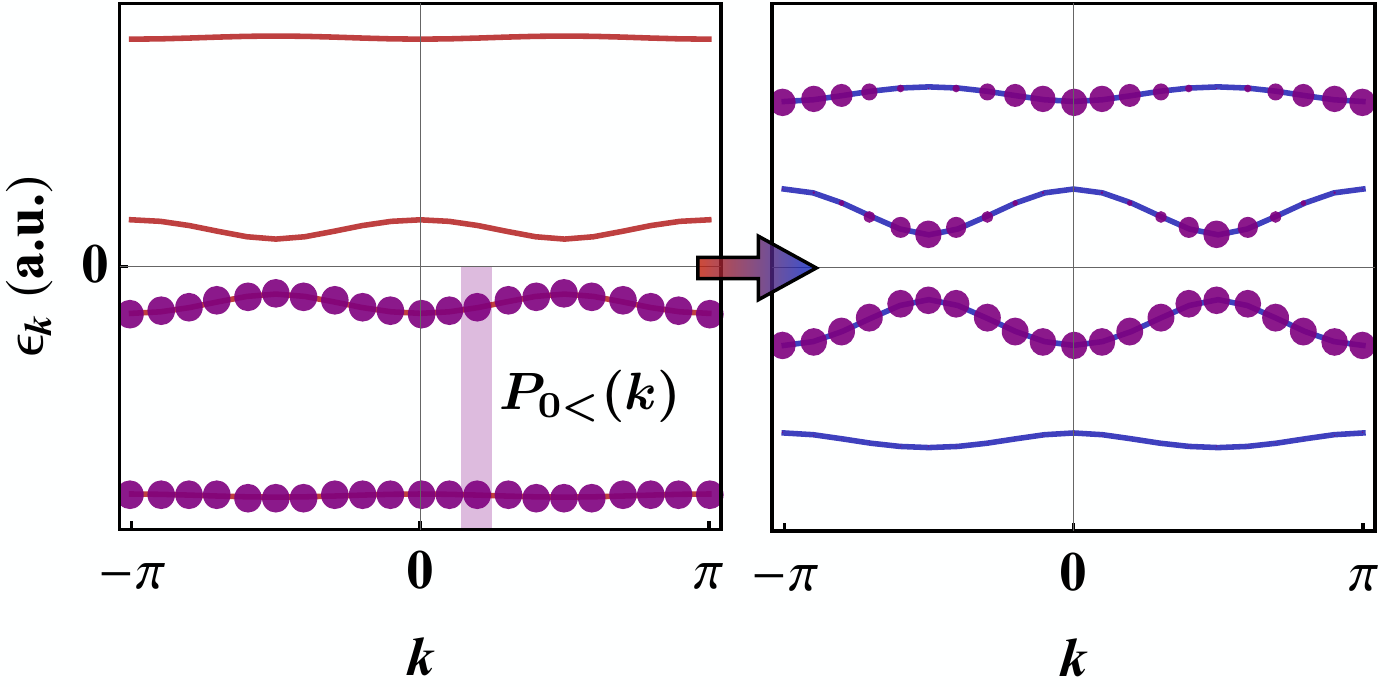}
       \end{center}
   \caption{Setting of Theorem~\ref{ThmspEG} --- quench in a free-fermion system with particle-hole symmetry. Initially, only the particle bands ($\epsilon_k<0$) are occupied and the Bloch projector is denoted as $P_{0<}(k)$ (see Eq.~(\ref{Blochproj})). The initial state is, in general, a highly excited state of the postquench Hamiltonian, whose hole bands ($\epsilon_k>0$) can be significantly occupied.}
   \label{fig4}
\end{figure}

For free-fermion systems, we examine all the Altland-Zirnbauer classes \cite{Altland1997}. In 1D, it is known that there are five nontrivial classes, including one complex class AIII and four real classes BDI, D, DIII and CII \cite{Ryu2008,Kitaev2009}. According to the previous analysis on dynamical symmetry breaking (especially Table~\ref{table1}), we know that both classes BDI and DIII reduce to class D, class CII reduces to class C and class AIII reduces to class A. Since classes C and A are both trivial, it suffices to consider \emph{class D}, which has an involutory particle-hole symmetry and is characterized by a $\mathbb{Z}_2$ topological number in 1D \cite{Kitaev2001}. This is the only Altland-Zirnbauer class in 1D that does not suffer from dynamical symmetry breaking so that the SPT order persists in the thermodynamic limit \cite{Cooper2018b}.

Since the energy spectrum of a free-fermion system in class D is paired as $(\epsilon,-\epsilon)$, the ES should be divided as $(\xi,1-\xi)$. Hence, we can define the single-particle entanglement gap as
\begin{equation}
\Delta^{\rm sp}_{\rm E}\equiv2\min_n\left|\xi_n-\frac{1}{2}\right|.
\label{spEGdef}
\end{equation}
In the presence of translation invariance, we can simply deal with the Bloch Hamiltonians and prove the following theorem.
\begin{theorem}[Free fermions]%[Lieb-Robinson bound on the single-particle EG]
%Consider a gapped infinite 1D translation-invariant lattice in class D. %with $d$ internal states per site.
Consider two 1D translation-invariant lattice systems in class D, whose Bloch Hamiltonians are given by $H_0(k)$ and $H(k)$ and the former is gapped and topologically nontrivial. We start from the ground state of $H_0(k)$ and make the following two assumptions: (i) the initial Bloch projector 
\begin{equation}
P_{0<}(k)\equiv\oint_{\gamma_<}\frac{dz}{2\pi i}\frac{1}{z-H_0(k)}
\label{Blochproj}
\end{equation}
is analytic on $\{k:|{\rm Im} k|\le\kappa,|{\rm Re} k|\le\pi\}$, where $\kappa>0$ and $\gamma_<$ is a loop that encircles all the particle bands; (ii) $H(k+i\kappa)$ is well-defined and diagonalizable for $\forall k\in[-\pi,\pi]$, i.e., 
\begin{equation}
H(k+i\kappa)=\sum^d_{\alpha=1}\epsilon_{k+i\kappa,\alpha}|u^{\rm R}_{k+i\kappa,\alpha}\rangle\langle u^{\rm L}_{k+i\kappa,\alpha}|,
\label{HnonHermi}
\end{equation}
where $|u^{\rm R}_{k+i\kappa, \alpha}\rangle$ ($\langle u^{\rm L}_{k+i\kappa, \alpha}|$) is the right (left) Bloch eigenstate of the $\alpha$-th band. Then the single-particle entanglement gap (\ref{spEGdef}) of a length-$l$ segment is upper bounded by
\begin{equation}
\Delta^{\rm sp}_{\rm E}\le%\frac{C\sinh^2\frac{\kappa l}{2}}{\sinh^2\frac{\kappa}{2}\sinh\kappa l}e^{-\kappa(l-vt)} < 
%\frac{C}{\sinh^2\frac{\kappa}{2}}e^{-\kappa (l-vt)}
%\frac{2C}{\sinh\kappa}
Ce^{-\kappa (l-vt)}
\label{spEG}
\end{equation}
during the time evolution governed by $H(k)$. %an arbitrary local and translation-invariant Hamiltonian in class D. 
Here %$C$, $\kappa$ and $v$ do not depend on $l$ or $t$. 
%Here $C$ and $v$ are given in Lemma~\ref{fimp}. %Eq.~(\ref{impvLR}).
\begin{equation}
C=%\sum_{\alpha,\beta}
\int^\pi_{-\pi}\frac{dk}{2\pi} \left(\sum^d_{\alpha=1} \| |u^{\rm R}_{k+i\kappa,\alpha}\rangle\| \||u^{\rm L}_{k+i\kappa,\alpha}\rangle\|\right)^2 \|P_{0<}(k+i\kappa)\|
%|u^{\rm R}_{k+i\kappa,\beta}\rangle\langle u^{\rm L}_{k+i\kappa,\beta}| \right\|,
\label{newC}
\end{equation}
with $\|\cdot\|$ being the operator norm and 
\begin{equation}
v=\kappa^{-1}\max_{k\in[-\pi,\pi],\alpha,\beta}{\rm Im}(\epsilon_{k+i\kappa,\alpha}-\epsilon_{k+i\kappa,\beta})
\label{impvLR}
\end{equation}
depends on neither $l$ nor $t$.
\label{ThmspEG}
\end{theorem}
As illustrated in Fig.~\ref{fig4}, this theorem depends crucially on the band picture of translation-invariant free-fermion systems. We can show that a positive $\kappa$ satisfying (i) and (ii) exists under quite general assumptions (see Appendix~\ref{BTCWN}). %The generalization to finite lattices is straightforward \cite{SME}.

\begin{figure}
\begin{center}
 \includegraphics[width=8cm, clip]{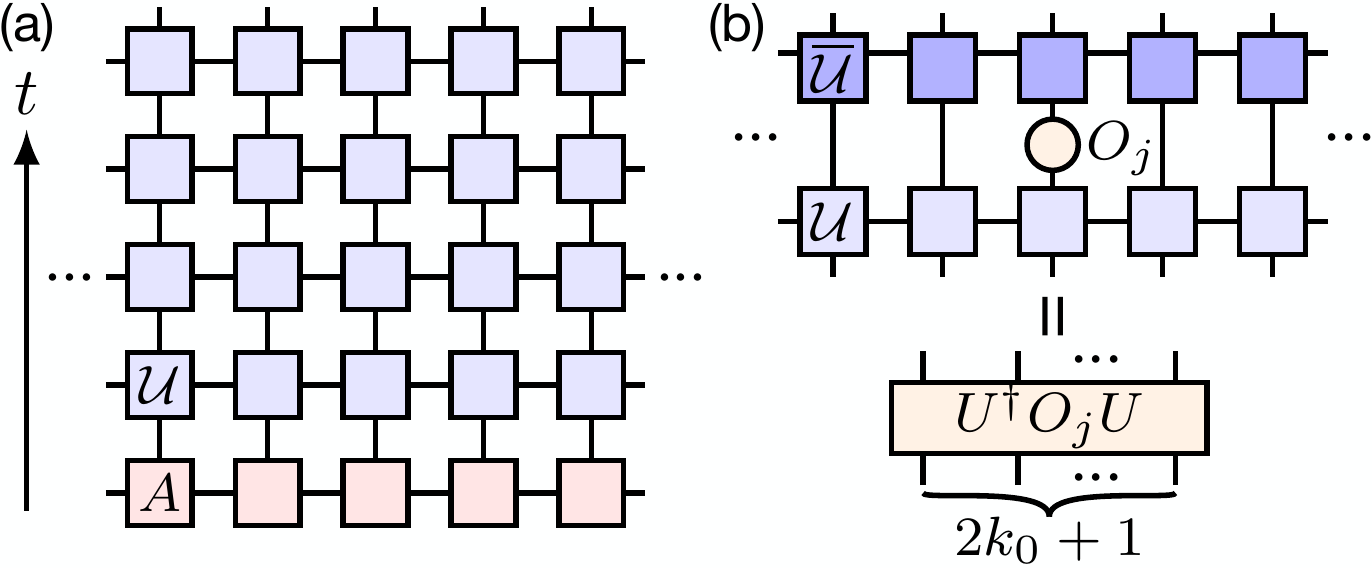}
       \end{center}
   \caption{(a) Setting of Theorem~\ref{MBEG} --- stroboscopic time evolution governed by an MPU (\ref{MPU}) generated by $\mathcal{U}_{jj'}$ starting from an MPS (\ref{MPS}) generated by $A_j$. (b) After evolution by an MPU, an on-site operator $O_j$ stays local but acts nontrivially on (at most) $2k_0+1$ sites. Here $k_0$ is the smallest integer $k$ such that the blocked tensor $\mathcal{U}_k$ in Eq.~(\ref{block}) is simple.}
   \label{fig5}
\end{figure}

For interacting SPT systems, we restrict ourselves to the tensor networks formalisms. That is, as illustrated in Fig.~\ref{fig5}(a), we start from an MPS and consider the \emph{stroboscopic} dynamics governed by a \emph{matrix-product unitary} (MPU) \cite{Po2016,Cirac2017,Gong2018c,Chen2018}
\begin{equation} 
U=\sum_{\{j_s,j'_s\}^L_{s=1}}\Tr[\mathcal{U}_{j_1j'_1}\mathcal{U}_{j_2j'_2}...\mathcal{U}_{j_Lj'_L}]|j_1j_2...j_L\rangle\langle j'_1j'_2...j'_L|,
\label{MPU}
\end{equation}
which is a special class of matrix-product operators \cite{Pirvu2010} satisfying $U^\dag U=\mathbb{1}^{\otimes L}$. %Similar to an MPS, the building block $\mathcal{U}$ of an MPU is a $D_U\times D_U$ matrix, with $D_U$ being the bond dimension. 
We assume that the MPU respects the same symmetries of the initial MPS, i.e., $[\rho^{\otimes L}_g,U]=0$ for $\forall g\in G$, and that it belongs to the trivial cohomology class so that the time-evolved MPS stays in the same SPT phase \cite{Gong2018c}. It is known that by putting together $k$ sites into one such that the building block $\mathcal{U}$ becomes $\mathcal{U}_k$:
\begin{equation}
\begin{tikzpicture}[scale=1.1]
%left
\draw[thick] (-0.15,0) -- (0.6,0) (0.225,-0.375) -- (0.225,0.375);
\draw[thick,fill=blue!10] (0,-0.225) rectangle (0.45,0.225);
\Text[x=0.25,y=0,fontsize=\footnotesize]{$\mathcal{U}_k$}
%left
\Text[x=1.1]{$\equiv$}
%right
\draw[thick] (1.6,0) -- (3.1,0) (3.65,0) -- (4.4,0);
\draw[thick] (1.975,-0.375) -- (1.975,0.375) (2.725,-0.375) -- (2.725,0.375) (4.025,-0.375) -- (4.025,0.375);
\draw[thick,fill=blue!10] (1.75,-0.225) rectangle (2.2,0.225) (2.5,-0.225) rectangle (2.95,0.225) (3.8,-0.225) rectangle (4.25,0.225);
\Text[x=3.4]{...}
\Text[x=1.975,y=0,fontsize=\footnotesize]{$\mathcal{U}$}
\Text[x=2.725,y=0,fontsize=\footnotesize]{$\mathcal{U}$}
\Text[x=4.025,y=0,fontsize=\footnotesize]{$\mathcal{U}$}
\draw[thick] (1.975,-0.45) .. controls (1.975,-0.5) and (3,-0.4) .. (3,-0.6);
\draw[thick] (4.025,-0.45) .. controls (4.025,-0.5) and (3,-0.4) .. (3,-0.6);
\Text[x=3,y=-0.75,fontsize=\footnotesize]{$k$}
%right
\Text[x=4.6,y=-0.2]{,}
\end{tikzpicture}
\label{block}
\end{equation}
the MPU can for sufficiently large $k$ be represented as a bilayer unitary circuit with each unitary operator acting on two adjacent blocked sites \footnote{Here we do not consider those anomalous MPUs with nontrivial chiral indices, since we use the MPUs to approximate genuine 1D dynamics generated by local Hamiltonians instead of the edge dynamics of 2D systems \cite{Po2016}.}: 
\begin{equation}
\begin{tikzpicture}[scale=1.2]
%left
\draw[thick] (0.3,0) -- (3.3,0);
\foreach \x in {0,...,4}
\draw[thick] (0.6*\x+0.6,-0.4) -- (0.6*\x+0.6,0.4);
\foreach \x in {0,...,4}
\draw[thick,fill=blue!10] (0.6*\x+0.4,-0.2) rectangle (0.6*\x+0.8,0.2);
\foreach \x in {0,...,4}
\Text[x=0.6*\x+0.65,fontsize=\footnotesize]{$\mathcal{U}_k$}
%left
\Text[x=3.6]{$=$}
%right
\foreach \x in {0,...,4}
\draw[thick] (0.5*\x+4.15,-0.65) -- (0.5*\x+4.15,0.65);
\foreach \x in {0,1}
\draw[thick,fill=yellow!10] (\x+4,-0.1) rectangle (\x+4.8,-0.45);
\foreach \x in {0,1}
\Text[x=\x+4.4,y=-0.275]{$u$}
\fill[yellow!10] (6,-0.1) rectangle (6.4,-0.45);
\draw[thick] (6.4,-0.1) -- (6,-0.1) -- (6,-0.45) -- (6.4,-0.45);
\foreach \x in {0,1}
\draw[thick,fill=orange!10] (\x+4.5,0.1) rectangle (\x+5.3,0.45);
\foreach \x in {0,1}
\Text[x=\x+4.9,y=0.275]{$v$}
\fill[orange!10] (3.9,0.1) rectangle (4.3,0.45);
\draw[thick] (3.9,0.1) -- (4.3,0.1) -- (4.3,0.45) -- (3.9,0.45);
%right
\Text[x=6.6,y=-0.2]{.}
\end{tikzpicture}
\end{equation}
Whenever such a representation is possible, we call the building-block tensor \emph{simple} \cite{Cirac2017}. In fact, given $\mathcal{U}$ and $k$ such that $\mathcal{U}_k$ is simple, $\mathcal{U}_{k'}$ is also simple for $\forall k'\ge k$. The smallest $k$ that makes $\mathcal{U}_k$ simple, which we denote as $k_0$, has a clear physical interpretation as the \emph{Lieb-Robinson length} --- as shown in Fig.~\ref{fig5}(b), any on-site operator evolved by the MPU generated by $\mathcal{U}$ acts nontrivially on at most $2k_0+1$ sites. Further details on MPUs can be found in Appendix~\ref{BPMPU}.

Unlike free-fermion systems in class D, which are characterized by a $\mathbb{Z}_2$ number so that there is at most one pair of stable topological entanglement modes near $\xi=\frac{1}{2}$ (leading to $2^2=4$-fold degeneracy in the many-body ES), the many-body ES of an SPT MPS can be $r^2$-fold degenerate for $\forall r=2,3,4,...$ in the thermodynamic limit. Here the square $r^2$ arises from the fact that a subsystem has %there are 
two edges. %under the periodic boundary condition. 
Minimal symmetries that realize these SPT MPSs are $\mathbb{Z}_r\times\mathbb{Z}_r$, whose second-order cohomology groups read $H^2(\mathbb{Z}_r\times\mathbb{Z}_r,{\rm U}(1))=\mathbb{Z}_r$. Given $r$ and a many-body ES $\{\zeta_n\}_n$ with $\zeta_{n}\ge\zeta_{n+1}$ (note that larger $\zeta$ corresponds to lower eigenvalue of the entanglement Hamiltonian), we define the many-body entanglement gap to be
\begin{equation}
\Delta^{\rm mb}_{\rm E}\equiv|\zeta_1-\zeta_{r^2}|.
\label{mbEGdef}
\end{equation}
Our main theorem is the following. %regarding this many-body entanglement gap reads
\begin{theorem}[Interacting systems]%Lieb-Robinson bound on the many-body entanglement gap]
Starting from an infinite SPT MPS with bond dimension $D$, the many-body entanglement gap (\ref{mbEGdef}) of a length-$l$ subsystem after $t$ steps of time evolution by a trivial symmetric MPU generated by $\mathcal{U}$ with bond dimension $D_U$ is bounded from above by
\begin{equation}
%\|\mathcal{E}^l_t-\mathcal{E}_\infty\|\le
%\max\{{\rm Cor}(X:Y)_t,\Delta_{{\rm S},t}\}
\Delta^{\rm mb}_{\rm E}\le C(l-2k_0t)^{D^2-1}e^{-\kappa(l-vt)}
\label{LRSG}
\end{equation}
for any $l-2k_0t\ge\frac{1+\mu}{1-\mu}$. Here $k_0$ is the smallest integer blocking number that makes $\mathcal{U}_{k_0}$ simple and $\mu$ is the spectrum radius of $\mathcal{E}-\mathcal{E}^{\infty}$, where $\mathcal{E}$ is defined in Eq.~(\ref{unicha}) for the initial MPS and $\mathcal{E}^\infty\equiv\lim_{n\to\infty}\mathcal{E}^n$, $\kappa=-\ln\mu$, $v=2k_0-\frac{\ln D_U}{\ln\mu}$, and the coefficient
\begin{equation}
C=4e^2D^2(D^2+1)\mu^{1-D^2}(1+\mu)^{D^2+\frac{1}{2}}(1-\mu)^{D^2-\frac{5}{2}} 
\label{Koeffizient}
\end{equation}
depends only on the initial MPS. 
%second largest eigenvalue of the unital channel associated with the initial MPS. 
\label{MBEG}
\end{theorem}

Let us explain why we focus on the stroboscopic dynamics generated by an MPU rather than a continuous dynamics generated by a local Hamiltonian. First, we expect this setting to be good enough because we can efficiently approximate a finite-time evolution generated by a local Hamiltonian as a bilayer unitary circuit \cite{Osborne2006}, which is equivalent to an MPU or a quantum cellular automaton \cite{Cirac2017}. The efficiency is ensured by the conventional Lieb-Robinson bound. By showing that the spectral shift is rigorously bounded by the approximation error, we expect a similar Lieb-Robinson bound on the many-body entanglement gap for continuous quench dynamics %from the discretized dynamics 
(see Appendix~\ref{ISCE}). Second, this formalism is of intrinsic interest for its own sake --- it gives exact descriptions for some Floquet systems \cite{Curt2016,Curt2016b,Khemani2016} and quantum circuits \cite{Nahum2017,Nahum2018,Curt2018,Curt2018c,Khemani2018,Chan2018b,Christoph2018,Pai2019,Tibor2019}, which have intensively been studied in the context of nonequilibrium phases of matter %entanglement growth 
and information scrambling. Moreover, this theorem exemplifies the power of tensor networks as \emph{analytical} methods for predicting long-time dynamical behaviors of interacting quantum many-body systems far from equilibrium, which are hardly accessible in numerics.

\section{Free fermions}
\label{FF} 
\begin{figure}
\begin{center}
 \includegraphics[width=8.5cm, clip]{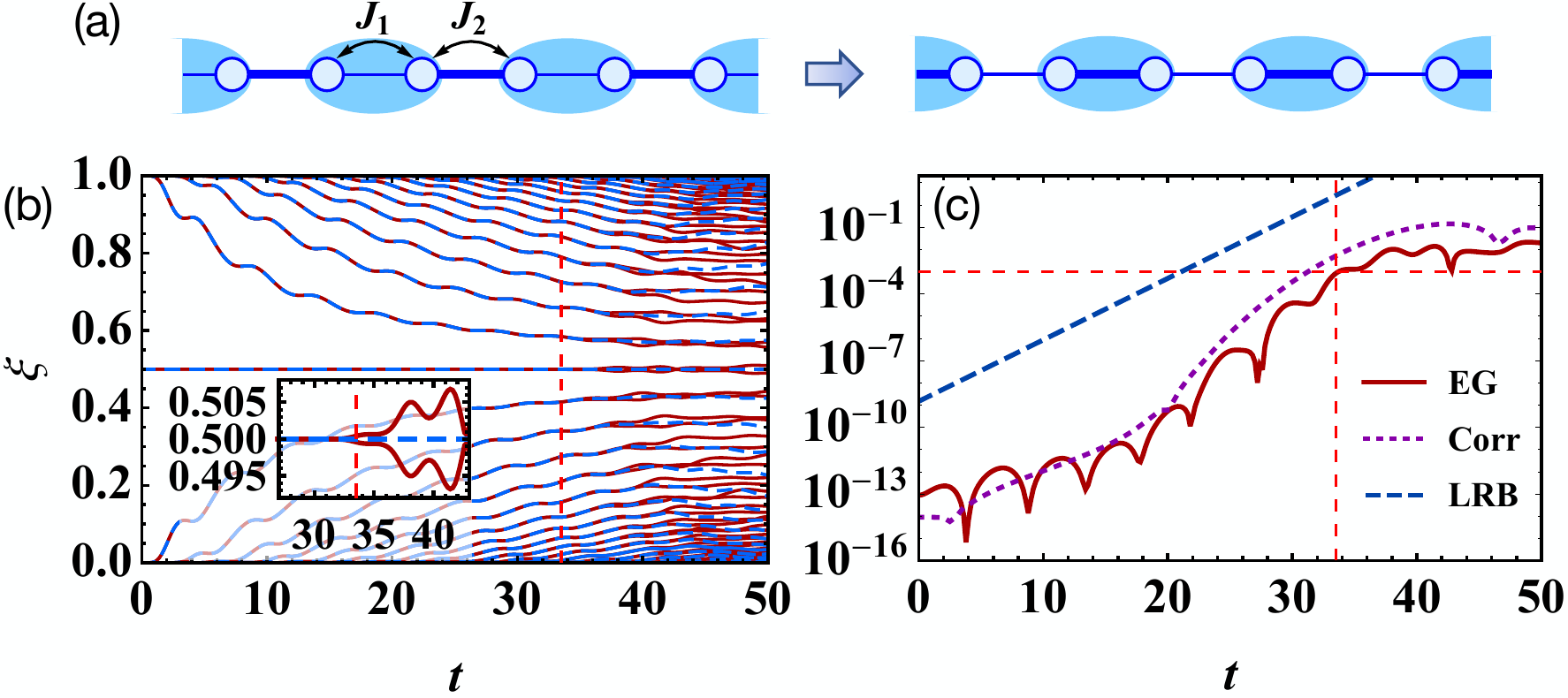}
       \end{center}
   \caption{(a) Quench protocol in the SSH model. (b) Single-particle ES dynamics after the quench. The parameters are quenched as $(J_1,J_2)=(0.5,1)\to(1,0.5)$ and $l=40$. The red solid and blue dashed curves correspond to $L=\infty$ and $L=2l$, respectively. In the former case, the single-particle ES splits after $t^*\sim33$ (indicated by the red dashed line and determined by threshold $\Delta^{\rm sp}_{\rm E}=10^{-3}$ as shown in (c)). Inset: Zoom in on the splitting of the topological entanglement modes. (c) Dynamics of the single-particle entanglement gap for $L=\infty$ (solid curve). The equal-time correlation $\|\langle j|P^\infty_<(t)|j+l\rangle\|$ (purple dotted curve) and the Lieb-Robinson bound (blue dashed curve) given by Theorem~\ref{ThmspEG} with $\kappa=0.6$ are superimposed for the sake of comparison.}
   \label{fig6}
\end{figure}

As mentioned above, for quench dynamics within the same Altland-Zirnbauer class \cite{Altland1997,Ryu2008,Kitaev2009}, the only nontrivial class in 1D that does not suffer from (partial) ``dynamical symmetry breaking" is class D \cite{Cooper2018b}. In fact, previous numerical studies on the Su-Schrieffer-Heeger (SSH) model \cite{Chung2013} and the Kitaev chain \cite{Chung2016} have revealed that the $\xi=\frac{1}{2}$ %the topological entanglement 
modes split after a characteristic time scale $t^*\sim\frac{l}{2v_{\max}}$, where $l$ is the length of the subsystem and $v_{\max}$ is the maximal group velocity of band dispersions. 
In Fig.~\ref{fig6}(b), we reproduce the splitting dynamics in the SSH model (Fig.~\ref{fig6}(a)), which shows that not only the topological entanglement mode but also the \emph{full} ES splits. In the following, we rigorously establish the underlying Lieb-Robinson bound on the ES splitting stated in Theorem~\ref{ThmspEG}.

\subsection{%Persistence of 
Zero entanglement gap for symmetric bipartition}
\label{zeroEG}
We first point out a crucial proposition --- for half-chain entanglement cut, i.e., $l=\frac{L}{2}$, the topological entanglement modes will be pinned \emph{exactly} at $\frac{1}{2}$ at any time %even in the long-time limit 
(see Fig.~\ref{fig6}(b)). Note that we can always choose the anti-unitary and involutory particle-hole-symmetry operator %, we can always choose the basis properly such that 
$\mathcal{C}$ to be %is represented as 
the complex conjugation $\mathcal{K}$ \cite{Wigner1960}. Under this basis, we have $H=iR$ with $R$ being a skew-symmetric real matrix. Suppose that $H$ is flattened so that $R^2=-\mathbb{I}^{\rm sp}$. If $\Pf R=-1$, where $\Pf$ denotes the Pfaffian, there will be a pair of topological entanglement modes exponentially close to $\xi=\frac{1}{2}$. With the translation invariance assumed, the system at least exhibits a \emph{half-chain translation symmetry}, leading to 
\begin{equation}
R=\begin{bmatrix} R_{\rm d} & R_{\rm o} \\ R_{\rm o} & R_{\rm d} \end{bmatrix}=\sigma^0\otimes R_{\rm d}+\sigma^x\otimes R_{\rm o}, 
\label{R}
\end{equation}
where $R_{\rm d}$ and $R_{\rm o}$ are both real and skew-symmetric and $\sigma^0=\begin{bsmallmatrix} 1 & 0 \\ 0 & 1 \end{bsmallmatrix}$, $\sigma^x=\begin{bsmallmatrix} 0 & 1 \\ 1 & 0 \end{bsmallmatrix}$. Moreover, the half-chain ES is given by the spectrum of $\frac{1}{2}(1-iR_{\rm d})$. From $R^2=-\mathbb{I}^{\rm sp}$ we obtain 
\begin{equation}
R^2_{\rm d}+R^2_{\rm o}=-\mathbb{I}^{\rm half},\;\;\;\;\{R_{\rm d},R_{\rm o}\}=\mathbb{0}^{\rm half}, 
\end{equation}
where $\mathbb{I}^{\rm half}$ and $\mathbb{0}^{\rm half}$ are the half-chain identity and zero operators within the single-particle sector. Provided that $R^2_{\rm o}<0$, we can find an anti-unitary operator 
\begin{equation}
\mathcal{A}\equiv i(-R^2_{\rm o})^{-\frac{1}{2}}R_{\rm o}\mathcal{K}, 
\label{mclA}
\end{equation}
such that 
\begin{equation}
\mathcal{A}^2=-\mathbb{I}^{\rm half},\;\;\;\;
[\mathcal{A},iR_{\rm d}]=0.
\end{equation}
Due to the interplay between the \emph{Kramers degeneracy} enforced by $\mathcal{A}$ and the nontrivial $\mathbb{Z}_2$ topology, two quasi-zero modes of $iR_{\rm d}$ must be pinned exactly at zero. Since the $\mathbb{Z}_2$ index stays unchanged in quench dynamics, the topological entanglement mode should always be pinned at $\frac{1}{2}$, leading to a persistent zero single-particle entanglement gap.

Even if $R_{\rm o}$ is not invertible so that $\mathcal{A}$ in Eq.~(\ref{mclA}) becomes ill-defined, we can still show that all the eigenvalues of $iR_{\rm d}$ falling in the range $(-1,1)$ are degenerate. For an arbitrary normalized eigenvector $\phi$ with $iR_{\rm d}\phi=\epsilon\phi$, $\epsilon\in(-1,1)$, we can construct 
\begin{equation}
\tilde \phi\equiv\frac{1}{\sqrt{1-\epsilon^2}} R_{\rm o}\bar\phi, 
\end{equation}
such that $\tilde\phi^\dag\tilde\phi=1$ %=-\frac{1}{1-\epsilon^2}\overline{\phi^\dag R^2_{\rm o}\phi}=\frac{1}{1-\epsilon^2}\overline{\phi^\dag (\mathbb{1}_L+R^2_{\rm d})\phi}=1$ 
and $iR_{\rm d}\tilde\phi%-\frac{1}{\sqrt{1-\epsilon^2}} \overline{iR_{\rm d}R_{\rm o}\phi}=\frac{1}{\sqrt{1-\epsilon^2}} \overline{R_{\rm o}(iR_{\rm d}\phi)}
=\epsilon\tilde\phi$. Moreover, we have
\begin{equation}
\begin{split}
\phi^\dag\tilde\phi&=\frac{1}{\sqrt{1-\epsilon^2}}\phi^\dag R_{\rm o}\bar\phi=\frac{1}{\sqrt{1-\epsilon^2}}(\phi^\dag R_{\rm o}\bar\phi)^{\rm T} \\
&=-\frac{1}{\sqrt{1-\epsilon^2}}\phi^\dag R_{\rm o}\bar\phi=-\phi^\dag\tilde\phi,
\end{split}
\end{equation}
which means $\tilde\phi$ is orthogonal to $\phi$.

It is worth mentioning that such an emergent symmetry in ES has systematically been studied in Ref.~\cite{Chang2014}. In addition to the above physical analysis, we also provide a rigorous proof in Appendix~\ref{EZM}.

\subsection{General idea}
To highlight the finite size of a periodic lattice, we will hereafter use $P^{(L)}_<$ instead of $P_<$ in Eq.~(\ref{FSproj}) as the single-particle projector onto the Fermi sea, where $L$ denotes the number of unit cells. As mentioned in Sec.~\ref{ESdeg}, with the single-particle projector onto a subsystem $S$ denoted as $P_S$, the single-particle ES coincides with the spectrum of $P_SP^{(L)}_<P_S$ \cite{Peschel2003}. We have shown in the previous subsection that for $L=2l$ with $l$ being the length of $S$, whenever $H^{\rm flat}$ in Eq.~(\ref{Hflat}) is characterized by a nontrivial $\mathbb{Z}_2$ number protected by the particle-hole symmetry, the spectrum of $P_SP^{(L)}_<P_S$ contains two degenerate eigenstates with eigenvalue $\frac{1}{2}$ and the entanglement gap exactly vanishes. To prove Theorem~\ref{ThmspEG}, it suffices to prove that the spectral shift from $P_SP^{(2l)}_<P_S$ to $P_SP^{(\infty)}_<P_S$ is exponentially small until a time scale proportional to $l$. To this end, a natural idea is to utilize the following Weyl's perturbation theorem. 
\begin{theorem}[Weyl's perturbation theorem]
Consider two Hermitian operators $O$ and $O'$ on a finite Hilbert space. %and their sum $O'=O+E$. 
Denoting the $j$th largest eigenvalue of $O$ and $O'$ as $\lambda_j$ and $\lambda'_j$, respectively, we have
\begin{equation}
|\lambda_j-\lambda'_j|\le\|O-O'\|.
\end{equation}
\label{Weylper}
\end{theorem}
For self-containedness, we provide a brief proof in Appendix~\ref{WPT}. According to this theorem, denoting the $n$th largest eigenvalue of $P_S P^{(L)}_< P_S$ as $\xi^{(L)}_n$, we have 
\begin{equation}
|\xi^{(L)}_n-\xi^{(\infty)}_n|\le\|P_S P^{(L)}_< P_S - P_SP^{(\infty)}_<P_S\|.
\end{equation}
According to Eq.~(\ref{spEGdef}), we have the following collorary: %follows that 
for a topological class D system, the single-particle entanglement gap is bounded from above as
\begin{equation}
\Delta^{\rm sp}_{\rm E}\le 2\|P_S P^{(2l)}_< P_S - P_SP^{(\infty)}_<P_S\|.
\label{spEGbound}
\end{equation}
It is thus sufficient to find a Lieb-Robinson bound on the right-hand side (rhs) of Eq.~(\ref{spEGbound}), which measures the finite-size correction to the correlation in a finite subsystem $S$.

Before going into rigorous proofs, let us first give an intuitive argument based on the \emph{Wannier-function} picture \cite{Marzari2012}. Note that the projector onto the Fermi sea can be expressed as
\begin{equation}
P^{(L)}_<=\sum_{j\in\mathbb{Z}_L,\alpha\in\mathcal{O}}|W^{(L)}_{j\alpha}\rangle\langle W^{(L)}_{j\alpha}|, 
\label{PLW}
\end{equation}
where $|W^{(L)}_{j\alpha}\rangle$ is a Wannier function of the $\alpha$th band centered at the $j$th site, $\mathbb{Z}_L\equiv\{1,2,...,L\}$ consists of all the sites and $\mathcal{O}$ consists of all the occupied bands. Since $|W^{(L)}_{j\alpha}\rangle$ is exponentially localized in real space \cite{Kohn1959,Cloizeaux1964,Marzari2007}, we expect an exponentially small difference between $|W^{(L)}_{j\alpha}\rangle$ and $|W^{(\infty)}_{j\alpha}\rangle$. By the same token, although $P^{(\infty)}_<$ contains infinitely more Wannier-function projectors than $P^{(L)}_<$, such a difference should again become exponentially small after being projected by $P_S$, provided that both $l$ and $L-l$ are sufficiently large compared with the localization length of a Wannier function. When the system is driven out of equilibrium, we expect the Wannier function to spread no faster than linearly, leading to a light-cone behavior of $\|P_S P^{(L)}_< P_S - P_SP^{(\infty)}_<P_S\|$. We will later make this argument rigorous for translation-invariant systems. Note that the above argument seems to be equally applicable to disordered systems with exponentially localized Wannier functions \cite{Kivelson1982}.

\begin{figure}
\begin{center}
       \includegraphics[width=8cm, clip]{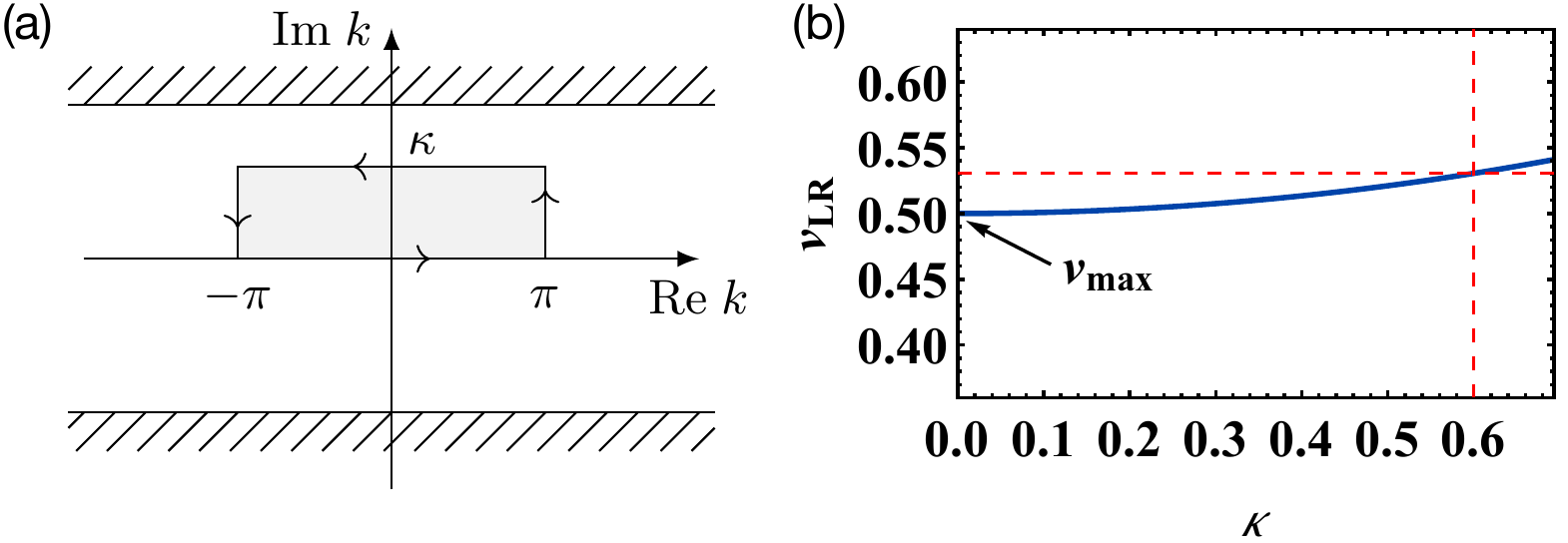}
       \end{center}
          \caption{(a) Contour deformation used in Eq.~(\ref{condef}). For sufficiently small $\kappa$, the shaded area inside the contour $\partial\{k:|{\rm Re}\;k|\le\pi,0\le{\rm Im}\;k\le\kappa\}$ contains no poles and thus $\int^\pi_{-\pi}=\int^{\pi+i\kappa}_{-\pi+i\kappa}$. Note that the two integrals along the vertical edges cancel out due to the $2\pi$-periodicity in ${\rm Re}\;k$. (b) $\kappa$ dependence of the Lieb-Robinson velocity $v_{\rm LR}=\frac{1}{2}v$ predicted by Lemma~\ref{fimp} for the SSH model with $(J_1,J_2)=(1,0.5)$. The maximal group velocity $v_{\max}=\min\{J_1,J_2\}$ naturally appears in the limit of $\kappa\to0$. The red dashed lines correspond to the specific choice $\kappa=0.6$ used in Fig.~\ref{fig6}(c). The monotonicity of $v_{\rm LR}$ with respect to $\kappa$ holds for general analytic Bloch Hamiltonians (see Appendix~\ref{monov}).}
          \label{fig7}
\end{figure}

\subsection{Lieb-Robinson bound on correlations in free-fermion systems}
An important ingredient in our proof is the conventional Lieb-Robinson bound on correlation functions. While the Lieb-Robinson bound for general interacting systems is certainly applicable to free-fermion systems, such a bound is usually very loose since it only involves a very limited amount of information about the system such as the hopping range and the maximal hopping amplitude. Here, we derive a greatly improved Lieb-Robinson bound on the correlation functions in translation-invariant free-fermion systems. Our result is almost optimal in the sense that the Lieb-Robinson velocity reaches the maximal (relative) group velocity of band dispersions in the semiclassical limit. %which we believe is of independent interest. 

A nice property of free-fermion systems is that, according to Wick's theorem, all the correlation functions can be decomposed into two-point correlation functions. Moreover, if the wave function is a particle-number eigenstate, only the correlators like $\langle c^\dag c\rangle$ are relevant. To compute such a correlator in the quench dynamics, we can use the formula
\begin{equation}
\langle\Psi_t|c^\dag_{j'a'}c_{ja}|\Psi_t\rangle=\langle j'a'|P_<(t)|ja\rangle,
\end{equation}
where $|ja\rangle=c^\dag_{ja}|{\rm vac}\rangle$ and $P_<(t)\equiv e^{-iHt}P_{0<}e^{iHt}$ is the time-evolved single-particle projector onto the Fermi sea. To measure the strength of correlation at the length scale $|j-j'|$, we can consider the norm of $\langle j|P_<(t)|j'\rangle$, which can be shown to obey a Lieb-Robinson bound given by the following lemma. 
%the correlation matrix $C_{jj'}(t)=C_{j-j'}(t)=C_{j'-j}(t)^\dag$, which is defined as
%\begin{equation}
%[C_{jj'}]_{aa'}=\langle\Psi_t|c^\dag_{j'a'}c_{ja}|\Psi_t\rangle
%\end{equation}

\begin{figure}
\begin{center}
       \includegraphics[width=7cm, clip]{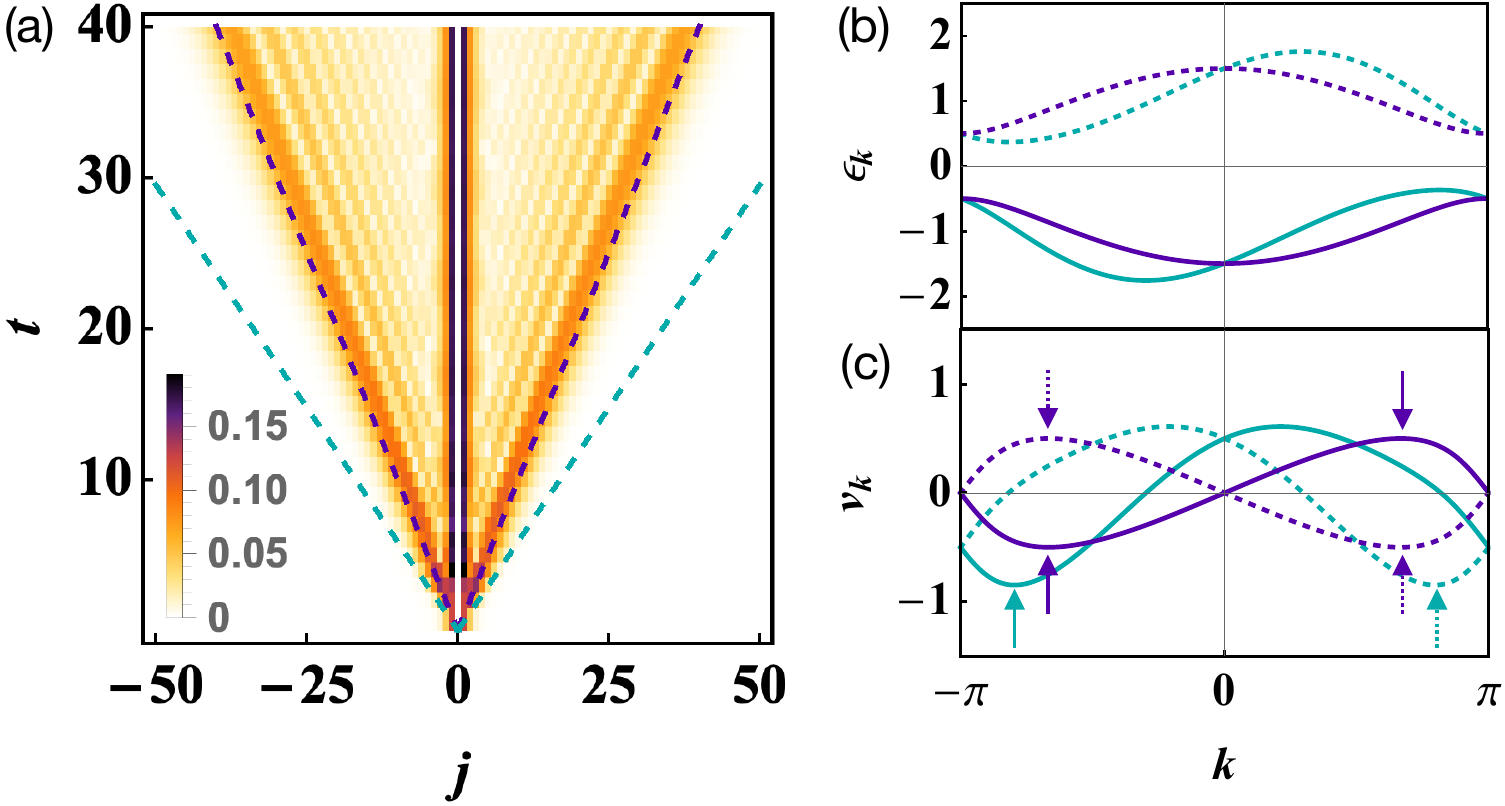}
       \end{center}
          \caption{(a) Dynamics of spatial correlations $\|\langle j|P_<(t)| 0\rangle\|$ ($j\neq0$) in an SSH chain involving $L=101$ unit cells. The parameters are the same as those in Fig.~\ref{fig6}. An additional particle-hole symmetric term $H_{\rm PHS}=\sum_{j,a=A,B}iJ(c^\dag_{j+1,a}c_{ja}-{\rm H.c.})$ does not change the dynamics of correlations. The dashed lines are given by $j=\pm v_{\max}t$, where $v_{\max}$ is the largest group velocity of the SSH model with (green) or without (purple) the additional term $H_{\rm PHS}$. (b) Band dispersions and (c) the corresponding group velocities of the SSH model (purple curves) and that with the additional term (green curves) with $J=0.5$. The arrows indicate where $v_{k\alpha}$ reaches its maximal absolute value. The solid and dotted curves correspond to the particle and hole bands, respectively.}
          \label{fig8}
\end{figure}

\begin{lemma} %[Lieb-Robinson bound on correlations] 
Consider a time-evolved projector $P^{(L)}_<(t)\equiv e^{-i Ht}P^{(L)}_{0<} e^{i Ht}$, where $P^{(L)}_{0<}$ is the projector onto the Fermi sea of $H_0$, which is gapped, and denote the Bloch Hamiltonians of $H_0$ and $H$ as $H_0(k)$ and $H(k)$, respectively. Then for $\forall \kappa>0$ such that (i) $P_{0<}(k)\equiv\sum_{\alpha\in\mathcal{O}}|u_{k\alpha}\rangle\langle u_{k\alpha}|$ (see also Eq.~(\ref{Blochproj})) is analytic over $\{k:|{\rm Re}\;k|\le\pi, |{\rm Im}\;k|\le\kappa\}$, where $|u_{k\alpha}\rangle,\alpha\in\mathcal{O}$ is the normalized Bloch eigenstate of the $\alpha$th occupied band, and (ii) $H(k+i\kappa)$ is well-defined and diagonalizable for $\forall k\in[-\pi,\pi]$, we have
\begin{equation}
\|\langle j| P^{(\infty)}_<(t) |j'\rangle\| \le Ce^{-\kappa(|j-j'|-vt)},\;\;\;\;\forall j,j'\in\mathbb{Z},%\;{\rm and}\;a,a'=1,2,...,d,
\label{Ptcorr}
\end{equation}
where $|j\rangle$ is the state localized at the $j$th site, and $C$ and $v$ are given in Eqs.~(\ref{newC}) and (\ref{impvLR}).
\label{fimp}
\end{lemma}
%The proof of Lemma~\ref{fimp} involves \emph{Cauchy's integral theorem}, a technique used to prove the exponential localization of Wannier functions \cite{Kohn1959,Cloizeaux1964,Kivelson1982,Marzari2007}. 
\emph{Proof.---} After analytic continuation, the Hermiticity constraints on the Bloch Hamiltonians are generalized to $H_0(k)^\dag=H_0(\bar k)$ and $H(k)^\dag=H(\bar k)$ (see Appendix~\ref{AC}). Accordingly, the Bloch projector $P_{0<}(k)$ satisfies
\begin{equation}
\begin{split}
P_{0<}(k)^\dag&\equiv-\oint_{\bar \gamma_<}\frac{d\bar z}{2\pi i}\frac{1}{\bar z-H_0(k)^\dag} \\&=\oint_{\gamma_<}\frac{d\bar z}{2\pi i}\frac{1}{\bar z-H_0(\bar k)}=P_{0<}(\bar k),
\end{split}
\end{equation}
%Since $\|A\|=\|A^\dag\|$ for $\forall A$, we have
%\begin{equation}
%\|P_<(k)\|=\|P_<(\bar k)\|,\;\;\;\;
%\|H'(k)\|=\|H'(\bar k)\|.
%\label{gleich}
%\end{equation}
where $\gamma_<$ is a loop that encircles the energies of all the occupied bands. Note that
\begin{equation}
\begin{split}
\langle j| P^{(\infty)}_< (t)|j'\rangle&=\int^\pi_{-\pi} \frac{dk}{2\pi} e^{ik(j-j')}P_<(k,t) \\
&=(\langle j'| P^{(\infty)}_< (t)|j\rangle)^\dag,%\;\;\;\;
\end{split}
\label{jPjp}
\end{equation}
where $P_<(k,t)\equiv e^{-iH(k)t} P_{0<}(k) e^{iH(k)t}$. This implies $\|\langle j| P^{(\infty)}_< (t)|j'\rangle\|=\|\langle j'| P^{(\infty)}_< (t)|j\rangle\|$, and therefore we can assume $j\ge j'$ without loss of generality. By deforming the contour of integration (see Fig.~\ref{fig7}(a)) and applying spectral decomposition to $H(k+i\kappa)$, which is generally non-Hermitian, we can bound the norm of Eq.~(\ref{jPjp}) from above by
\begin{widetext}
\begin{equation}
\begin{split}
&\|\langle j| P^{(\infty)}_< (t)|j'\rangle\| = e^{-\kappa|j-j'|}\left\|\int^\pi_{-\pi}\frac{dk}{2\pi}e^{ik(j-j')}e^{-iH(k+i\kappa)t} P_{0<}(k+i\kappa) e^{iH(k+i\kappa)t}\right\| \\
&= e^{-\kappa|j-j'|}\left\|\sum_{\alpha,\beta}\int^\pi_{-\pi}\frac{dk}{2\pi}e^{ik(j-j')-i(\epsilon_{k+i\kappa,\alpha}-\epsilon_{k+i\kappa,\beta})t}|u^{\rm R}_{k+i\kappa,\alpha}\rangle\langle u^{\rm L}_{k+i\kappa,\alpha}|P_{0<}(k+i\kappa) |u^{\rm R}_{k+i\kappa,\beta}\rangle\langle u^{\rm L}_{k+i\kappa,\beta}| \right\| \\
&\le\sum_{\alpha,\beta}e^{-\kappa|j-j'|}
\int^\pi_{-\pi}\frac{dk}{2\pi}e^{{\rm Im}(\epsilon_{k+i\kappa,\alpha}-\epsilon_{k+i\kappa,\beta})t}\left\| |u^{\rm R}_{k+i\kappa,\alpha}\rangle\langle u^{\rm L}_{k+i\kappa,\alpha}|P_{0<}(k+i\kappa) |u^{\rm R}_{k+i\kappa,\beta}\rangle\langle u^{\rm L}_{k+i\kappa,\beta}| \right\| \\
&\le e^{-\kappa|j-j'|+\max_{k\in[-\pi,\pi],\alpha,\beta}{\rm Im}(\epsilon_{k+i\kappa,\alpha}-\epsilon_{k+i\kappa,\beta})t}\sum_{\alpha,\beta}\int^\pi_{-\pi}\frac{dk}{2\pi}\| |u^{\rm R}_{k+i\kappa,\alpha}\rangle\langle u^{\rm L}_{k+i\kappa,\alpha}|\| \|P_{0<}(k+i\kappa)\| \||u^{\rm R}_{k+i\kappa,\beta}\rangle\langle u^{\rm L}_{k+i\kappa,\beta}|\|,
\end{split}
\label{condef}
\end{equation}
\end{widetext}
which completes the proof of Lemma~\ref{fimp}. $\square$
%Remarkably, the above proof does not rely on translation invariance and Eq.~(\ref{expdecay}) has already been justified for general 1D disordered systems \cite{Kivelson1982}.

Let us discuss how our result is related to the conventional group velocity \cite{Ashcroft1976}
\begin{equation}
v_{k\alpha}=\frac{d\epsilon_{k\alpha}}{dk}. 
\label{vg}
\end{equation}
In the presence of the sublattice symmetry, as is the case of the SSH model, the eigenenergies are paired as $\pm\epsilon_k$ for $\forall k\in[-\pi,\pi]$ (see purple curves in Fig.~\ref{fig8}(b)), and $v$ in Eq.~(\ref{impvLR}) becomes 
\begin{equation}
v=2\kappa^{-1}\max_{k\in[-\pi,\pi],\alpha}|{\rm Im}\;\epsilon_{k+i\kappa,\alpha}|.
\label{vSsym}
\end{equation}
The maximal group velocity 
%\begin{equation}
\begin{equation}
v_{\max}\equiv\max_{k\in[-\pi,\pi],\alpha}\left|\frac{d\epsilon_{k\alpha}}{dk}\right|
\label{vmax}
\end{equation}
%=\max_{k,\alpha}\left|\frac{d\epsilon_{k\alpha}}{dk}\right|
%\end{equation}
thus naturally emerges when $\kappa\to0$ (see Fig.~\ref{fig7}(b)) due to the relation $\epsilon_{k+i\kappa,\alpha}=\epsilon_{k\alpha}+i\frac{d\epsilon_{k\alpha}}{dk}\kappa+O(\kappa^2)$. Since the bound in Eq.~(\ref{Ptcorr}) can be made rather small by a sufficiently large $l$ %\gg\kappa^{-1}$ 
for a given small $\kappa$, we expect that the Lieb-Robinson velocity $v_{\rm LR}=\frac{1}{2}v$ \cite{Bravyi2006} is essentially given by $v_{\max}$ at large length scales. More precisely, suppose that we scale up $l$ and $t$ simultaneously while keeping $\frac{l}{t}$ fixed; then, as long as $\frac{l}{t}<2v_{\rm max}$, we can always choose $\kappa>0$ such that the bound in Eq.~(\ref{Ptcorr}) scales like $e^{-O(l)}$ and thus vanishes in the thermodynamic limit. This is quite reasonable since the group velocity in Eq.~(\ref{vg}) is derived in the semiclassical regime, where the length scale is much larger than the lattice constant \cite{Ashcroft1976}. 

In general, however, the energy dispersions are not paired at each $k$. This can be the case even if there is a particle-hole symmetry, which only requires $\epsilon_{k\alpha}=-\epsilon_{-k\bar\alpha}$ with $\bar\alpha$ being the particle-hole conjugation of $\alpha$ (see, for example, green curves in Fig.~\ref{fig8}(b)). In the limit of $\kappa\to0$, $v$ in Eq.~(\ref{impvLR}) generally reaches the maximal \emph{relative group velocity} 
\begin{equation}
v_{\rm mr}\equiv\max_{k\in[-\pi,\pi],\alpha,\beta}\left(\frac{d\epsilon_{k\alpha}}{dk}-\frac{d\epsilon_{k\beta}}{dk}\right),
\label{vmr}
\end{equation}
which is smaller than twice of the maximal group velocity (\ref{vmax}) and thus gives a tighter bound on the propagation of correlation. For example, as shown in Fig.~\ref{fig8}(a), the dynamics of correlation in the SSH model does not change in the presence of an additional particle-hole symmetric term, which enhances the maximal group velocity (see Fig.~\ref{fig8}(c)) but leaves the relative group velocity invariant over the entire Brillouin zone. 

One may ask whether $v$ in Eq.~(\ref{impvLR}) can be made smaller than Eq.~(\ref{vmr}) for some $\kappa>0$ so that the Lieb-Robinson velocity can be even tighter. However, by employing a continuous version of the \emph{majorization} technique \cite{Arnold2010}, we can prove that Eq.~(\ref{impvLR}) \emph{increases monotonically} with respect to $\kappa$ (see Fig.~\ref{fig7}(b) for example and Appendix~\ref{monov} for the general proof). Therefore, our rigorous bound does \emph{not} lead to any tighter bound than the maximal relative group velocity. This is physically reasonable since a completely destructive interference between the modes with maximal relative group velocities seems impossible due to the difference in wave numbers. Our result thus quantitatively justifies and refines the widely used quasiparticle picture on the propagation of correlation in the quench dynamics \cite{Fagotti2016}, which, to our knowledge, has analytically been confirmed only in specific situations \cite{Calabrese2006,Cazalilla2006,Calabrese2011}.

\subsection{Proof of Theorem~\ref{ThmspEG}}
We are now in a position to prove the first main result. To bound the single-particle entanglement gap (\ref{spEG}) from the exponential decay in the correlation function, we need the following lemma.
\begin{lemma}
Denoting $P^{(L)}_<$ as the projector onto the Fermi sea of a length-$L$ lattice system, we have
\begin{equation}
\langle j| P^{(L)}_<|j'\rangle - \langle j|P^{(\infty)}_< |j'\rangle=\sum_{n\in\mathbb{Z}\backslash\{0\}}\langle j| P^{(\infty)}_< |j'+nL\rangle %,\;\;\;\;\forall j,j'\in\mathbb{Z}_L.
\label{PLPinf}
\end{equation}
for $\forall j,j'\in\mathbb{Z}_L$. %and arbitrary internal states $a$ and $a'$.
\label{finitesize}
\end{lemma}
This result arises from the combination of Eq.~(\ref{PLW}) and the relation between the Wannier functions on finite and infinite lattices (see Appendix~\ref{lem2}). Lemma~\ref{finitesize} can be used to derive a bound on the finite-size correction to the correlation matrix, which determines the ES.
\begin{lemma}
Consider a length-$l$ segment embedded in a gapped translation invariant 1D lattice system with length $L$ ($1\le l<L\le\infty$) under the periodic boundary condition. We denote the projector onto the Fermi sea as $P^{(L)}_<$. In the thermodynamic limit ($L\to\infty$), we assume that (justified in Lemma~\ref{fimp})
\begin{equation}
\|\langle j| P^{(\infty)}_< |j'\rangle\| \le Ce^{-\kappa|j-j'|},\;\;\;\;\forall j,j'\in\mathbb{Z}, %\;{\rm and}\;a,a'=1,2,...,d,
\label{expdecay}
\end{equation}
where %the constants 
$C$ and $\kappa>0$ do not depend on $j$ and $j'$. Then, we have
\begin{equation}
%|\xi^{(L)}_n - \xi^{(\infty)}_n|
\|P_SP^{(L)}_<P_S - P_SP^{(\infty)}_<P_S\|\le \frac{2C\sinh\kappa l}{(e^{\kappa L}-1)\sinh\kappa},
%\;\;\;\;\forall n=1,2,...,ld.
\label{deltalam}
\end{equation}
where $P_S\equiv\sum^l_{j=1}|j\rangle\langle j|\otimes\mathbb{1}_{\rm I}$ is the projector onto the segment.
\label{fses}
\end{lemma}
%The main idea for proving Lemma~\ref{fses} is to use \emph{Weyl's perturbation theorem} \cite{Bhatia1997}, i.e., with the subsystem projector denoted by $P_S$, the shift in the finite-size ES $|\xi^{(L)}_n - \xi^{(\infty)}_n|$ is bounded by $\|P_SP^{(L)}_<P_S-P_SP^{{(\infty)}}_<P_S\|$, which can be shown from Eq.~(\ref{expdecay}) to be exponentially small \cite{SME}. 
%While we do not give the detailed proof  
\emph{Proof.---} According to Eq.~(\ref{PLPinf}), the norm of $\langle j|P^{(L)}_<|j'\rangle - \langle j|P^{(\infty)}_<|j'\rangle$ is bounded from above by
\begin{equation}
\begin{split}
&\;\;\;\;\;\;\|\langle j|P^{(L)}_<|j'\rangle - \langle j|P^{(\infty)}_<|j'\rangle\| \\
&\le \sum_{n\in\mathbb{Z}\backslash\{0\}}\|\langle j| P^{(\infty)}_< |j'+nL\rangle\| \\
&\le C\sum_{n\in\mathbb{Z}\backslash\{0\}}e^{-\kappa|j-j'-nL|} \\
&=C\sum^{\infty}_{n=1}[e^{-\kappa(nL+j'-j)}+e^{-\kappa(nL+j-j')}] \\
&=\frac{2C\cosh\kappa(j-j')}{e^{\kappa L}-1},
\end{split}
\label{entryPLPinf}
\end{equation}
where Eq.~(\ref{expdecay}) has been used. %Let us denote the projector onto the segment as . 
Using Eq.~(\ref{entryPLPinf}) and the norm inequality \cite{Bhatia1990}
\begin{equation}
\|O\|^2\le\sum_{j,j'}\|O_{jj'}\|^2,
\end{equation}
with $O_{jj'}\equiv P_jOP_{j'}$, $P_jP_{j'}=\delta_{jj'}P_j$ and $\sum_j P_j=\mathbb{1}$ for an arbitrary bounded partitioned operator $O=\sum_{j,j'}O_{jj'}$, we obtain
\begin{equation}
\begin{split}
&\;\;\;\;\;\;\|P_SP^{(L)}_<P_S - P_SP^{(\infty)}_<P_S \|^2 \\
&\le %\|P_{\rm S}P^{(L)}_<P_{\rm S} - P_{\rm S}P^{(\infty)}_<P_{\rm S} \|^2_2=
\sum^{l-1}_{j,j'=0}\|\langle j| P^{(L)}_< |j'\rangle- \langle j|P^{(\infty)}_< |j'\rangle\|^2 \\
&\le 4C^2 \sum^{l}_{j,j'=1} \frac{\cosh^2\kappa(j-j')}{(e^{\kappa L}-1)^2} \\
&=\frac{2C^2[(\sum^{l}_{j=1}e^{2\kappa j})(\sum^{l}_{j'=1}e^{-2\kappa j'})+l^2]}{(e^{\kappa L}-1)^2}\\
&\le\left[\frac{2C\sinh\kappa l}{(e^{\kappa L}-1)\sinh\kappa}\right]^2,
\end{split}
\end{equation}
where we have used $\sinh\kappa l\ge l\sinh\kappa$ for $l\ge 1$. $\square$ 

The remaining step to prove Theorem~\ref{ThmspEG} is simply to combine  Lemmas~\ref{fses} and  \ref{fimp} with Eq.~\ref{spEGbound}. By identifying $C$ in Eq.~(\ref{expdecay}) with $Ce^{\kappa v t}$ in Eq.~(\ref{Ptcorr}), we find that Eq.~(\ref{spEGbound}) leads to Theorem~\ref{ThmspEG}. 

Finally, let us discuss how to generalize Theorem~\ref{ThmspEG} to a finite lattice with length $L>l$. The existence of such a Lieb-Robinson bound is already clear from the triangle inequality $|\xi^{(L)}_n-\xi^{(2l)}_n|\le|\xi^{(2l)}_n-\xi^{(\infty)}_n|+|\xi^{(L)}_n-\xi^{(\infty)}_n|$, the rhs of which can be further bounded by Lemma~\ref{fses}. However, this bound is too loose since the exact degeneracy is not reproduced when $L=2l$. To obtain a tighter bound, we use the following generalization of Lemma~\ref{finitesize}:
\begin{equation}
\begin{split}
&\langle j| P^{(L_1)}_<|j'\rangle - \langle j|P^{(L_2)}_< |j'\rangle \\
=&\sum_{n_1\in\mathbb{Z}\backslash\frac{{\rm LCM}(L_1,L_2)}{L_1}\mathbb{Z}}\langle j| P^{(\infty)}_< |j'+n_1L_1\rangle \\
-&\sum_{n_2\in\mathbb{Z}\backslash\frac{{\rm LCM}(L_1,L_2)}{L_2}\mathbb{Z}}\langle j| P^{(\infty)}_< |j'+n_2L_2\rangle,
\end{split}
\label{PL12}
\end{equation}
where ${\rm LCM}$ denotes the least common multiple. Following the calculations in Lemma~\ref{fses}, we can use Eq.~(\ref{PL12}) to derive
\begin{equation}
\begin{split}
\Delta^{\rm sp}_{\rm E}&\le\frac{4C\sinh\kappa l}{\sinh\kappa}e^{\kappa vt} \\
&\times\left(\frac{1}{e^{\kappa L}-1}+\frac{1}{e^{2\kappa l}-1}-\frac{2}{e^{\kappa{\rm LCM}(L,2l)}-1}\right),
\end{split}
\end{equation}
which reproduces Eq.~(\ref{spEG}) for $L=\infty$ and $\Delta^{\rm sp}_{\rm E}=0$ for $L=2l$. Moreover, the bound is $O(1)$ when $L$ is close to $l$. This is reasonable because the many-body ES of a length-$l$ segment should be the same as its complement with length $L-l$. If $L-l$ is comparable to or even smaller than the localization length of the entanglement edge modes, both many-body and single-particle entanglement gaps will be significantly nonzero.

\section{Interacting systems} 
\label{Interaktion}
Let us move on to interacting systems. It suffices to consider spin (bosonic) systems since interacting fermions can be mapped onto spin systems via the Jordan-Wigner transformation, which preserves the locality in the presence of the fermion-parity superselection rule \cite{Verstraete2005}. The only thing we should be cautious about is that fermionic SPT states with Majorana modes will be transformed into spontaneous symmetry broken states, as will be discussed in details in Sec.~\ref{AFSPT}. Although the very notion of the band is no longer applicable, we can employ MPSs to efficiently describe the ground state of a gapped local Hamiltonian \cite{Verstraete2006}. While the formalism is different, we can again upper bound the entanglement gap by a quantity closely related to the correlation of two local observables at the boundaries of the subsystem, as detailed in the following.

%This Lemma arises from the \emph{locality-preserving} property of an MPU. As shown in Fig.~\ref{fig3}(c), $k_0$ is essentially the smallest $k$ such that $U^\dag O_jU$ acts trivially outside $[j-k,j+k]$ for $\forall\;O_j$ acting only on the $j$th site. Combining Lemma~\ref{minpoly} with the main result in Ref.~\cite{Wolf2015}, we obtain

\subsection{ES of an MPS} %Entanglement spectrum of a matrix-product state}
We consider a translation-invariant normal MPS in the canonical form \footnote{In the thermodynamic limit, an MPS being normal is equivalent to that its associated channel has a unique fixed point.}, as given in Eq.~(\ref{MPS}). %We loose the assumption of being normal into that the associated channel given in Eq.~(\ref{unicha}) has a \emph{unique} fixed point that may not be full-ranked. 
For an arbitrary orthonormal basis $\{|\alpha\rangle\}^D_{\alpha=1}$ on the virtual level, we can decompose Eq.~(\ref{MPS}) into
\begin{equation}
|\Psi\rangle=\sum_{\alpha,\beta}|\psi_{\alpha\beta}\rangle|\Phi_{\beta\alpha}\rangle,
\label{PsipsiPhi}
\end{equation}
where
\begin{equation}
\begin{split}
|\psi_{\alpha\beta}\rangle&=\sum_{\{j_s\}^l_{s=1}}\langle\alpha|A_{j_1}A_{j_2}...A_{j_l}|\beta\rangle|j_1j_2...j_l\rangle,\\
|\Phi_{\beta\alpha}\rangle&=\sum_{\{j_s\}^L_{s=l+1}}\langle\beta|A_{j_{l+1}}A_{j_{l+2}}...A_{j_L}|\alpha\rangle|j_{l+1}j_{l+2}...j_L\rangle.
\end{split}
\end{equation}
In particular, if $|\alpha\rangle$'s are chosen to be the eigenstates of $\Lambda$ in Eq.~(\ref{Lam}) with eigenvalues $\lambda_\alpha$'s ($\sum^D_{\alpha=1}\lambda_\alpha=1$), we have in the thermodynamic limit $L\to\infty$ \cite{Perez2007}
\begin{equation}
\begin{split}
\langle\Phi_{\beta\alpha}|\Phi_{\beta'\alpha'}\rangle
&=\lim_{L\to\infty}\langle\beta'|\mathcal{E}^{L-l}(|\alpha'\rangle\langle\alpha|)|\beta\rangle\\
&=\langle\beta'|\mathcal{E}^\infty(|\alpha'\rangle\langle\alpha|)|\beta\rangle \\
&=\langle\beta'|\Tr[\Lambda|\alpha'\rangle\langle\alpha|]\mathbb{1}_{\rm v}|\beta\rangle\\
&=\lambda_\alpha\delta_{\beta,\beta'}\delta_{\alpha,\alpha'}.
\end{split}
\end{equation}
where $\mathcal{E}^\infty(\cdot)\equiv\lim_{L\to\infty}\mathcal{E}^L(\cdot)=\Tr[\Lambda\;\cdot\;]\mathbb{1}_{\rm v}$. Therefore, the reduced density matrix of subsystem $[1,l]\subset\mathbb{Z}$ (i.e., the segment consisting of sites $1,2,...,l$) can be written as
\begin{equation}
\rho_{[1,l]}=\sum_{\alpha,\beta}\lambda_\alpha|\psi_{\alpha\beta}\rangle\langle \psi_{\alpha\beta}|.
\label{rho1l}
\end{equation}
Unlike $|\Phi_{\beta\alpha}\rangle$'s, $|\psi_{\alpha\beta}\rangle$'s are not strictly orthogonal to each other:
\begin{equation}
\begin{split}
\langle\psi_{\alpha\beta}|\psi_{\alpha'\beta'}\rangle=\lambda_\beta\delta_{\alpha,\alpha'}\delta_{\beta,\beta'}+\epsilon_{\alpha\beta,\alpha'\beta'},\\
%\sum^{D^2-1}_{n=1}\nu^l_n\langle\beta'|\sigma^{\rm L\dag}_n|\beta\rangle\langle\alpha|\sigma^{\rm R}_n|\alpha'\rangle
\epsilon_{\alpha\beta,\alpha'\beta'}=\langle\alpha'|(\mathcal{E}^l-\mathcal{E}^\infty)(|\beta'\rangle\langle\beta|)|\alpha\rangle.
\end{split}
\label{epab}
\end{equation}
Defining $|\phi_{\alpha,\beta}\rangle\equiv\sqrt{\lambda_\alpha}|\psi_{\alpha,\beta}\rangle$, we can simplify Eq.~(\ref{rho1l}) as
\begin{equation}
\rho_{[1,l]}=\sum_{\alpha\beta}|\phi_{\alpha\beta}\rangle\langle \phi_{\alpha,\beta}|
\label{rhosimp}
\end{equation}
with
\begin{equation}
\langle\phi_{\alpha\beta}|\phi_{\alpha'\beta'}\rangle=\lambda_\alpha\lambda_\beta\delta_{\alpha,\alpha'}\delta_{\beta,\beta'}+\sqrt{\lambda_\alpha\lambda_{\alpha'}}\epsilon_{\alpha\beta,\alpha'\beta'}.
\label{phiab}
\end{equation}

\begin{figure}
\begin{center}
 \includegraphics[width=5cm, clip]{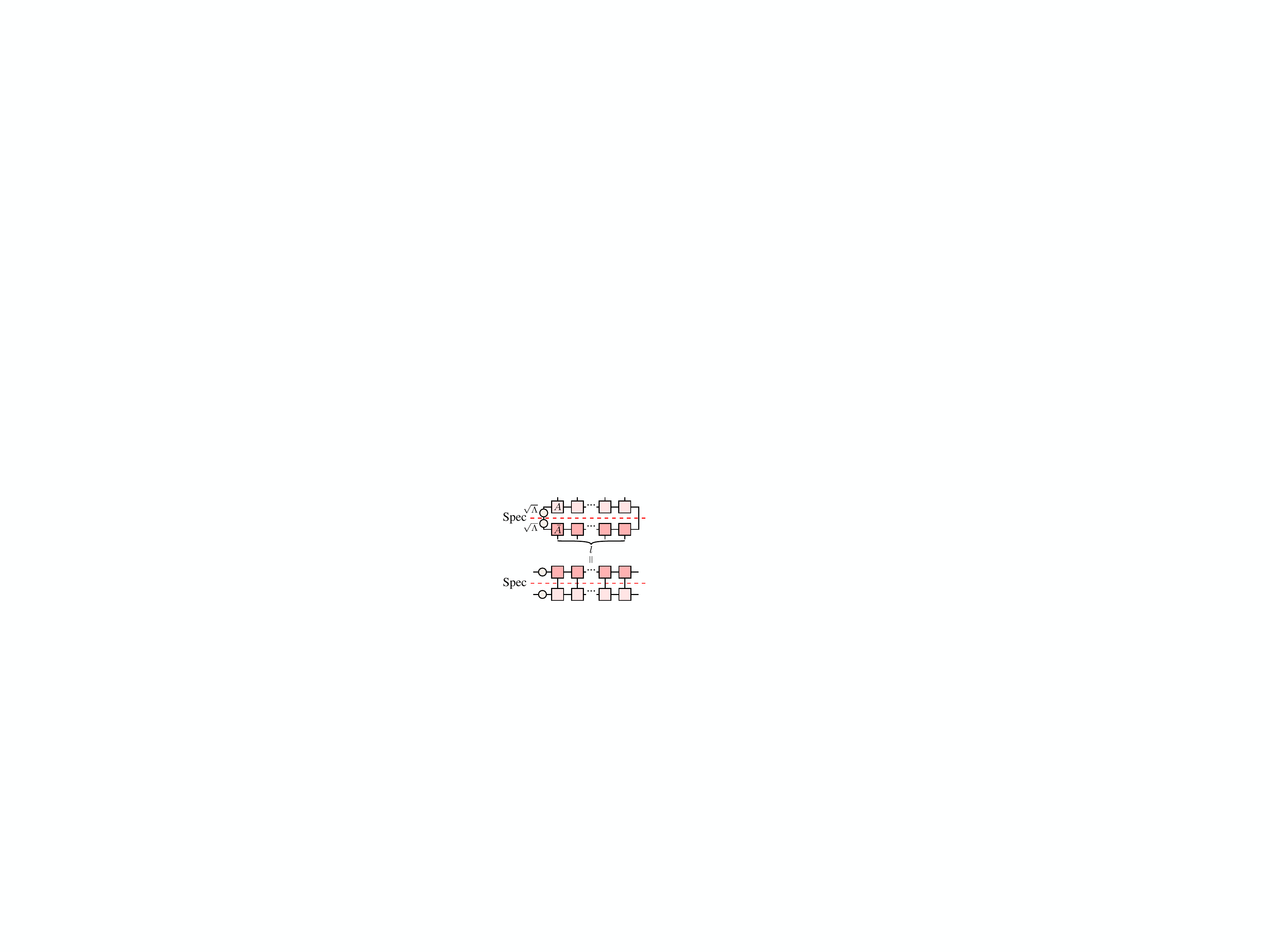}
       \end{center}
   \caption{The spectrum of Eq.~(\ref{rhosimp}) (top), which is nothing but the many-body ES of a length-$l$ segment embedded in an infinite MPS, coincides with that of Eq.~(\ref{MpMP}) (bottom) according to Lemma~\ref{rhoma}.}
   \label{fig9}
\end{figure}

To estimate the spectrum of $\rho_{[1,l]}$, we need the following lemma.
\begin{lemma}
Given a Hermitian operator $\rho=\sum^J_{j=1}|\phi_j\rangle\langle\phi_j|>0$, where $|\phi_j\rangle$'s are generally neither normalized nor orthogonal to each other, the spectrum of $\rho$ coincides with the nonzero part of the spectrum of $M\in\mathbb{C}^{J\times J}$ with $M_{jj'}=\langle\phi_j|\phi_{j'}\rangle$.
\label{rhoma}
\end{lemma}
\emph{Proof.---} Denoting $\mathcal{V}={\rm span}\{|\phi_j\rangle\}^J_{j=1}$ on which $\rho$ acts, we can expand any $|\psi\rangle\in\mathcal{V}$ as $|\psi\rangle=|\boldsymbol{\phi}\rangle\boldsymbol{c}=\sum^J_{j=1}c_j|\phi_j\rangle$, where $|\boldsymbol{\phi}\rangle\equiv[|\phi_1\rangle,|\phi_2\rangle,...,|\phi_J\rangle]$ and $\boldsymbol{c}=[c_1,c_2,...,c_J]^{\rm T}$. Defining $\langle\boldsymbol{\phi}|=[\langle\phi_1|,\langle\phi_2|,...,\langle\phi_J|]^{\rm T}$, we have $\rho=|\boldsymbol{\phi}\rangle\langle\boldsymbol{\phi}|$ and $M=\langle\boldsymbol{\phi}|\boldsymbol{\phi}\rangle$. %The claim of Lemma~\ref{rhoma} is equivalent to 
Now we prove the following equivalence.
\begin{quote}
\emph{There are $r$ linearly independent eigenstates $\{|\psi_j\rangle\}^r_{j=1}$ of $\rho$ with the same eigenvalue $\lambda\neq0$. $\Leftrightarrow$ There are $r$ linearly independent eigenvectors $\{\boldsymbol{v}_j\}^r_{j=1}$ of $M$ with eigenvalue $\lambda\neq0$}. 
\end{quote}%\newline
This statement implies Lemma~\ref{rhoma}.

$\Rightarrow$: Expanding $|\psi_j\rangle$ as $|\boldsymbol{\phi}\rangle\boldsymbol{v}_j$, by assumption we have 
\begin{equation}
\rho|\psi_j\rangle=|\boldsymbol{\phi}\rangle\langle \boldsymbol{\phi}|\boldsymbol{\phi}\rangle\boldsymbol{v}_j=|\boldsymbol{\phi}\rangle M\boldsymbol{v}_j=\lambda|\psi_j\rangle=\lambda|\boldsymbol{\phi}\rangle\boldsymbol{v}_j.
\label{psic}
\end{equation}
Multiplying $\langle\boldsymbol{\phi}|$ from the left of Eq.~(\ref{psic}), we obtain $M^2\boldsymbol{v}_j=\lambda M\boldsymbol{v}_j$, implying that $M\boldsymbol{v}_j\neq\boldsymbol{0}$ (otherwise $\rho|\psi_j\rangle=0$) is an eigenvector of $M$ with eigenvalue $\lambda$. Suppose that $M\boldsymbol{v}_j$'s are not linearly independent, which means 
\begin{equation}
\sum^r_{j=1}k_jM\boldsymbol{v}_j=\boldsymbol{0}
\label{nolind}
\end{equation}
for some $k_j$'s with $\sum^r_{j=1}|k_j|^2\neq0$. Operating Eq.~(\ref{nolind}) on $|\boldsymbol{\phi}\rangle$ gives $\sum^r_{j=1}k_j\rho|\psi_j\rangle=\lambda\sum^r_{j=1}k_j|\psi_j\rangle=0$, which contradicts the linear independence of $\{|\psi_j\rangle\}^r_{j=1}$. 

$\Leftarrow$: By assumption, we have $M\boldsymbol{v}_j=\lambda\boldsymbol{v}_j$ for $j=1,2,...,r$. Defining $|\psi_j\rangle=|\boldsymbol{\phi}\rangle\boldsymbol{v}_j$, we again obtain Eq.~(\ref{psic}), implying that $|\psi_j\rangle$ is an eigenstate of $\rho$ with eigenvalue $\lambda$. Suppose that $|\psi_j\rangle$'s are not linearly independent, which means
\begin{equation}
\sum^r_{j=1}k_j|\psi_j\rangle=\sum^r_{j=1}k_j|\boldsymbol{\phi}\rangle\boldsymbol{v}_j=0
\label{nolindpsi}
\end{equation}
for some $k_j$'s with $\sum^r_{j=1}|k_j|^2\neq0$. Operating %Left multiplying $\langle\boldsymbol{\phi}|$ to 
Eq.~(\ref{nolindpsi}) on $\langle\boldsymbol{\phi}|$ gives $\sum^r_{j=1}k_jM\boldsymbol{v}_j=\lambda\sum^r_{j=1}k_j\boldsymbol{v}_j=\boldsymbol{0}$, which contradicts the linear independence of $\{\boldsymbol{v}_j\}^r_{j=1}$. $\square$ \newline
According to Lemma~\ref{rhoma}, with the unimportant zero part neglected, the ES is nothing but the spectrum of $M_{\alpha\beta,\alpha'\beta'}=\langle\phi_{\alpha\beta}|\phi_{\alpha'\beta'}\rangle$, whose entries are given in Eq.~(\ref{phiab}).

\subsection{Exact ES degeneracy in the thermodynamic limit}
\label{EESD}
In the limit of $l\to\infty$, $\epsilon_{\alpha\beta,\alpha'\beta'}$ in Eq.~(\ref{epab}) vanishes. Hence, $M_{\alpha\beta,\alpha'\beta'}=\langle\phi_{\alpha\beta}|\phi_{\alpha'\beta'}\rangle$ becomes diagonalized and the ES is simply given by $\{\lambda_\alpha\lambda_\beta\}^D_{\alpha,\beta=1}$, which is the spectrum of $\Lambda^{\otimes2}$. In this case, SPT order enforces the ES to be exactly degenerate, as a result of symmetry fractionalization on the virtual level.
\begin{lemma}
The ES of any topologically nontrivial %(normal) 
MPS protected by unitary symmetries is at least four-fold degenerate in the thermodynamic limit of a subsystem.
\label{minESdeg}
\end{lemma}
\emph{Proof.---} Denoting the symmetry group as $G$ and its projective representation on the virtual level as $V_g$, we have \cite{Perez2008}
\begin{equation}
[V_g,\Lambda]=0,\;\;\;\;\forall g\in G,
\label{VgLam}
\end{equation}
where $\Lambda$ is the unique left fixed point of the unital channel associated with the MPS. Suppose that there is a non-degenerate eigenvalue $\lambda$ associated with eigenstate $|\lambda\rangle$. From Eq.~(\ref{VgLam}) we have
\begin{equation}
V_g|\lambda\rangle=\nu_g|\lambda\rangle,\;\;\nu_g\in{\rm U}(1),\;\;\;\;\forall g\in G. 
\label{Vglamnu}
\end{equation}
Since $V_gV_h=\omega_{g,h}V_{gh}$ for $\forall g,h\in G$, Eq.~(\ref{Vglamnu}) implies 
\begin{equation}
\nu_g\nu_h=\omega_{g,h}\nu_{gh}\;\;\Leftrightarrow\;\;\omega_{g,h}=\frac{\nu_g\nu_h}{\nu_{gh}},\;\;\;\;\forall g,h\in G.
\end{equation}
Hence, $\omega_{g,h}$ belongs to the trivial cohomology class, contradicting the assumption that $|\Psi\rangle$ is topological. Therefore, the ES in the thermodynamic limit of a subsystem, which is nothing but the spectrum of $\Lambda^{\otimes2}$, must be at least four-fold degenerate. $\square$ 

We mention that anti-unitary symmetries can also support nontrivial interacting SPT phases with the exactly degenerate ES in the thermodynamic limit. A prototypical example is the involutary ($\mathcal{T}^2=1$) time-reversal symmetry, which may be fractionalized into an anti-involutary ($\mathcal{T}^2_{\rm v}=-1$) anti-unitary symmetry on the virtual level, leading to the Kramers degeneracy in $\Lambda$ \cite{Pollmann2010}. However, since anti-unitary symmetries suffer dynamical symmetry breaking \cite{Cooper2018b}, the corresponding SPT order cannot be dynamically stable.

\subsection{Bound on the many-body entanglement gap} %at equilibrium}
%Weyl's perturbation theorem that the lift of many-body ES degeneracy is rigorously upper bounded by Eq.~(11) in the main text.
In general, $M_{\alpha\beta,\alpha'\beta'}$ for a finite $l$ can be decomposed into $[\Lambda^{\otimes2}]_{\alpha\beta,\alpha'\beta'}$ and a perturbative term: 
\begin{equation}
%\begin{split}
M_{\alpha\beta,\alpha'\beta'}=[\Lambda^{\otimes2}]_{\alpha\beta,\alpha'\beta'}+P_{\alpha\beta,\alpha'\beta'},
%=\lambda_\alpha\lambda_\beta\delta_{\alpha,\alpha'}\delta_{\beta,\beta'}+\sqrt{\lambda_\alpha\lambda_{\alpha'}}\epsilon_{\alpha\beta,\alpha'\beta'}, 
%\end{split}
\label{MpMP}
\end{equation}
where 
\begin{equation}
P_{\alpha\beta,\alpha'\beta'}=\sqrt{\lambda_\alpha\lambda_{\alpha'}}\langle\alpha'|(\mathcal{E}^l-\mathcal{E}^\infty)(|\beta'\rangle\langle\beta|)|\alpha\rangle.
\end{equation}
It is clear from Eq.~(\ref{MpMP}) that the exact eigenvalue degeneracy in $\Lambda^{\otimes2}$ is generally lifted by $P$. However, according to Weyl's perturbation theorem, the many-body entanglement gap should be upper bounded by twice the norm of $P$: 
%We denote the spectrum of $\rho_{[1,l]}$ as $\{\zeta_n\}^{d^l}_{n=1}$ with $\zeta_n\ge\zeta_{n+1}$ and define the many-body entanglement gap as $\Delta^{\rm mb}_{\rm E}\equiv|\zeta_1-\zeta_r|$ with $r$ being the largest integer such that $\zeta_r=\zeta_1$ in the limit of $l\to\infty$. 
\begin{equation}
\Delta^{\rm mb}_{\rm E}\le 2\|P\|.
\label{mbP}
\end{equation}
To proceed further, we upper bound $\|P\|$ by $\|P\|_2\equiv\sqrt{\Tr[P^\dag P]}$, which is the Schatten 2-norm and its square takes a rather simple form: 
\begin{widetext}
\begin{equation}
\begin{split}
\|P\|^2_2&=\sum^D_{\alpha,\alpha',\beta,\beta'=1}\lambda_\alpha\lambda_{\alpha'}\langle\alpha'|(\mathcal{E}^l-\mathcal{E}^\infty)(|\beta'\rangle\langle\beta|)|\alpha\rangle\langle\alpha|(\mathcal{E}^l-\mathcal{E}^\infty)(|\beta\rangle\langle\beta'|)|\alpha'\rangle \\
&=\sum^D_{\alpha,\alpha',\beta,\beta'=1}\lambda_\alpha\lambda_{\alpha'}\langle\alpha'\alpha|(\mathcal{E}^l-\mathcal{E}^\infty)^{\otimes2}(|\beta'\beta\rangle\langle\beta\beta'|)|\alpha\alpha'\rangle\\
&=\sum^D_{\alpha,\alpha',\beta,\beta'=1}\Tr[\Lambda^{\otimes2}|\alpha\alpha'\rangle\langle\alpha'\alpha|(\mathcal{E}^l-\mathcal{E}^\infty)^{\otimes2}(|\beta'\beta\rangle\langle\beta\beta'|)]\\
&=\Tr[\Lambda^{\otimes2}\mathbb{S}(\mathcal{E}^l-\mathcal{E}^\infty)^{\otimes2}(\mathbb{S})], 
\end{split}
\end{equation}
\end{widetext}
where we have used $[\mathcal{E}(O)]^\dag=\mathcal{E}(O^\dag)$ and $\mathbb{S}\equiv\sum^D_{\alpha,\beta=1}|\alpha\beta\rangle\langle\beta\alpha|$ is the swap operator acting on two copies of virtual Hilbert spaces. By defining the norm of a superoperator as
$\|\mathcal{L}\|\equiv\max_{\|O_1\|_2=\|O_2\|_2=1}|\Tr[O^\dag_2\mathcal{L}(O_1)]|$, we can bound $\|P\|^2_2$ by
\begin{equation}
\begin{split}
\|P\|^2_2&\le\|\Lambda^{\otimes2}\mathbb{S}\|_2\|\mathbb{S}\|_2\|(\mathcal{E}^l-\mathcal{E}^\infty)^{\otimes2}\| \\
&\le \frac{D}{2}\|\mathcal{E}^l-\mathcal{E}^\infty\|^2,
\end{split}
\label{P22}
\end{equation}
where $D$ arises from $\|\mathbb{S}\|_2$, and $\frac{1}{2}$ upper bounds $\|\Lambda^{\otimes2}\mathbb{S}\|_2=\Tr[\Lambda^2]$ since the spectrum of $\Lambda$ is at least two-fold degenerate. Combining Eq.~(\ref{P22}) with Eq.~(\ref{mbP}), we obtain
\begin{equation}
\Delta^{\rm mb}_{\rm E}\le\sqrt{2D}\|\mathcal{E}^l-\mathcal{E}^\infty\|.
\label{ElEinf}
\end{equation}
%It is well-known 
Note that $\mathcal{E}^l-\mathcal{E}^\infty$ appears routinely in the correlation functions of MPSs and leads to an exponential decay \cite{Werner1992,Perez2007,Verstraete2008}. Quantitatively, $\|\mathcal{E}^l-\mathcal{E}^\infty\|$ can be upper bounded by $c_\epsilon(\mu+\epsilon)^l$ for $\forall \epsilon>0$, where $\mu$ is the spectral radius of $\mathcal{E}-\mathcal{E}^\infty$ and $c_\epsilon$ does not depend on $l$ \cite{Werner1992}, implying that $\Delta^{\rm mb}_{\rm E}$ is also exponentially small just like the correlation functions mentioned above. We emphasize that %due to the non-Hermiticity of $\mathcal{E}$, 
$\|\mathcal{E}^l-\mathcal{E}^\infty\|$ may scale like ${\rm poly}(l)\mu^l$ \cite{Wolf2012}, in which case a nonzero $\epsilon$ is \emph{necessary} for giving a rigorous bound like $c_\epsilon(\mu+\epsilon)^l$.

\subsection{Proof of Theorem~\ref{MBEG}}
\label{MBMR}
%To study the impact of time evolution, we focus on the case in which, as illustrated by Fig.~\ref{fig3}(b), the dynamics is stroboscopic and governed by an MPU. We assume the MPU to be symmetric, i.e., $[\rho^{\otimes L}_g,U]=0$ for $\forall g\in G$, and in the trivial cohomology class so that the MPS stays in the same SPT phase \cite{Gong2018c}. 
Let us analyze how the bound given in Eq.~(\ref{ElEinf}) changes when an MPS evolves according to a symmetric and trivial MPU. Denoting the bond dimension of the MPU as $D_U$, %the bond dimension 
that of the MPS after $t$ steps of evolutions is no more than $DD^t_U$.  Moreover, the spectrum of the associated unital channel \emph{stays invariant} during the time evolution as a result of unitarity \cite{Gong2018c} (see also Lemma~\ref{invspe} in Appendix~\ref{BPMPU}), and so does $\mu$. Having in mind that $\|\mathcal{E}^l-\mathcal{E}^\infty\|$ is bounded by $c_\epsilon(\mu+\epsilon)^l$, it is natural to expect a Lieb-Robinson bound from Eq.~(\ref{ElEinf}). However, this expectation may fail --- given $\epsilon$, $c_\epsilon$ may grow faster than exponentially in time. To rule out this possiblity, we utilize the \emph{function-algebra}-based techniques developed in Ref.~\cite{Wolf2015} to carefully estimate the growth of $\|\mathcal{E}^l-\mathcal{E}^\infty\|$.

The main idea of Ref.~\cite{Wolf2015} is represented by the following lemma.
\begin{lemma}
Given an operator $\mathcal{M}$ on a finite linear space generating a bounded semigroup $\{\mathcal{M}^n\}_{n\in\mathbb{N}}$, i.e., $\|\mathcal{M}^n\|\le C_{\mathcal{M}}$ %($C$ is a $n$-independent constant) 
for $\forall n\in\mathbb{N}$, and an element $f$ in the Wiener algebra $W\equiv\{f\in{\rm Hol}(\mathbb{D}):f(z)=\sum_{p\in\mathbb{N}}f_pz^p,\|f\|_W\equiv\sum_{p\in\mathbb{N}}|f_p|<\infty\}$ (${\rm Hol}(\mathbb{D})$: set of holomorphic functions within the unit disk $\mathbb{D}\equiv\{z\in\mathbb{C}:|z|<1\}$), we have
\begin{equation}
\|f(\mathcal{M})\|\le C_{\mathcal{M}}\|f\|_{W/m_\mathcal{M}W},
\label{fWmW}
\end{equation}
where $\|f\|_{W/m_\mathcal{M}W}\equiv\inf\{\|g\|_W:g=f+m_\mathcal{M}h,\;h\in W\}$ and $m_{\mathcal{M}}\in W$ is the minimal polynomial of $\mathcal{M}$, i.e., the nonzero polynomial with the lowest degree such that $m_{\mathcal{M}}(\mathcal{M})=0$ and the leading coefficient (that of the highest degree) is equal to $1$.
\label{fnorm}
\end{lemma}
\emph{Proof.---} Due to $\|\mathcal{M}^n\|\le C_{\mathcal{M}}$ for $\forall n\in\mathbb{N}$, for $\forall f\in W$, the norm of $f(\mathcal{M})$ can be upper bounded by
\begin{equation}
\|f(\mathcal{M})\|\le \sum_{p\in\mathbb{N}} |f_p| \|\mathcal{M}^p\|\le C_{\mathcal{M}}\sum_{p\in\mathbb{N}}|f_p|=\|f\|_W.
\label{ffW}
\end{equation}
Since $m_{\mathcal{M}}$ is the minimal polynomial of $\mathcal{M}$, for $\forall h\in W$, we have
\begin{equation}
f(\mathcal{M})=(f+m_{\mathcal{M}}h)(\mathcal{M}).
\label{fMgM}
\end{equation}
Applying Eq.~(\ref{ffW}) to the rhs of Eq.~(\ref{fMgM}) gives
\begin{equation}
\|f(\mathcal{M})\|\le C_{\mathcal{M}}\|f+m_{\mathcal{M}}h\|_W,\;\;\;\;\forall h\in W.
\label{fhW}
\end{equation}
Equation (\ref{fWmW}) then follows from minimization of the rhs of Eq.~(\ref{fhW}). $\square$ \newline
An arbitrary unital channel $\mathcal{E}$ satisfies the condition in %the above 
Lemma~\ref{fnorm} due to the fact that $\mathcal{E}^n$ is again a unital channel and $\|\mathcal{E}^n\|\le\sqrt{\frac{D}{2}}$, where $D$ is the Hilbert-space dimension \cite{Perez2006}. %and so does $\mathcal{E}-\mathcal{E}^\infty$. 
Accordingly, $\mathcal{E}-\mathcal{E}^\infty$ also satisfies the condition. 

Regarding the minimal polynomial of $\mathcal{E}-\mathcal{E}^\infty$ during time evolution, we have the following lemma. 
%With all the previous results in hand, we are ready to prove the Lieb-Robinson bound on the many-body EG. To this end, we need the following lemma.
\begin{lemma}
Suppose that a single step of evolution by an MPU generated by $\mathcal{U}$ changes the associated unital channel of an MPS from $\mathcal{E}$ into $\mathcal{E}'$. Denoting the minimal polynomials of $\mathcal{E}-\mathcal{E}^\infty$ and $\mathcal{E}'-\mathcal{E}'^\infty$ as $m$ and $m'$, %($\mathcal{E}$ and $\mathcal{E}'$ follow the notations in Lemma~\ref{invspe}), 
respectively, we have
\begin{equation}
m'(z)|z^{2k_0} m(z), 
\label{EpdE}
\end{equation}
namely, $m'(z)$ is a divisor of $z^{2k_0}m(z)$, where $k_0$ is the smallest integer such that $\mathcal{U}_{k_0}$ is simple. %(see Theorem~\ref{MPUsimple}).
\label{minpoly}
\end{lemma}
\emph{Proof.---} By assumption, %According to Eqs.~(\ref{Uid}) and (\ref{simple}), 
for $\forall k\ge2k_0$, the open boundary tensor $\overline{\mathcal{U}}_k\mathcal{U}_k$ can be decomposed into
\begin{equation}
\begin{tikzpicture}[scale=1.1]
%left
\draw[thick] (-0.15,0.375) -- (1.35,0.375) (1.65,0.375) -- (2.4,0.375) (-0.15,-0.375) -- (1.35,-0.375) (1.65,-0.375) -- (2.4,-0.375);
\draw[thick] (0.225,-0.75) -- (0.225,0.75) (0.975,-0.75) -- (0.975,0.75) (2.025,-0.75) -- (2.025,0.75);
\draw[thick,fill=blue!30] (0,0.15) rectangle (0.45,0.6) (0.75,0.15) rectangle (1.2,0.6) (1.8,0.15) rectangle (2.25,0.6); 
\draw[thick,fill=blue!10] (0,-0.15) rectangle (0.45,-0.6) (0.75,-0.15) rectangle (1.2,-0.6) (1.8,-0.15) rectangle (2.25,-0.6);
\Text[x=1.5]{...}
\Text[x=0.225,y=0.375,fontsize=\footnotesize]{$\overline{\mathcal{U}}$}
\Text[x=0.225,y=-0.375,fontsize=\footnotesize]{$\mathcal{U}$}
\Text[x=0.975,y=0.375,fontsize=\footnotesize]{$\overline{\mathcal{U}}$}
\Text[x=0.975,y=-0.375,fontsize=\footnotesize]{$\mathcal{U}$}
\Text[x=2.025,y=0.375,fontsize=\footnotesize]{$\overline{\mathcal{U}}$}
\Text[x=2.025,y=-0.375,fontsize=\footnotesize]{$\mathcal{U}$}
\draw[thick] (0.225,-0.8) .. controls (0.225,-0.85) and (1.125,-0.75) .. (1.125,-0.95);
\draw[thick] (2.025,-0.8) .. controls (2.025,-0.85) and (1.125,-0.75) .. (1.125,-0.95);
\Text[x=1.125,y=-1.1,fontsize=\footnotesize]{$k$}
%left
\Text[x=2.75]{$\equiv$}
%middle
\draw[thick] (3.1,0.375) -- (3.85,0.375) (3.1,-0.375) -- (3.85,-0.375);
\draw[thick] (3.475,-0.75) -- (3.475,0.75);
\draw[thick,fill=blue!30] (3.25,0.15) rectangle (3.7,0.6);
\draw[thick,fill=blue!10] (3.25,-0.15) rectangle (3.7,-0.6);
\Text[x=3.5,y=0.375,fontsize=\footnotesize]{$\overline{\mathcal{U}}_k$}
\Text[x=3.5,y=-0.375,fontsize=\footnotesize]{$\mathcal{U}_k$}
%middle
\Text[x=4.2]{$=$}
%right
\draw[thick] (4.55,0.375) -- (5.4,0.375) -- (5.4,-0.375) -- (4.55,-0.375);
\draw[thick] (4.925,-0.75) -- (4.925,0.75);
\draw[thick,fill=brown!10] (5.4,0) circle (0.15);
\draw[thick,fill=blue!30] (4.7,0.15) rectangle (5.15,0.6);
\draw[thick,fill=blue!10] (4.7,-0.15) rectangle (5.15,-0.6);
\draw[thick] (5.65,-0.75) -- (5.65,0.75);
\Text[x=4.95,y=-0.38,fontsize=\scriptsize]{$\mathcal{U}_{k_0}$}
\Text[x=4.95,y=0.38,fontsize=\scriptsize]{$\overline{\mathcal{U}}_{k_0}$}
\Text[x=5.4,fontsize=\scriptsize]{$\rho$}
\draw[thick] (6.75,-0.375) -- (5.9,-0.375) -- (5.9,0.375) -- (6.75,0.375);
\draw[thick] (6.375,-0.75) -- (6.375,0.75);
\draw[thick,fill=blue!30] (6.15,0.15) rectangle (6.6,0.6);
\draw[thick,fill=blue!10] (6.15,-0.15) rectangle (6.6,-0.6);
\Text[x=6.4,y=-0.38,fontsize=\scriptsize]{$\mathcal{U}_{k_0}$}
\Text[x=6.4,y=0.38,fontsize=\scriptsize]{$\overline{\mathcal{U}}_{k_0}$}
\Text[x=6,y=-1,fontsize=\footnotesize]{$\mathbb{1}^{\otimes{k-2k_0}}$}
%right
\Text[x=7,y=-0.2]{.}
\end{tikzpicture}
\label{einfach}
\end{equation}
%where $\rho$ is the unique fixed point of the quantum channel $\sum^d_{i,j=1}\mathcal{U}_{ij}\cdot\mathcal{U}^\dag_{ij}$ associated with the MPU. 
We express $m$ explicitly as $\sum_kc_kz^k$, which satisfies $c_0=0$ (due to that the spectrum of $\mathcal{E}-\mathcal{E}^\infty$ contains zero) and 
\begin{equation}
\begin{split}
m(\mathcal{E}-\mathcal{E}^\infty)=\sum_kc_k(\mathcal{E}^k-\mathcal{E}^\infty)=0\\
\Leftrightarrow\;\;
\sum_k c_k\mathcal{E}^k=\sum_k c_k\mathcal{E}^\infty.
\end{split}
\label{mcalE}
\end{equation}
Then, using Eq.~(\ref{einfach}), we can evaluate $z^{2k_0}m(z)|_{z=\mathcal{E}'}$ as
\begin{widetext}
\begin{equation}
\begin{tikzpicture}[scale=1.1]
%left
\draw[thick] (-0.15,1.125) -- (1.35,1.125) (1.65,1.125) -- (2.4,1.125) (-0.15,0.375) -- (1.35,0.375) (1.65,0.375) -- (2.4,0.375) (-0.15,-0.375) -- (1.35,-0.375) (1.65,-0.375) -- (2.4,-0.375) (-0.15,-1.125) -- (1.35,-1.125) (1.65,-1.125) -- (2.4,-1.125);
\draw[thick] (0.225,-1.3) -- (0.225,1.3) (0.975,-1.3) -- (0.975,1.3) (2.025,-1.3) -- (2.025,1.3);
\draw[thick,fill=blue!30] (0,0.15) rectangle (0.45,0.6) (0.75,0.15) rectangle (1.2,0.6) (1.8,0.15) rectangle (2.25,0.6); 
\draw[thick,fill=red!30] (0,0.9) rectangle (0.45,1.35) (0.75,0.9) rectangle (1.2,1.35) (1.8,0.9) rectangle (2.25,1.35); 
\draw[thick,fill=blue!10] (0,-0.15) rectangle (0.45,-0.6) (0.75,-0.15) rectangle (1.2,-0.6) (1.8,-0.15) rectangle (2.25,-0.6);
\draw[thick,fill=red!10] (0,-0.9) rectangle (0.45,-1.35) (0.75,-0.9) rectangle (1.2,-1.35) (1.8,-0.9) rectangle (2.25,-1.35); 
\Text[x=-0.75]{$\sum_k c_k$}
\Text[x=1.5]{...}
\Text[x=0.225,y=1.125,fontsize=\footnotesize]{$\bar A$}
\Text[x=0.225,y=0.375,fontsize=\footnotesize]{$\overline{\mathcal{U}}$}
\Text[x=0.225,y=-0.375,fontsize=\footnotesize]{$\mathcal{U}$}
\Text[x=0.225,y=-1.125,fontsize=\footnotesize]{$A$}
\Text[x=0.975,y=1.125,fontsize=\footnotesize]{$\bar A$}
\Text[x=0.975,y=0.375,fontsize=\footnotesize]{$\overline{\mathcal{U}}$}
\Text[x=0.975,y=-0.375,fontsize=\footnotesize]{$\mathcal{U}$}
\Text[x=0.975,y=-1.125,fontsize=\footnotesize]{$A$}
\Text[x=2.025,y=1.125,fontsize=\footnotesize]{$\bar A$}
\Text[x=2.025,y=0.375,fontsize=\footnotesize]{$\overline{\mathcal{U}}$}
\Text[x=2.025,y=-0.375,fontsize=\footnotesize]{$\mathcal{U}$}
\Text[x=2.025,y=-1.125,fontsize=\footnotesize]{$A$}
\draw[thick] (0.225,-1.45) .. controls (0.225,-1.5) and (1.125,-1.4) .. (1.125,-1.6);
\draw[thick] (2.025,-1.45) .. controls (2.025,-1.5) and (1.125,-1.4) .. (1.125,-1.6);
\Text[x=1.25,y=-1.75,fontsize=\footnotesize]{$k+2k_0$}
%left
%markmiddle
\draw[thick,purple,dashed,fill=purple!5] (3,-0.25) rectangle (3.9,0.25);
\draw[thick,purple,dashed,fill=purple!5] (5.05,-0.7) rectangle (5.7,0.7);
%markmiddle
\Text[x=3.25]{$=\;\sum_kc_k$}
%middle
\draw[thick] (4,0.375) -- (4.85,0.375) -- (4.85,-0.375) -- (4,-0.375);
\draw[thick] (4.375,-1.3) -- (4.375,1.3);
\draw[thick] (5.375,-0.4) -- (5.375,0.4);
\draw[thick] (6.375,-1.3) -- (6.375,1.3);
\draw[thick] (6.75,0.375) -- (5.9,0.375) -- (5.9,-0.375) -- (6.75,-0.375);
\draw[thick] (4,1.125) -- (5,1.125) -- (5,0.375) -- (5.75,0.375) -- (5.75,1.125) -- (6.75,1.125) 
(4,-1.125) -- (5,-1.125)  -- (5,-0.375) -- (5.75,-0.375) -- (5.75,-1.125) -- (6.75,-1.125);
\draw[thick,fill=red!30] (4.15,0.9) rectangle (4.6,1.35);
\draw[thick,fill=blue!30] (4.15,0.15) rectangle (4.6,0.6);
\draw[thick,fill=blue!10] (4.15,-0.15) rectangle (4.6,-0.6);
\draw[thick,fill=red!10] (4.15,-0.9) rectangle (4.6,-1.35);
\draw[thick,fill=brown!10] (4.85,0) circle (0.15);
\draw[thick,fill=red!30] (5.15,0.15) rectangle (5.6,0.6);
\draw[thick,fill=red!10] (5.15,-0.15) rectangle (5.6,-0.6);
\draw[thick,fill=red!30] (6.15,0.9) rectangle (6.6,1.35);
\draw[thick,fill=blue!30] (6.15,0.15) rectangle (6.6,0.6);
\draw[thick,fill=blue!10] (6.15,-0.15) rectangle (6.6,-0.6);
\draw[thick,fill=red!10] (6.15,-0.9) rectangle (6.6,-1.35);
\Text[x=4.4,y=1.125,fontsize=\scriptsize]{$\bar A_{k_0}$}
\Text[x=4.4,y=0.375,fontsize=\scriptsize]{$\overline{\mathcal{U}}_{k_0}$}
\Text[x=4.85,fontsize=\scriptsize]{$\rho$}
\Text[x=4.4,y=-0.375,fontsize=\scriptsize]{$\mathcal{U}_{k_0}$}
\Text[x=4.4,y=-1.125,fontsize=\scriptsize]{$A_{k_0}$}
\Text[x=5.4,y=0.375,fontsize=\footnotesize]{$\bar A_k$}
\Text[x=5.4,y=-0.375,fontsize=\footnotesize]{$A_k$}
\Text[x=6.4,y=1.125,fontsize=\scriptsize]{$\bar A_{k_0}$}
\Text[x=6.4,y=0.375,fontsize=\scriptsize]{$\overline{\mathcal{U}}_{k_0}$}
\Text[x=6.4,y=-0.375,fontsize=\scriptsize]{$\mathcal{U}_{k_0}$}
\Text[x=6.4,y=-1.125,fontsize=\scriptsize]{$A_{k_0}$}
%middle
%markright
\draw[thick,purple,dashed,fill=purple!5] (7.25,-0.25) rectangle (9.3,0.25);
\draw[thick,purple,dashed,fill=purple!5] (10.55,-0.7) rectangle (11.2,0.7);
%markright
\Text[x=8.1]{$=\;\lim_{l\to\infty}\sum_kc_k$}
%right
\draw[thick] (9.5,0.375) -- (10.35,0.375) -- (10.35,-0.375) -- (9.5,-0.375);
\draw[thick] (9.875,-1.3) -- (9.875,1.3);
\draw[thick] (10.875,-0.4) -- (10.875,0.4);
\draw[thick] (11.875,-1.3) -- (11.875,1.3);
\draw[thick] (12.25,0.375) -- (11.4,0.375) -- (11.4,-0.375) -- (12.25,-0.375);
\draw[thick] (9.5,1.125) -- (10.5,1.125) -- (10.5,0.375) -- (11.25,0.375) -- (11.25,1.125) -- (12.25,1.125) 
(9.5,-1.125) -- (10.5,-1.125)  -- (10.5,-0.375) -- (11.25,-0.375) -- (11.25,-1.125) -- (12.25,-1.125);
\draw[thick,fill=red!30] (9.65,0.9) rectangle (10.1,1.35);
\draw[thick,fill=blue!30] (9.65,0.15) rectangle (10.1,0.6);
\draw[thick,fill=blue!10] (9.65,-0.15) rectangle (10.1,-0.6);
\draw[thick,fill=red!10] (9.65,-0.9) rectangle (10.1,-1.35);
\draw[thick,fill=brown!10] (10.35,0) circle (0.15);
\draw[thick,fill=red!30] (10.65,0.15) rectangle (11.1,0.6);
\draw[thick,fill=red!10] (10.65,-0.15) rectangle (11.1,-0.6);
\draw[thick,fill=red!30] (11.65,0.9) rectangle (12.1,1.35);
\draw[thick,fill=blue!30] (11.65,0.15) rectangle (12.1,0.6);
\draw[thick,fill=blue!10] (11.65,-0.15) rectangle (12.1,-0.6);
\draw[thick,fill=red!10] (11.65,-0.9) rectangle (12.1,-1.35);
\Text[x=9.9,y=1.125,fontsize=\scriptsize]{$\bar A_{k_0}$}
\Text[x=9.9,y=0.375,fontsize=\scriptsize]{$\overline{\mathcal{U}}_{k_0}$}
\Text[x=10.35,fontsize=\scriptsize]{$\rho$}
\Text[x=9.9,y=-0.375,fontsize=\scriptsize]{$\mathcal{U}_{k_0}$}
\Text[x=9.9,y=-1.125,fontsize=\scriptsize]{$A_{k_0}$}
\Text[x=10.9,y=0.375,fontsize=\footnotesize]{$\bar A_l$}
\Text[x=10.9,y=-0.375,fontsize=\footnotesize]{$A_l$}
\Text[x=11.9,y=1.125,fontsize=\scriptsize]{$\bar A_{k_0}$}
\Text[x=11.9,y=0.375,fontsize=\scriptsize]{$\overline{\mathcal{U}}_{k_0}$}
\Text[x=11.9,y=-0.375,fontsize=\scriptsize]{$\mathcal{U}_{k_0}$}
\Text[x=11.9,y=-1.125,fontsize=\scriptsize]{$A_{k_0}$}
%right
\end{tikzpicture}
\label{vanish}
\end{equation}
\begin{equation*}
\begin{tikzpicture}[scale=1.1]
\Text[x=-1.6]{$=(\sum_kc_k)\lim_{l\to\infty}$}
%left
\draw[thick] (-0.15,1.125) -- (3,1.125) (-0.15,-1.125) -- (3,-1.125);
\draw[thick] (-0.15,0.375) -- (0.7,0.375) -- (0.7,-0.375) -- (-0.15,-0.375);
\draw[thick] (3,0.375) -- (2.15,0.375) -- (2.15,-0.375) -- (3,-0.375);
\draw[thick] (0.95,0.375) -- (1.9,0.375) -- (1.9,-0.375) -- (0.95,-0.375) -- cycle;
\draw[thick] (0.225,1.2) -- (0.225,-1.2) (1.425,1.2) -- (1.425,-1.2) (2.625,1.2) -- (2.625,-1.2);
\draw[thick,fill=red!30] (0,0.9) rectangle (0.45,1.35) (1.2,0.9) rectangle (1.65,1.35) (2.4,0.9) rectangle (2.85,1.35);
\draw[thick,fill=blue!30] (0,0.15) rectangle (0.45,0.6) (1.2,0.15) rectangle (1.65,0.6) (2.4,0.15) rectangle (2.85,0.6); 
\draw[thick,fill=blue!10] (0,-0.15) rectangle (0.45,-0.6) (1.2,-0.15) rectangle (1.65,-0.6) (2.4,-0.15) rectangle (2.85,-0.6);
\draw[thick,fill=red!10] (0,-0.9) rectangle (0.45,-1.35) (1.2,-0.9) rectangle (1.65,-1.35) (2.4,-0.9) rectangle (2.85,-1.35);
\draw[thick,fill=brown!10] (0.7,0) circle (0.15);
\draw[thick,fill=brown!10] (1.9,0) circle (0.15);
\Text[x=0.25,y=0.375,fontsize=\scriptsize]{$\overline{\mathcal{U}}_{k_0}$}
\Text[x=0.25,y=-0.375,fontsize=\scriptsize]{$\mathcal{U}_{k_0}$}
\Text[x=1.45,y=0.375,fontsize=\footnotesize]{$\overline{\mathcal{U}}_l$}
\Text[x=1.45,y=-0.375,fontsize=\footnotesize]{$\mathcal{U}_l$}
\Text[x=2.65,y=0.375,fontsize=\scriptsize]{$\overline{\mathcal{U}}_{k_0}$}
\Text[x=2.65,y=-0.375,fontsize=\scriptsize]{$\mathcal{U}_{k_0}$}
\Text[x=0.25,y=1.125,fontsize=\scriptsize]{$\bar A_{k_0}$}
\Text[x=0.25,y=-1.125,fontsize=\scriptsize]{$A_{k_0}$}
\Text[x=1.45,y=1.125,fontsize=\footnotesize]{$\bar A_l$}
\Text[x=1.45,y=-1.125,fontsize=\footnotesize]{$A_l$}
\Text[x=2.65,y=1.125,fontsize=\scriptsize]{$\bar A_{k_0}$}
\Text[x=2.65,y=-1.125,fontsize=\scriptsize]{$A_{k_0}$}
\Text[x=0.7,fontsize=\scriptsize]{$\rho$}
\Text[x=1.9,fontsize=\scriptsize]{$\rho$}
%left
\Text[x=4.55]{$=(\sum_kc_k)\lim_{l\to\infty}$}
%right
\draw[thick] (6.375,1.1) -- (6.375,-1.1);
\draw[thick] (6,1.125) -- (6.75,1.125) (6,0.375) -- (6.75,0.375) (6,-0.375) -- (6.75,-0.375) (6,-1.125) -- (6.75,-1.125);
\draw[thick,fill=red!30] (6.15,0.9) rectangle (6.6,1.35); 
\draw[thick,fill=blue!30] (6.15,0.15) rectangle (6.6,0.6); 
\draw[thick,fill=blue!10] (6.15,-0.15) rectangle (6.6,-0.6); 
\draw[thick,fill=red!10] (6.15,-0.9) rectangle (6.6,-1.35); 
\Text[x=6.375,y=1.125,fontsize=\footnotesize]{$\bar A_l$}
\Text[x=6.375,y=0.375,fontsize=\footnotesize]{$\overline{\mathcal{U}}_l$}
\Text[x=6.375,y=-0.375,fontsize=\footnotesize]{$\mathcal{U}_l$}
\Text[x=6.375,y=-1.125,fontsize=\footnotesize]{$A_l$}
%right
\Text[x=7,y=-0.2]{,}
\end{tikzpicture}
\end{equation*}
\end{widetext}
where we have used Eq.~(\ref{mcalE}) (enclosed in dashed rectangles) and Eq.~(\ref{einfach}). Lemma~\ref{minpoly} follows immediately from Eq.~(\ref{vanish}), which is the tensor-network representation of $\mathcal{E}'^{2k_0}m(\mathcal{E}')=\sum_kc_k\mathcal{E}'^\infty\Leftrightarrow z^{2k_0}m(z)|_{z=\mathcal{E}'-\mathcal{E}'^\infty}=0$. $\square$

A direct corollary of Lemma~\ref{minpoly} (applying $t$ times) is $m_t(z)|z^{2k_0t} m(z)$, where $m_t(z)$ ($m(z)\equiv m_0(z)$) is the minimal polynomial of $\mathcal{E}_t-\mathcal{E}^\infty_t$, with $\mathcal{E}_t$ determined from the MPS at time $t$. Using this fact, as long as $l>2k_0t$, we have
%This fact gives an upper bound of $\|z^l\|_{W/m_{\mathcal{E}_t}W}$:
\begin{equation}
\begin{split}
&\|z^l\|_{W/m_tW}\\
\le&\inf\{\|g\|_W:g=z^l+z^{2k_0t} m h,\;h\in W\} \\
=&\|z^{l-2k_0t}\|_{W/mW},
\end{split}
\end{equation}
where we have used $\|g\|_W=\|z^ng\|_W$ for $\forall n\in\mathbb{N}$. According to the main result of Ref.~\cite{Wolf2015}, which is summarized as Theorem~\ref{mainwolf} in Appendix~\ref{uniconv}, when $l-2k_0t>\frac{\mu}{1-\mu}$, we can bound $\|\mathcal{E}^l_t-\mathcal{E}^{\infty}_t\|$ from above by 
\begin{equation}
\begin{split}
%\|\mathcal{E}^l_t-\mathcal{E}^{(\infty)}_t\|\le
\mu^{l-2k_0t+1}\frac{4C_te^2\sqrt{|m_\mathcal{E}|}(|m_\mathcal{E}|+1)}{(l-2k_0t)[1-(1+\frac{1}{l-2k_0t})\mu]^{\frac{3}{2}}} \\
\times\sup_{|z|=\mu(1+\frac{1}{l-2k_0t})}\frac{1}{|B(z)|},
\end{split}
\end{equation}
where $C_t\equiv\sup_{n\in\mathbb{N}}\|\mathcal{E}^n_t\|$ and $B(z)$ is the
%$|m_{\mathcal{E}}|$ is the degree of the minimal polynomial $m_{\mathcal{E}}$, 
Blaschke product (\ref{Bz}) with respect to the spectrum of $\mathcal{E}\equiv\mathcal{E}_0$. 
%\begin{equation}
%B(z)=\prod_{z-\lambda|m_{\mathcal{E}}}\frac{z-\lambda}{1-\bar\lambda z}.
%\end{equation}
If we further require $l-2k_0t\ge\frac{1+\mu}{1-\mu}$, even in the worst case, $\|\mathcal{E}^l_t-\mathcal{E}^{\infty}_t\|$ is bounded by an exponentially small quantity up to polynomial corrections (see Eq.~(20) in Ref.~\cite{Wolf2015}): 
\begin{equation}
\begin{split}
\|\mathcal{E}^l_t-\mathcal{E}^{\infty}_t\|\le4e^2C_t\sqrt{|m|}(|m|+1)\left(\frac{1+\mu}{1-\mu}\right)^{\frac{3}{2}}\\
\times\left[\frac{1-\mu^2}{\mu}(l-2k_0t)\right]^{|m|-1}\mu^{l-2k_0t}.
\end{split}
\label{wirstbound}
\end{equation}
Denoting the bond dimension of the initial MPS $|\Psi_0\rangle$ and that of $|\Psi_t\rangle=U^t|\Psi_0\rangle$ after $t$ steps of time evolutions %and that of the MPU 
as $D$ and $D_t$, respectively, we have 
\begin{equation}
D_t\le DD^t_U=De^{t\ln D_U},
\label{Dt} 
\end{equation}
and hence \cite{Perez2006}
\begin{equation}
C_t\le%({\rm dim}\;\mathcal{E}_t)^{\frac{1}{4}}=
\sqrt{\frac{D_t}{2}}\le \sqrt{\frac{D}{2}}e^{\frac{\ln D_U}{2}t}. 
\label{Ct}
\end{equation}
Combing Eq.~(\ref{ElEinf}) and  Eqs.~(\ref{wirstbound})-(\ref{Ct}) with $|m|\le D^2$ ($D\ge2$, otherwise $|\Psi\rangle$ is a %trivial 
product state), we obtain Theorem~\ref{MBEG}.

Our theorem rigorously establishes the dynamical stability for general SPT systems in 1D. Yet another important implication of Theorem~\ref{MBEG} is that the \emph{topological discrete time-crystalline oscillation}, which is a toggle between different SPT phases \cite{Potter2017} generated by an MPU with nontrivial cohomology \cite{Gong2018c}, persists up to a time scale which increases at least linearly with respect to the system size. To see this, we have only to apply the theorem to $U^p$ with trivial cohomology, where $p$ is the order of the cohomology group.

\subsection{Applicability to interacting fermions}
\label{AFSPT}
As mentioned in the beginning of this section, a fermionic system in 1D can always be mapped into a spin chain through the Jordan-Wigner transformation: %to map the system into a $Ld$-spin (bosonic) system:
\begin{equation}
c^\dag_{ja}=%\left(
\sigma^+_{(j-1)d+a}\prod_{l<(j-1)d+a}\sigma^z_l,%\right).
\end{equation}
where $\sigma^z=\begin{bsmallmatrix}1 & 0\\0 & -1\end{bsmallmatrix}$ and $\sigma^+\equiv\frac{1}{2}(\sigma^x+i\sigma^y)=\begin{bsmallmatrix}0 & 1\\0 & 0\end{bsmallmatrix}$. Here we follow the setup in Sec.~\ref{DefES}, i.e., we consider $d$ internal states in each site of the original fermion system, so that the length of the corresponding spin chain is $Ld$. The fermion-parity operator reads
\begin{equation}
P_{\rm f}=(-)^{\sum_{j,a}c^\dag_{ja}c_{ja}}=\prod_{j,a}\sigma^z_{jd+a},
\end{equation}
which sets the superselection rule and is thus a $\mathbb{Z}_2$ symmetry of the Hamiltonian. Rigorously speaking, to keep the locality of the obtained spin Hamiltonian, we have to assume the open boundary condition. However, as long as the chain is sufficiently long, its subsystem property should not depend on the boundary condition. While a fermionic system never breaks the $\mathbb{Z}_2$ symmetry, the corresponding spin system may spontaneously break the $\mathbb{Z}_2$ symmetry, whenever the fermionic system exhibits a Majorana mode at the open boundaries \cite{Fidkowski2011,Chen2011b}. If the $\mathbb{Z}_2$ symmetry is not broken, the proof for 1D bosonic SPT systems applies directly to the fermionic SPT systems, which have even parities. Remarkably, after a few modifications, essentially the same proof applies even to fermionic SPT systems with odd parities.

While it is possible to directly describe fermionic SPT phases using fermionic MPSs \cite{Bultinck2017,Kapustin2018}, where the anti-commutativity of fermionic operators is encoded in an additional $\mathbb{Z}_2$-graded algebraic structure of the tensors, we would rather like to follow the approach in Refs.~\cite{Fidkowski2011,Chen2011b} and work in the spin picture after the Jordan-Wigner transformation. In the spin picture, denoting $|\Psi_{\rm SB}\rangle$ as a ``physical" symmetry-broken ground state, the corresponding exact (cat) ground states read \cite{Fidkowski2011}
\begin{equation}
\begin{split}
|\Psi\rangle&=\frac{1}{\sqrt{2}}(|\Psi_{\rm SB}\rangle+(-)^\eta P_{\rm f}|\Psi_{\rm SB}\rangle) \\
&=\frac{1}{\sqrt{2}}\sum_{\{j_s\}^L_{s=1}}\Tr[Z^{\eta}A_{j_1}...A_{j_L}]|j_1...j_L\rangle,
\end{split}
\label{cat}
\end{equation}
where $\eta=0,1$, $|\Psi_{\rm SB}\rangle$ is generated by a normal tensor $B_j$ without $P_{\rm f}$ symmetry and
\begin{equation}
Z=\sigma^z\otimes\mathbb{1}_{\rm v},\;\;\;\;A_j=(\sigma^z)^{|j|}\otimes B_j,\;\;\;\;j=1,2,...,2^d,
\end{equation}
with $|j|$ being the parity of state $|j\rangle$, i.e., $P_{\rm f}|j\rangle=(-)^{|j|}|j\rangle$. Let us consider the spectral property of the associated quantum channel: 
%\begin{equation}
$\mathcal{E}(\rho)\equiv\sum^{2^d}_{j=1}A_j\rho A_j^\dag=\lambda\rho$,
%\end{equation}
where $\lambda$ is an eigenvalue and $\rho$ can generally be expressed as $\rho=\sum_{w=0,x,y,z}\sigma^w\otimes\rho_w$.  
%($\sigma^y=\begin{bsmallmatrix}0 & -i\\i & 0\end{bsmallmatrix}$). 
%Since $zx=-xz$, $zy=-yz$, 
Defining $|w|\in\mathbb{Z}_2$ from $\sigma^w\sigma^z=(-)^{|w|}\sigma^z\sigma^w$, we obtain
\begin{equation}
\begin{split}
\mathcal{E}(\rho)&=\sum_{w=0,x,y,z}\sigma^w\otimes\sum^{2^d}_{j=1}(-)^{|w||j|}B_j\rho_wB^\dag_j \\
&=\sum_{w=0,x,y,z}\sigma^w\otimes\lambda\rho_w,
\end{split}
\end{equation}
implying that the spectrum of $\mathcal{E}$ is the union of those of $\mathcal{E}_\pm$ defined by $\mathcal{E}_\pm(\cdot)=\sum^{2^d}_{j=1}(\pm)^{|j|}B_j(\cdot)B^\dag_j$, with each eigenvalue doubled. The doubling arises from the fact that, given $\rho_+$ ($\rho_-$) as an eigenstate of $\mathcal{E}_+$ ($\mathcal{E}_-$) with eigenvalue $\lambda_+$ ($\lambda_-$), $\sigma^0\otimes\rho_+$ and $\sigma^z\otimes\rho_+$ ($\sigma^x\otimes\rho_+$ and $\sigma^y\otimes\rho_+$) are two degenerate eigenstates of $\mathcal{E}$ with the same eigenvalue $\lambda_+$ ($\lambda_-$). Recalling that $B_j$ is a normal tensor, we know that, after normalization, the spectral radius of $\mathcal{E}_+$ is one and that it only has a single eigenvalue equal to one. Let $\mu_-$ be the spectral radius of $\mathcal{E}_-$. Since $B_j$ is not $P_{\rm f}$-symmetric, we have $\mu_-<1$ \cite{Perez2008}. By gauge transforming $B_j$'s such that $\mathcal{E}_+(\mathbb{1}_{\rm v})=\mathbb{1}_{\rm v}$ and $\mathcal{E}_+(\Lambda_+)=\Lambda_+$, we have
\begin{equation}
\begin{split}
\mathcal{E}^\infty(\cdot)\equiv\lim_{l\to\infty}\mathcal{E}^l(\cdot)&=\frac{1}{2}\sigma^0\otimes\mathbb{1}_{\rm v}\Tr[\sigma^0\otimes\Lambda_+(\cdot)] \\
&+\frac{1}{2}\sigma^z\otimes\mathbb{1}_{\rm v}\Tr[\sigma^z\otimes\Lambda_+(\cdot)],
\end{split}
\end{equation}
which satisfies $Z\mathcal{E}^\infty(Z\cdot Z)Z=\mathcal{E}^\infty(\cdot)$. According to Fig.~\ref{fig9}, the ES in the thermodynamic limit of the subsystem is given by the spectrum of $\frac{1}{4}(\sigma^0\otimes\sigma^0+\sigma^z\otimes\sigma^z)\otimes\Lambda^{\otimes 2}_+$, which does not depend on $\eta$ in Eq.~(\ref{cat}) and is at least two-fold degenerate. If there are additional symmetries, then following the analysis for symmetric normal MPSs we know that a nontrivial projective representation on the virtual level may enforce an $r$-fold degeneracy in the spectrum of $\Lambda_+$, leading to a total of $2r^2$-fold degeneracy in the ES.

When the parity symmetry-broken state is evolved by a parity symmetric MPU $U$, we find that the cat-state structure (\ref{cat}) stays valid since $P_{\rm f}U=UP_{\rm f}$ by assumption. For an individual $|\Psi_{\rm SB}\rangle$, we know that the degeneracy in the time-evolved fixed point $\Lambda_+(t)$ determined from $U^t|\Psi_{\rm SB}\rangle$ should stay unchanged. Since the convergence bound given in Ref.~\cite{Wolf2015} applies equally to the quantum channels with multiple steady states, the Lieb-Robinson bound on the entanglement gap given in Theorem~\ref{MBEG} is also applicable to fermionic SPT states with Majorana modes. Moreover, the coeffecient $C$ in Eq.~(\ref{Koeffizient}) can be tightened by a factor of $2$ due to the normalization prefactor $\frac{1}{\sqrt{2}}$ in Eq.~(\ref{cat}). Also, denoting the spectral radius of $\mathcal{E}_+-\mathcal{E}^\infty_+$ as $\mu_+$, we have $\mu=\min\{\mu_+,\mu_-\}$.

\section{Discussions}
\label{Diskussionen}
While the Lieb-Robinson bound places a rigorous upper bound on the maximal splitting between degenerate ES values, it is usually a highly nontrivial problem to determine the degree of degeneracy, especially when the system undergoes partial symmetry breaking. In addition, the derivation of the rigorous Lieb-Robinson bounds on entanglement gaps makes full use of the translation invariance and the results are meaningful only up to $t^*\sim\frac{l}{v}$. It is natural to ask what happens when the translation invariance breaks down or/and in longer time scales. Here we give some heuristic arguments to address these issues. 

\subsection{Partial symmetry breaking}
\subsubsection{Free fermions}
For free-fermion systems with Altland-Zirnbauer symmetries, we have explained the effect of dynamical symmetry breaking and the reduction of symmetry classes \cite{Cooper2018b}. The only two classes whose reduced class is nontrivial are classes BDI and DIII, both of which reduce to class D. In stark contrast, the former gives a surjective group homomorphism $\mathbb{Z}\to\mathbb{Z}_2$, while the latter gives a trivial one $\mathbb{Z}_2\to0$ ($\subset\mathbb{Z}_2$). If we look at the ES dynamics for a quenched class BDI system with winding number $W\in\mathbb{Z}^+$, up to the Lieb-Robinson time $t^*$, we expect two persistent $\xi=\frac{1}{2}$ modes if $W$ is odd, while all the topological modes at $\xi=\frac{1}{2}$ immediately multifurcate if $W$ is even. For a nontrivial class DIII system, however, the entanglement gap should immediately open after a quench.

\begin{figure}
\begin{center}
       \includegraphics[width=6cm, clip]{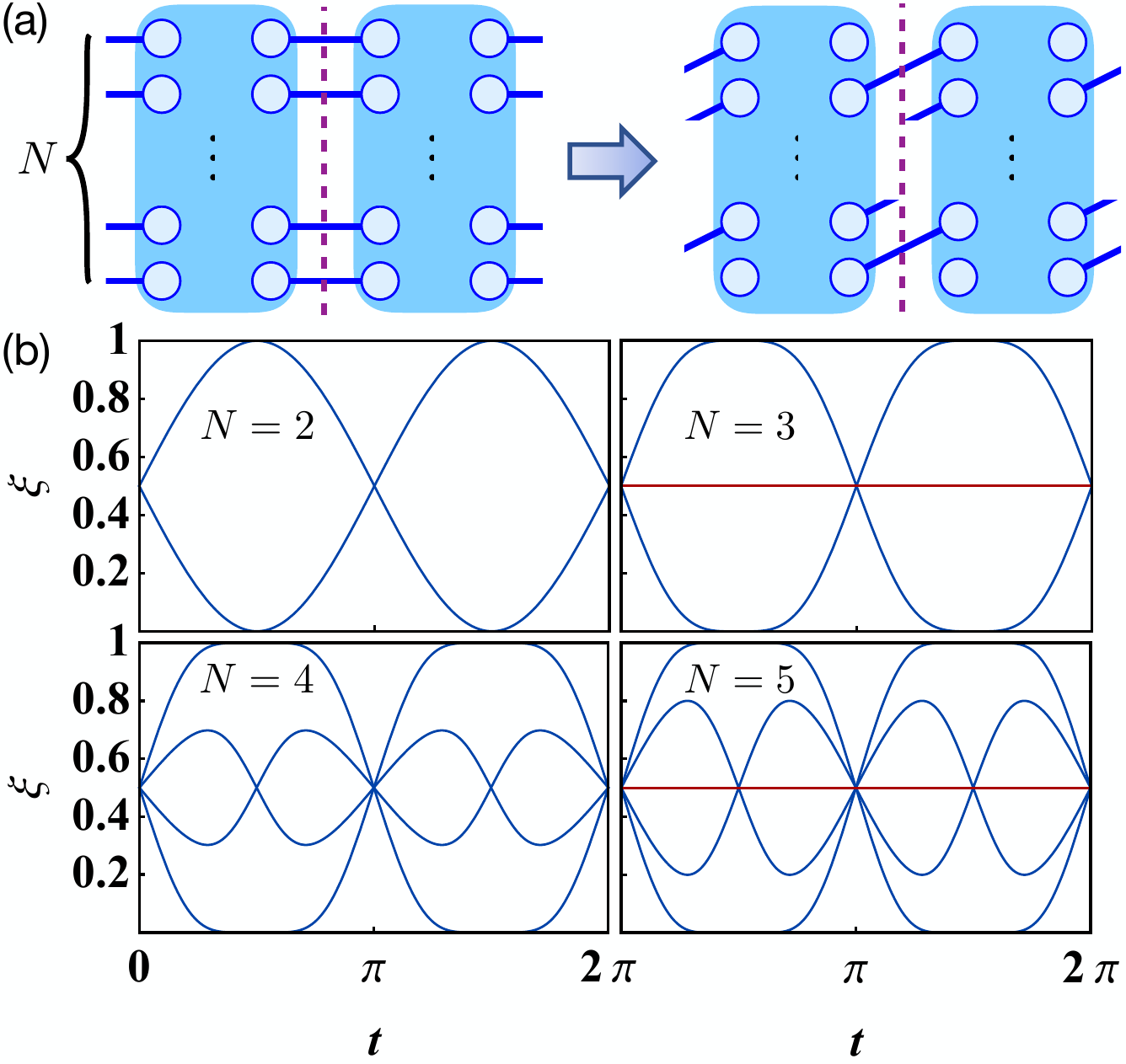}
       \end{center}
   \caption{(a) Quench from $N$ decoupled dimer chains described by $H_0$ (\ref{H0N}) to another flat-band Hamiltonian $H$ (\ref{HN}) with inter-chain couplings. Both $H_0$ and $H$ belong to class BDI. (b) Dynamics of the single-particle ES for $N=2,3,4,5$ demonstrating the nontrivial reduction $\mathbb{Z}\to\mathbb{Z}_2$. The red lines correspond to the topological entanglement modes at $\xi=\frac{1}{2}$, which persist only for odd $N$. Here we set $J=1$ in all panels.}
   \label{fig10}
\end{figure}

We give an analytically solvable \emph{flat-band} model to demonstrate the nontrivial $\mathbb{Z}\to\mathbb{Z}_2$ reduction. For simplicity, we focus on the ES under the open boundary condition, whose double gives the exact ES under the periodic boundary condition due to the zero correlation length imposed by the band flatness. Consider $N$ copies of dimer chains described by the Hamiltonian
\begin{equation}
H_0=-J_0\sum_j\sum^N_{\alpha=1}(c^\dag_{2j+1,\alpha}c_{2j,\alpha}+{\rm H.c.}),
\label{H0N}
\end{equation}
where $c_{x\alpha}$ denotes the fermion annihilation operator at the $x$th site of the $\alpha$th chain. Starting from the ground state of $H_0$, we suddenly quench the Hamiltonian into (see Fig.~\ref{fig10}(a))
\begin{equation}
H=-J\sum_j\sum^{N-1}_{\alpha=1}(c^\dag_{2j+1,\alpha}c_{2j,\alpha+1}+{\rm H.c.}).
\label{HN}
\end{equation}
As shown in Fig.~\ref{fig10}(b), the ES dynamics indeed demonstrates the expected surjective group homomorphism $\mathbb{Z}\to\mathbb{Z}_2$. The detailed calculations are given in Appendix~\ref{DESDF}, where we prove the existence/absence of $\xi=\frac{1}{2}$ mode for odd/even $N$. We also note that the $\mathbb{Z}\to\mathbb{Z}_2$ reduction here should be distinguished from that in Ref.~\cite{Gong2018b} which concerns the \emph{spatial-temporal} topology of quench dynamics starting from a topologically \emph{trivial} state according to a topological Hamiltonian.

\subsubsection{Interacting systems}
\label{subsubIS}
The Haldane phase \cite{Haldane1983}, which is a prototypical SPT phase corresponding to the nontrivial element in $H^2(\mathbb{Z}_2\times\mathbb{Z}_2,{\rm U}(1))=\mathbb{Z}_2$, always becomes trivial upon any symmetry breaking quench since $H^2(\mathbb{Z}_2,{\rm U}(1))=H^2(0,{\rm U}(1))=0$. See Ref.~\cite{Cooper2018b} for a numerical demonstration for an immediate opening of the entanglement gap. Here, we consider a more general situation where the initial SPT is protected by some unitary symmetries that form a group $G$, and the postquench Hamiltonian $H$ only respects the symmetries in a subgroup $\tilde G<G$.

\begin{table}[tbp]
\caption{Two possible quench protocols that partially break the symmetry into $\tilde G=\mathbb{Z}_2\times\mathbb{Z}_2$ and  $\tilde G=\mathbb{Z}_3\times\mathbb{Z}_3$ starting from a SPT state protected by $G=\mathbb{Z}_6\times\mathbb{Z}_6$. Here $\nu$ and $\tilde\nu$ label the SPT phases protected by $G$ and $\tilde G$, whose (minimal) degeneracies in the open-boundary ES are denoted by $r$ and $\tilde r$, respectively. In addition, $s\equiv\frac{r}{\tilde r}$ is the number of splitting ($s=1$ means no splitting), as numerically verified in Figs.~\ref{fig16}(b) and (c).}
\begin{center}
\begin{tabular}{cc|ccc|ccc}
\hline\hline
\multicolumn{2}{c|}{$G=\mathbb{Z}_6\times\mathbb{Z}_6$} & \multicolumn{3}{c}{$\tilde G=\mathbb{Z}_2\times\mathbb{Z}_2$
} & \multicolumn{3}{|c}{$\tilde G=\mathbb{Z}_3\times\mathbb{Z}_3$} \\ \hline
\;$\nu\in\mathbb{Z}_6$\; & \;\;\;$r$\;\;\; & \;$\tilde\nu\in\mathbb{Z}_2$\; & \;\;\;$\tilde r$\;\;\; & \;\;\;$s$%Split
\;\;\; & \;$\tilde\nu\in\mathbb{Z}_3$\; & \;\;\;$\tilde r$\;\;\; & \;\;\;$s$%Split
\;\;\; \\
\hline
$0$ & $1$ & $0$ & $1$ & $1$ & $0$ & $1$ & $1$  \\
$1$ & $6$ & $1$ & $2$ & $3$ & $2$ & $3$ & $2$  \\
$2$ & $3$ & $0$  & $1$ & $3$ & $1$ & $3$ & $1$  \\
$3$ & $2$ & $1$ & $2$ & $1$ & $0$ & $1$ & $2$  \\
$4$ & $3$ & $0$ & $1$ & $3$ & $2$ & $3$ & $1$  \\
$5$ & $6$ & $1$ & $2$ & $3$ & $1$ & $3$ & $2$ \\
\hline\hline
\end{tabular}
\end{center}
\label{table2}
\end{table}

It is, in general, very difficult to determine the reduced $\tilde G$-symmetric SPT phase and the corresponding ES degeneracy from a given $G$-symmetric SPT phase. See Appendix~\ref{KSPT} for an exact mathematical formulation of this problem in terms of category-theoretic languages. Here, let us consider a class of minimal but yet nontrivial examples --- $G=\mathbb{Z}_N\times\mathbb{Z}_N\to\tilde G=\mathbb{Z}_n\times\mathbb{Z}_n$, where $N=np$ with $p\in\mathbb{Z}^+$. Suppose that the initial state correspond to $\nu\in\mathbb{Z}_N=H^2(\mathbb{Z}_N\times\mathbb{Z}_N,{\rm U}(1))$. By choosing the coboundary gauge properly, the projective representation $V$ on the virtual level satisfies
\begin{equation}
V_{(a,b)}V_{(a',b')}=\omega^{\nu a'b}_NV_{(a+a',b+b')},%\;\;\;\;\forall(a,b),(a',b')\in\mathbb{Z}_N\times\mathbb{Z}_N.
\end{equation}
where $(a,b),(a',b')\in\mathbb{Z}_N\times\mathbb{Z}_N$. Regarding the elements in the subgroup $\mathbb{Z}_n\times\mathbb{Z}_n$, we find
\begin{equation}
\begin{split}
\tilde V_{a,b}\tilde V_{a',b'}&=V_{ap,bp}V_{a'p,b'p}\\
&=\omega^{\nu a'bp^2}_NV_{(a+a')p,(b+b')p} \\
&=\omega^{p\nu a'b}_n\tilde V_{a+a',b+b'},%\;\;\;\;\forall(a,b),(a',b')\in\mathbb{Z}_n\times\mathbb{Z}_n,
\end{split}
\end{equation}
where we have used the fact that $(a,b)\in\mathbb{Z}_n\times\mathbb{Z}_n$ corresponds to $(ap,bp)\in\mathbb{Z}_N\times\mathbb{Z}_N$. This implies that the group homomorphism from $\mathbb{Z}_N=H^2(\mathbb{Z}_N\times\mathbb{Z}_N,{\rm U}(1))$ to $\mathbb{Z}_n=H^2(\mathbb{Z}_n\times\mathbb{Z}_n,{\rm U}(1))$ is given by 
\begin{equation}
\nu\to \tilde\nu=p\nu\mod n. 
\end{equation}
Note that such a group homomorphism can be trivial, as is the case for $N=4$ and $n=2$ due to $\tilde\nu=2\nu\mod2=0$. The minimal realizations of nontrivial group homomorphisms turn out to be $N=6$ and $n=3$, which gives rise to $\mathbb{Z}_6\to\mathbb{Z}_3:\nu\to\tilde\nu=-\nu\mod3$, and $N=6$ and $n=2$, which gives rise to $\mathbb{Z}_6\to\mathbb{Z}_2:\nu\to\tilde\nu=\nu\mod2$. Using the fact that the ES of a $\mathbb{Z}_N\times\mathbb{Z}_N$-symmetric SPT state characterized by $\nu$ is (at least) $\frac{N}{{\rm GCD}(\nu,N)}$-fold degenerate under the open boundary condition \cite{Gong2018c}, where ${\rm GCD}$ is the greatest common divisor, we can figure out how many branches the original ES should split into. The results are summarized in Table~\ref{table2} and are numerically verified by some minimal models (see Appendix~\ref{MMSPT}).

\subsection{Effects of disorder} 
\label{Stoerung}
As schematically illustrated in the right top panel in Fig.~\ref{fig2}, disorder can dramatically alter the universal dynamical behavior of the entanglement growth. In this subsection, we qualitatively address the impact of disorder on the ES dynamics in 1D SPT systems.

\begin{figure}
\begin{center}
       \includegraphics[width=8cm, clip]{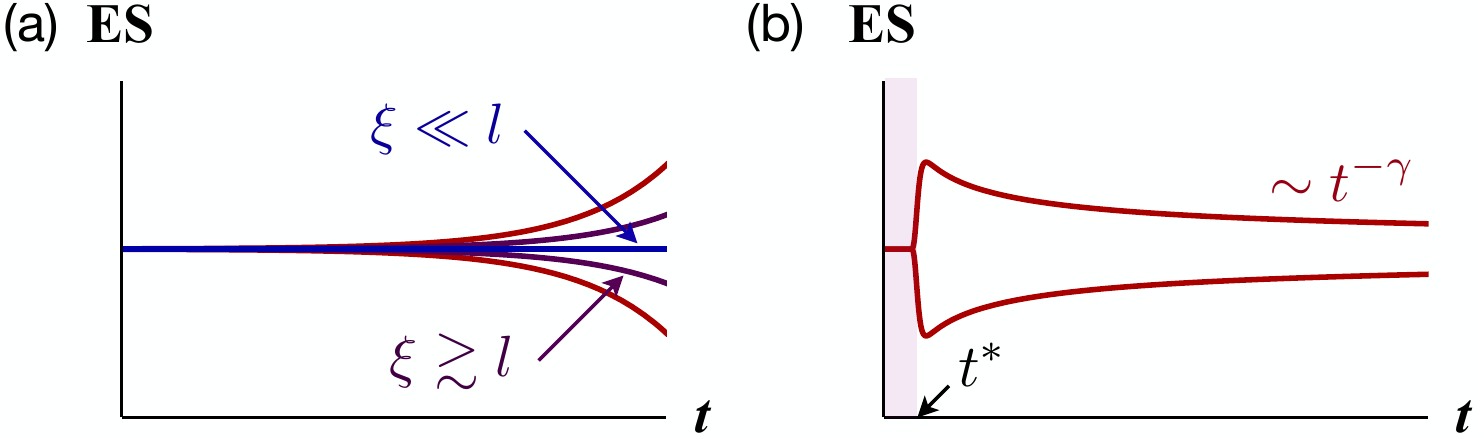}
       \end{center}
          \caption{(a) Expected influence of disorder on the single-particle-ES dynamics starting from a topological state. When the localization length $\xi$ is larger or comparable with the subsystem length $l$, we can still observe the splitting (purple curve) just like the clean case (red curve). On the other hand, if $l$ is much larger than $\xi$, the ES splitting will become invisible. (b) Power-law relaxation of the single-particle ES for $t\gg t^*$, where $t^*$ is the time scale set by the Lieb-Robinson bound. This phenomenon is discovered in Refs.~\cite{Chung2013,Chung2016,Chung2017}, where it is also conjectured that the $\xi=\frac{1}{2}$ modes will revive in the limit of $t\to\infty$ if $\bar H_\infty$ given in Eq.~(\ref{Havg}) is topologically nontrivial.}
          \label{fig11}
\end{figure}

Since the entanglement-gap opening is ultimately related to operator spreading and propagation of correlation, we expect disorder in the postquench Hamiltonian $H$ to \emph{stabilize} the SPT order during quench dynamics, provided that the disorder does respect the symmetry. For free fermions in 1D, the Anderson localization occurs at an arbitrarily small disorder strength and introduces a new length scale $\xi$ \cite{Mirlin2008}, which is a typical localization length of the eigenfunctions of $H$ and measures how far a wave packet can diffuse. As illustrated in Fig.~\ref{fig11}(a), for $l\gg\xi$, we expect that $\Delta^{\rm sp}_{\rm E}$ is saturated at $e^{-O(l)}$ and the SPT order survives even in the limit of $t\to\infty$ (see numerical signatures in Appendix~\ref{DSSH}). On the other hand, for very weak disorder, $l$ may be comparable to or even smaller than $\xi$, and we can still observe the opening of the entanglement gap. This occurs even for a half-chain bipartition, unless the disorder configuration happens to respect the half-chain translation symmetry.

In the presence of interactions, we may expect qualitatively different behavior because 
the entanglement-entropy growth %and correlation propagation 
is \emph{unbounded} in the many-body localized phase \cite{Nandkishore2015}, although it is logarithmical and hence extremely slow \cite{Znidaric2008,Pollmann2012,Altman2013,Abanin2013,Abanin2015}. In this case, we conjecture that, after a possible transient similar to the noninteracting case, $\Delta^{\rm mb}_{\rm E}$ grows no faster than 
%following 
a \emph{power law}, which is similar to the behavior of out-of-time-order correlators \cite{Chen2017,Swingle2017,Fan2017}. This is supported by the numerics on a phenomenological model (see Appendix~\ref{PMMBL}).

\subsection{Longer time scales}
While our rigorous results ensure the persistence of (approximate) ES degeneracy until a time scale that grows linearly long with the subsystem size, these results are not useful for describing the dynamical behavior at longer time scales. In fact, previous numerical studies on some two-band free-fermion models have revealed the possibility for a single-particle topological entanglement mode to return to $\xi=\frac{1}{2}$ in the long-time limit \cite{Chung2013,Chung2016,Chung2017}. In these studies, a conjecture is made to the effect that such behavior is determined by the topology of the time-averaged Hamiltonian 
\begin{equation}
\bar H_\infty\equiv\lim_{T\to\infty}\frac{1}{T}\int^T_0dt e^{-iHt}H_0e^{iHt}, 
\label{Havg}
\end{equation}
at least for two-band systems in class D and class BDI. It would be interesting to examine whether this is indeed true and, if yes, to what extent (e.g., multiple bands and other symmetry classes). Moreover, it is found in Ref.~\cite{Chung2017} that the long-time relaxation dynamics obey some universal \emph{power law}, i.e., $\Delta^{\rm sp}_{\rm E}(t)-\Delta^{\rm sp}_{\rm E}(\infty)\sim t^{-\gamma}$ ($\gamma>0$) for sufficiently large $t$ (see Fig.~\ref{fig11}(b)). This behavior can quantitatively be understood from steepest descent calculations, which require the lattice system to be infinite. Remarkably, the power-law relaxation of local observables has been rigorously proved for general finite single-band systems in arbitrary dimensions starting from any state with local correlations (not necessarily Gaussian) and within the revival time \cite{Eisert2016,Eisert2018,Murthy2018}. These rigorous results raise another interesting question concerning whether the signature of free-fermion topology could emerge in quench dynamics starting from an interacting quantum many-body state.

As for the long-time behavior of interacting systems, we generally do not have an extensive number of conserved quantities and the steady state should resemble an excited state of the posquench Hamiltonian $H$ with energy $\langle\Psi_0|H|\Psi_0\rangle$, provided that the eigenstate thermalization hypothesis holds true \cite{Rigol2008,Polkovnikov2011,Nandkishore2015}. Even if the ground state of $H$ is in an SPT phase, the topological features generally disappear in the excited states. Hence, the retrieval of ES degeneracy in the steady state seems unlikely. The scenario should be very simple if $H$ is many-body localized --- the topological information in the initial state, i.e., the degeneracy in the many-body ES, is expected to persist in the long-time limit without a transient loss or revival.

\section{Summary and outlook} 
\label{Zusammenfassung}
%Rigorous results in quantum many-body systems, represented by 
The Lieb-Schultz-Mattis-Oshikawa-Hastings theorem \cite{Lieb1961,Oshikawa2000,Hastings2004}, the Lieb-Robinson bound \cite{Lieb1972} and the entanglement area law \cite{Hastings2007} are of critical importance in quantum many-body systems. These rigorous results are of great current interest in light of theoretical insights from topological material science \cite{Parameswaran2013,Watanabe2015,Cheng2016} and quantum information \cite{Bravyi2006,Brandao2013,Brandao2015,Cho2018}, and of rapidly growing experimental relevance \cite{Bloch2012b,Monroe2014,Roos2014,Greiner2015,Greiner2016,Greiner2019}. In addition to these fundamental achievements, we have established a general principle for the dynamics of entanglement gaps in SPT systems that are driven out of equilibrium by a local Hamiltonian or MPU. For free fermions, we have extensively used both the band and the Wannier-function pictures to derive a Lieb-Robinson bound on the single-particle entanglement gap (Theorem~\ref{ThmspEG}). As a byproduct, we have clarified the relation between the Lieb-Robinson velocity and the group velocity of band dispersions. For interacting SPT systems, we have employed the tensor-network approaches and the techniques of function algebra to derive the Lieb-Robinson bound on the many-body entanglement gap (Theorem~\ref{MBEG}). This result illustrates that the shortcoming of tensor-network approaches as numerical tools does not hinder its powerfulness as analytical tools for dealing with long-time quantum dynamics. We have also considered the impact of partial symmetry breaking, in which case the ES degeneracy may immediately be lifted partially or completely. Possible effects of disorder and relaxation behaviors at longer time scales have also been discussed. 

As future studies, it would be of fundamental importance to go beyond the tensor-network formalism and prove a Lieb-Robinson bound on the many-body entanglement gap for continuous quench dynamics starting from exact ground states of local Hamiltonians. As discussed in Appendix~\ref{ISCE}, while improving MPUs to continuously generated unitaries alone seems plausible, it is far from clear how we can improve the assumption of MPS initial states to exact ground states. It is also natural to consider the generalization to higher dimensions \cite{Cirac2011b}, where we may have to use other indicators to measure the sharpness of SPT order. Moreover, some techniques used here may break down. For example, we cannot construct an exponentially localized Wannier function for a two-dimensional Chern insulator \cite{Marzari2007}. Finally, we note that considerable efforts have recently been made for generalizing the notions of Lieb-Robinson bounds and topological phases to lattice systems with long-range hoppings and interactions \cite{Gorshkov2014,Gorshkov2014b,Gorshkov2015,Vodola2016}. It would also be interesting to relax the locality assumption and consider long-range systems, which can naturally be implemented with trapped ions \cite{Monroe2014,Roos2014,Zhang2017,Zhang2017b}, Rydberg atoms \cite{Bernien2017}, polar molecules \cite{Ye2016} and nitrogen-vacancy centers \cite{Choi2017}.

\acknowledgements
We acknowledge Y. Ashida, M. A. Cazalilla, M.-C. Chung, I. Danshita, K. Fujimoto, R. Hamazaki, K. Kawabata, N. Matsumoto, and M. McGinley for valuable discussions. In particular, Z. G. appreciates Z. Wang for providing a simple picture about the monotonicity of Eq.~(\ref{impvLR}). This work was supported by KAKENHI Grant No. JP18H01145, No. JP17H02922 and a Grant-in-Aid for Scientific Research on Innovative Areas ``Topological Materials Science” (KAKENHI Grant No. JP15H05855) from the Japan Society for the Promotion of Science. Z. G. was supported by MEXT. N. K. was supported by the Leading Graduate Schools ``ALPS”. The authors thank the Yukawa Institute for Theoretical Physics at Kyoto University, where this work was initiated during the International Molecule Program YITP-T-18-01 on ``Floquet Theory: Fundamentals and Applications".

\appendix

\section{Band theory with a complex wave number}%Some results on 1D Bloch Hamiltonians with complex wave numbers}
\label{BTCWN}
In this appendix, we discuss how to define 1D Bloch Hamiltonians with complex wave numbers through analytic continuation. This is always possible for quasi-local hoppings, which decay exponentially with respect to the hopping range. We argue that the analytically continued Hamiltonian should generally be diagonalizable, even if the original one respects certain anti-unitary symmetries. In addition, we prove and explain why the Lieb-Robinson velocity in Eq.~(\ref{impvLR}) is monotonic with respect to the imaginary wave number and thus upper bounds the maximal relative group velocity. Finally, we provide an example of the SSH model.

\subsection{Analytic continuation}
\label{AC}
A Bloch Hamiltonian $H(k)$ can generally be expanded as
\begin{equation}
H(k)=\sum_{n\in\mathbb{Z}}e^{ikn}H_n,
\label{HkF}
\end{equation}
where each Fourier component can be obtained as
\begin{equation}
H_n=\int^\pi_{-\pi}\frac{dk}{2\pi} H(k) e^{-ikn}=H^\dag_{-n}.
\label{Hn}
\end{equation}
A sufficient condition for the rhs of Eq.~(\ref{HkF}) to converge is 
\begin{equation}
\sum_{n\in\mathbb{Z}}\|H_n\|<\infty,
\label{Hnconv}
\end{equation}
which is valid even if $H(k)$ is non-Hermitian and is obviously satisfied by any model with a finite hopping range $R$, i.e., 
\begin{equation}
H_n=H_{-n}=\mathbb{0},\;\;\;\;\forall n>R. 
\end{equation}
For such a class of models, the analytic continuation to the complex wave number
\begin{equation}
H(k+i\kappa)\equiv\sum_{n\in\mathbb{Z}} e^{ikn-\kappa n}H_n
\label{Hkappa}
\end{equation}
is well-defined for $\forall \kappa\in\mathbb{R}$ and satisfies $H(k+i\kappa+2\pi)=H(k+i\kappa)$ and $H(k+i\kappa)^\dag=H(k-i\kappa)$ for $\forall k\in[-\pi,\pi]$. 

The analytic continuation can be applied to a wider class of Bloch Hamiltonians whose Fourier components satisfy
\begin{equation}
\|H_n\|\le C_0 e^{-\kappa_0|n|},\;\;\;\;\forall n\in\mathbb{Z},
\label{Hnexp}
\end{equation}
where $C_0,\kappa_0\in\mathbb{R}^+$ do not depend on $n$. One can easily check the validity of Eq.~(\ref{Hnconv}). In this case, for $\forall \kappa\in(-\kappa_0,\kappa_0)$, we have
\begin{equation}
\|e^{-\kappa n}H_n\|\le C_0 e^{-(\kappa_0-|\kappa|)|n|},\;\;\;\;\forall n\in\mathbb{Z},
\end{equation}
so that $H(k+i\kappa)$ in Eq.~(\ref{Hkappa}) converges.  
%Moreover, for $\forall p\in\mathbb{Z}^+$, the $p$th-order derivative
%\begin{equation}
%\left.\frac{d^pH(K)}{dK^p}\right|_{K=k+i\kappa}\equiv\sum_{n\in\mathbb{Z}} (in)^pe^{ikn-\kappa n}H_n
%\end{equation}
%also converges due to
%\begin{equation}
%\|n^p e^{-\kappa n} H_n\|\le
%\end{equation}
Now let us show that if $H(k)$ is gapped and satisfies Eq.~(\ref{Hnexp}), then the flattened Hamiltonian or the Bloch projector (\ref{Blochproj}) also satisfies Eq.~(\ref{Hnexp}) but with $C_0$ and $\kappa_0$ modified. To see this, we note that the $n$th Fourier component of the Bloch projector
%\begin{equation}
%P_n=\int^\pi_{-\pi}\frac{dk}{2\pi}\int_{\gamma_<}\frac{dz}{2\pi i}\frac{e^{-ikn}}{z-H(k)}
%\end{equation}
can be written as
\begin{equation}
P_n=\int^\pi_{-\pi}\frac{dk}{2\pi}\oint_{\gamma_<}\frac{dz}{2\pi i}\frac{e^{-ikn-\kappa n}}{z-H(k-i\kappa)}
\label{Pn}
\end{equation}
by deforming the integral contour. With length of $\gamma_<$ denoted as $l_{\gamma_<}\equiv\oint_{\gamma_<}|dz|$, such an integral (\ref{Pn}) can be bounded from above by
\begin{equation}
\begin{split}
\|P_n\|&\le\int^\pi_{-\pi}\frac{dk}{2\pi}\oint_{\gamma_<}\frac{|dz|}{2\pi}\left\|\frac{1}{z-H(k-i\kappa)}\right\|e^{-\kappa n} \\
&\le \frac{l_{\gamma_<}}{2\pi}\max_{k\in[-\pi,\pi],z\in\gamma_<}\left\|\frac{1}{z-H(k-i\kappa)}\right\|e^{-\kappa n},
\end{split}
\end{equation}
provided that $\det[z-H(k-i\kappa)]\neq0$ for $\forall z\in\gamma_<$, which can always be satisfied by sufficiently small $\kappa$ due to a finite gap. This result ensures the existence of $\kappa>0$ such that $P_<(k)$ can be analytically continued to $\{k:|{\rm Re}\;k|\le\pi,|{\rm Im}\;k|\le\kappa\}$, implying that condition (i) in Theorem~\ref{ThmspEG} can always be satisfied.

On the other hand, the analytic continuation cannot be applied to long-range hopping, i.e., $\|H_n\|\sim O(n^{-\gamma})$ with $\gamma\ge 0$ --- while $H(k)$ stays well-defined for $\gamma>1$, Eq.~(\ref{Hkappa}) diverges whenever $\kappa\neq0$. Therefore, Theorem~\ref{ThmspEG} cannot be applied to long-range systems.

%\subsection{Bloch projector and its analyticity}
\subsection{Diagonalizability}
We argue that condition (ii) in Theorem~\ref{ThmspEG} is also satisfied in general --- that is, there always exists a nonzero $\kappa_0$ such that $H(k+i\kappa)$ is diagonalizable for $\forall |\kappa|<\kappa_0$. To show this, it is sufficient to show that $H(k+i\kappa)$ becomes non-diagonalizable only at a discrete set of complex wave numbers. This is expected to be true if $H(k+i\kappa)$ does not respect any symmetries, since an obvious mechanism for being nondiagonalizable is the emergence of a second-order exceptional point \cite{Heiss2012}, which requires the fine-tuning of two parameters. Here, these two parameters are $k$ and $\kappa$. However, if $H(k+i\kappa)$ respects an anti-unitary symmetry or anti-symmetry $\mathcal{S}$ for given $k$ and $\kappa$, i.e.,
\begin{equation}
\mathcal{S}H(k+i\kappa)\mathcal{S}^{-1}=\pm H(k+i\kappa),
\label{SHS}
\end{equation}
we only need to fine tune a single parameter to create an exceptional point. This type of Hamiltonians have been actively studied in the context of non-Hermitian topological systems \cite{,Gong2018,Kawabata2018,Zhou2019,Kawabata2019}, where $\mathcal{S}$ is typically the parity-time symmetry and $H(k+i\kappa)$ may not be analytic and thus $\kappa$ is no more than a control parameter.  In the following, we will show that $H(k+i\kappa)$, which is an analytic continuation of $H(k)$ with an anti-unitary symmetry $\mathcal{S}$, satisfies
\begin{equation}
\mathcal{S}H(k+i\kappa)\mathcal{S}^{-1}=\pm H(k-i\kappa)
\end{equation}
instead of Eq.~(\ref{SHS}), so we still need to fine-tune both $k$ and $\kappa$ to make $H(k+i\kappa)$ non-diagonalizable.

By imposing Eq.~(\ref{SHS}) for $\kappa=0$, we can use Eq.~(\ref{Hn}) to obtain
\begin{equation}
\begin{split}
\mathcal{S}H_n\mathcal{S}^{-1}&=\int^\pi_{-\pi}\frac{dk}{2\pi} \mathcal{S}H(k)\mathcal{S}^{-1} e^{ikn}\\&=\pm\int^\pi_{-\pi}\frac{dk}{2\pi}H(k)e^{ikn}=\pm H_{-n}.
\end{split}
\end{equation}
Therefore, the action of $\mathcal{S}$ on $H(k+i\kappa)$ gives
\begin{equation}
\begin{split}
\mathcal{S}H(k+i\kappa)\mathcal{S}^{-1}&=\sum_{n\in\mathbb{Z}} e^{-ikn-\kappa n}\mathcal{S}H_n\mathcal{S}^{-1} \\
&=\pm\sum_{n\in\mathbb{Z}} e^{i(k-i\kappa)(-n)}H_{-n}\\
&=\pm H(k-i\kappa).
\end{split}
\end{equation}
Similarly, we can show that $\mathcal{S}H(k)\mathcal{S}^{-1}=\pm H(-k)$, which is the case of the particle-hole symmetry, gives $\mathcal{S}H(k+i\kappa)\mathcal{S}^{-1}=\pm H(-k+i\kappa)$ by analytic continuation and thus $H(k+i\kappa)$ is expected to stay diagonalizable for sufficiently small $\kappa$.

%nevertheless, Jordan block, polynomial prefactor
Finally, we note that even if $H(k+i\kappa)$ happens to be non-diagonalizable, Eq.~(\ref{spEG}) should still be valid with $C$ replaced by a polynomial of $t$. This is because $H(k+i\kappa)$ can always be transformed into the Jordan normal form and a nontrivial Jordan block with size $s\ge2$ contributes a polynomial prefactor with degree $s-1$ in $e^{-iH(k+i\kappa)t}$. %Such a polynomial correction is unimportant compared with the exponential part. 

\subsection{Monotonicity of Eq.~(\ref{impvLR})}
\label{monov}
We prove that Eq.~(\ref{impvLR}) is monotonic in terms of $\kappa$ and thus reaches its minimum at $\kappa=0$. To simplify the notation, we first introduce
\begin{equation}
\Delta_{\alpha\beta}(k)\equiv\epsilon_{k\alpha}-\epsilon_{k\beta},
\end{equation}
which satisfies $\overline{\Delta_{\alpha\beta}(k+i\kappa)}=\Delta_{\alpha\beta}(k-i\kappa)$ for $k,\kappa\in\mathbb{R}$. For further simplification, we omit the subscript ``$\alpha\beta$" and define
\begin{equation}
v(k,\kappa)\equiv{\rm Re}\Delta'(k+i\kappa).
\end{equation}
We then have $v(k,\kappa)=v(k,-\kappa)$ and 
\begin{equation}
\begin{split}
V(k,\kappa)\equiv\frac{1}{2\kappa}\int^\kappa_{-\kappa}d\kappa'v(k,\kappa')
=\frac{{\rm Im}\Delta(k+i\kappa)}{\kappa}.
%|^{\kappa'=\kappa}_{\kappa'=-\kappa}
\end{split}
\label{Vkkappa}
\end{equation}
To show the monotonicity of $\max_{k\in[-\pi,\pi]}|V(k,\kappa)|$, it is sufficient to show that $V(k,\kappa)$ as a function of $k$ is \emph{majorized} by $V(k,\kappa')$ for $\forall \kappa'>\kappa$. By the statement that an integrable real function $f_1:I\equiv[a,b]\to\mathbb{R}$ majorizes $f_2:I\to\mathbb{R}$, we mean that there exists a kernel $K(x;x'):I\times I\to\mathbb{R}^+\bigcup\{0\}$ satisfying
\begin{equation}
\int_I dxK(x;x')=\int_I dx'K(x;x')=1
\label{Kx1}
\end{equation}
and
\begin{equation}
f_2(x)=\int_Idx'K(x;x')f_1(x'). 
\end{equation}
Such a definition is a straightforward generalization of the majorization for real vectors, which can also be regarded as functions with $I$ being a discrete set. That is, a real vector $\boldsymbol{a}$ is majorized by $\boldsymbol{b}$ if there exists a \emph{doubly stochastic matrix} $M_{\rm ds}$, whose entries are all non-negative and the sum of each row/column equals to one, such that $\boldsymbol{a}=M_{\rm ds}\boldsymbol{b}$ \cite{Arnold2010}. Note that $K(x;x')$ is a continuous version of $M_{\rm ds}$. 
%In analogy, we expect that $f_2$ is majorized by $f_1$ if and only if  
%If $f_1$ and $f_2$ are both continuous, 
If $f_1$ majorizes $f_2$, for $\forall x\in I$, we have
\begin{equation}
f_2(x)\le\int_Idx'K(x;x')\max_{x\in I}f_1(x)=\max_{x\in I} f_1(x)
\label{f2f1}
\end{equation}
due to $K(x;x')\ge0$, $f_1(x')\le\max_{x\in I}f_1(x)$ for $\forall x'\in I$ and Eq.~(\ref{Kx1}). Since Eq.~(\ref{f2f1}) is true for $\forall x\in I$, we have $\max_{x\in I}f_2(x)\le\max_{x\in I}f_1(x)$. Similarly, we have $\min_{x\in I}f_2(x)\le\min_{x\in I}f_1(x)$ and thus $\max_{x\in I}|f_1(x)|\ge \max_{x\in I}|f_2(x)|$. 

Now let us consider the kernel that transforms $V(k,\kappa_1)$ into $V(k,\kappa_2)$ with $\kappa_1\ge\kappa_2$. Since $v(k,\kappa)$ is a \emph{harmonic function} which is periodic in $k$ and even in $\kappa$, it can generally be expanded as 
\begin{equation}
v(k,\kappa)=\sum^\infty_{n=1}[a_n\cos(nk)+b_n\sin(nk)]\cosh(n\kappa),
\end{equation}
where $a_n,b_n\in\mathbb{R}$. Accordingly, we can obtain a general form of $V(k,\kappa)$ in Eq.~(\ref{Vkkappa}):
\begin{equation}
V(k,\kappa)=\sum^\infty_{n=1}[A_n\cos(nk)+B_n\sin(nk)]\frac{\sinh(n\kappa)}{\kappa},
\label{Vgen}
\end{equation}
where we have redefined the coefficients as $A_n=\frac{a_n}{n}$ and $B_n=\frac{b_n}{n}$. This general form (\ref{Vgen}) implies the following kernel that transforms $V(k,\kappa_1)$ into $V(k,\kappa_2)$:
\begin{equation}
K(k,\kappa_2;k',\kappa_1)=\frac{1}{2\pi}+\frac{1}{\pi}\sum^\infty_{n=1}\frac{\kappa_1\sinh(n\kappa_2)}{\kappa_2\sinh(n\kappa_1)}\cos[n(k-k')],
\label{K12}
\end{equation}
where the constant term is determined by Eq.~(\ref{Kx1}). To see why the kernel takes this form, we have only to note that $\cos[n(k-k')]=\cos(nk)\cos(nk')+\sin(nk)\sin(nk')$ and $\int^\pi_{-\pi}\frac{dk'}{\pi}\cos(nk')V(k',\kappa)$ ($\int^\pi_{-\pi}\frac{dk'}{\pi}\cos(nk')V(k',\kappa)$) gives the Fourier coefficient before $\cos(nk)$ ($\sin(nk)$), which should be modified in an $n$-dependent manner following the change of $\kappa$. We can rewrite Eq.~(\ref{K12}) into a more compact form
\begin{equation}
K(k,\kappa_2;k',\kappa_1)=\frac{1}{2\pi}\sum_{n\in\mathbb{Z}}\frac{\kappa_1\sinh(n\kappa_2)}{\kappa_2\sinh(n\kappa_1)}e^{in(k-k')}.
\label{KF}
\end{equation}
We can use Eq.~(\ref{KF}) to check that 
\begin{equation}
K(k,\kappa;k',\kappa)=\frac{1}{2\pi}\sum_{n\in\mathbb{Z}}e^{in(k-k')}=\delta(k-k'). 
\end{equation}
Another special case that transforms $V(k,\kappa)$ into $V(k,0)=v(k,0)$ reads
\begin{equation}
K(k,0;k',\kappa)=\frac{1}{2\pi}\sum_{n\in\mathbb{Z}}\frac{n\kappa}{\sinh(n\kappa)}e^{in(k-k')}.
\end{equation}
In general, whenever $\kappa_1>\kappa_2$, the Fourier coefficients in Eq.~(\ref{KF}) decays as $e^{-(\kappa_1-\kappa_2)|n|}$ for large $|n|$, implying the convergence.

\begin{figure}
\begin{center}
       \includegraphics[width=7cm, clip]{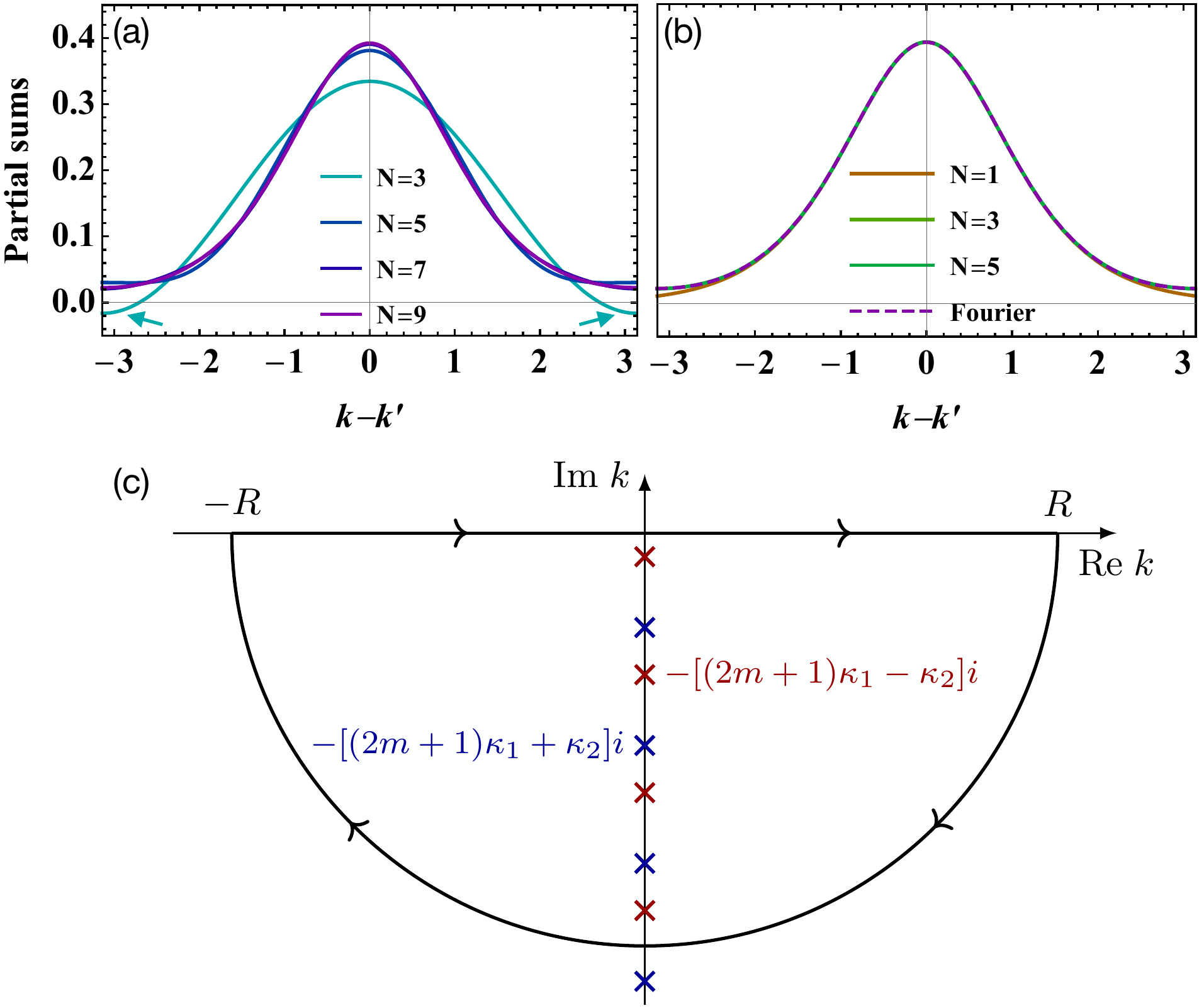}
       \end{center}
          \caption{Partial sums $\sum^{(N-1)/2}_{n=-(N-1)/2}$ of the kernel function $K(k,\kappa_2;k',\kappa_1)$ with $\kappa_1=2$ and $\kappa_2=0$ expanded as in (a) Eq.~(\ref{KF}) and (b) Eq.~(\ref{KY}). The curve labeled by ``Fourier" in (b) is nothing but the curve in (a) for $N=9$. Note that the partial sum of the Fourier series may be not positive for certain $k-k'$ (see arrows in (a)), but $K(k,\kappa_2;k',\kappa_1)$ is clearly positive from Eq.~(\ref{KY}). (c) Contour integral and poles of $Y(k;\kappa_2,\kappa_1)$. In the limit of $R\to\infty$, as in the case in Eq.~(\ref{FY}), we should sum up all the residues with respect to the poles on the lower-half imaginary axis.}
          \label{fig12}
\end{figure}

The remaining problem is to confirm whether Eq.~(\ref{KF}) is non-negative for $\forall\kappa_1>\kappa_2=0$. This is not clear at first glance since the partial sum in Eq.~(\ref{KF}) generally contains negative parts (see Fig.~\ref{fig12}(a)), especially when $\kappa_2$ is close to $\kappa_1$. Nevertheless, the positivity becomes clear in a different expansion:  
\begin{equation}
K(k,\kappa_2;k',\kappa_1)=\sum_{n\in\mathbb{Z}} Y(k-k'+2n\pi;\kappa_2,\kappa_1),
\label{KY}
\end{equation}
where
\begin{equation}
Y(k;\kappa_2,\kappa_1)=\frac{\sin(\frac{\kappa_2}{\kappa_1}\pi)}{2\kappa_2[\cosh(\frac{\pi }{\kappa_1}k)+\cos(\frac{\kappa_2}{\kappa_1}\pi)]}
\label{Y}
\end{equation}
is positive over $\mathbb{R}$. In perticular, when $\kappa_2=0$, Eq.~(\ref{Y}) becomes
\begin{equation}
Y(k;0,\kappa)=\frac{\pi}{2\kappa[\cosh(\frac{\pi }{\kappa}k)+1]}.
\end{equation}
To prove Eq.~(\ref{KY}), we have only to calculate the $n$th Fourier coefficient of the rhs:
\begin{equation}
\begin{split}
&\int^\pi_{-\pi}\frac{dk}{2\pi}\sum_{m\in\mathbb{Z}}Y(k+2m\pi;\kappa_2,\kappa_1)e^{-ink} \\
=&\int^\infty_{-\infty}\frac{dk}{2\pi}Y(k;\kappa_2,\kappa_1)e^{-ink}\\
%=&\int^\infty_{-\infty}\frac{dk}{2\pi i}\frac{(e^{i\frac{\kappa_2}{\kappa_1}\pi}-e^{-i\frac{\kappa_2}{\kappa_1}\pi})e^{\frac{\pi }{\kappa_1}k-ink}}{2\kappa_2(e^{\frac{\pi }{\kappa_1}k}+e^{i\frac{\kappa_2}{\kappa_1}\pi})(e^{\frac{\pi }{\kappa_1}k}+e^{-i\frac{\kappa_2}{\kappa_1}\pi})} \\
=&\int^\infty_{-\infty}\frac{dk}{4\pi i\kappa_2}\left[\frac{e^{-ink}}{e^{-\frac{\pi(k+i\kappa_2)}{\kappa_1}}+1}-\frac{e^{-ink}}{e^{-\frac{\pi(k-i\kappa_2)}{\kappa_1}}+1}\right] \\
=&\frac{\kappa_1}{2\pi\kappa_2}\sum^\infty_{m=0}(e^{-n[(2m+1)\kappa_1-\kappa_2]}-e^{-n[(2m+1)\kappa_1+\kappa_2]}) \\
=&\frac{\kappa_1}{2\pi\kappa_2}\frac{\sinh(n\kappa_2)}{\sinh(n\kappa_1)},
\end{split}
\label{FY}
\end{equation} 
where we have used the residue theorem in deriving the fourth equality (see Fig.~\ref{fig12}(c)). The final result in Eq.~(\ref{FY}) indeed coincides with that in Eq.~(\ref{KF}).

\begin{figure}
\begin{center}
       \includegraphics[width=7cm, clip]{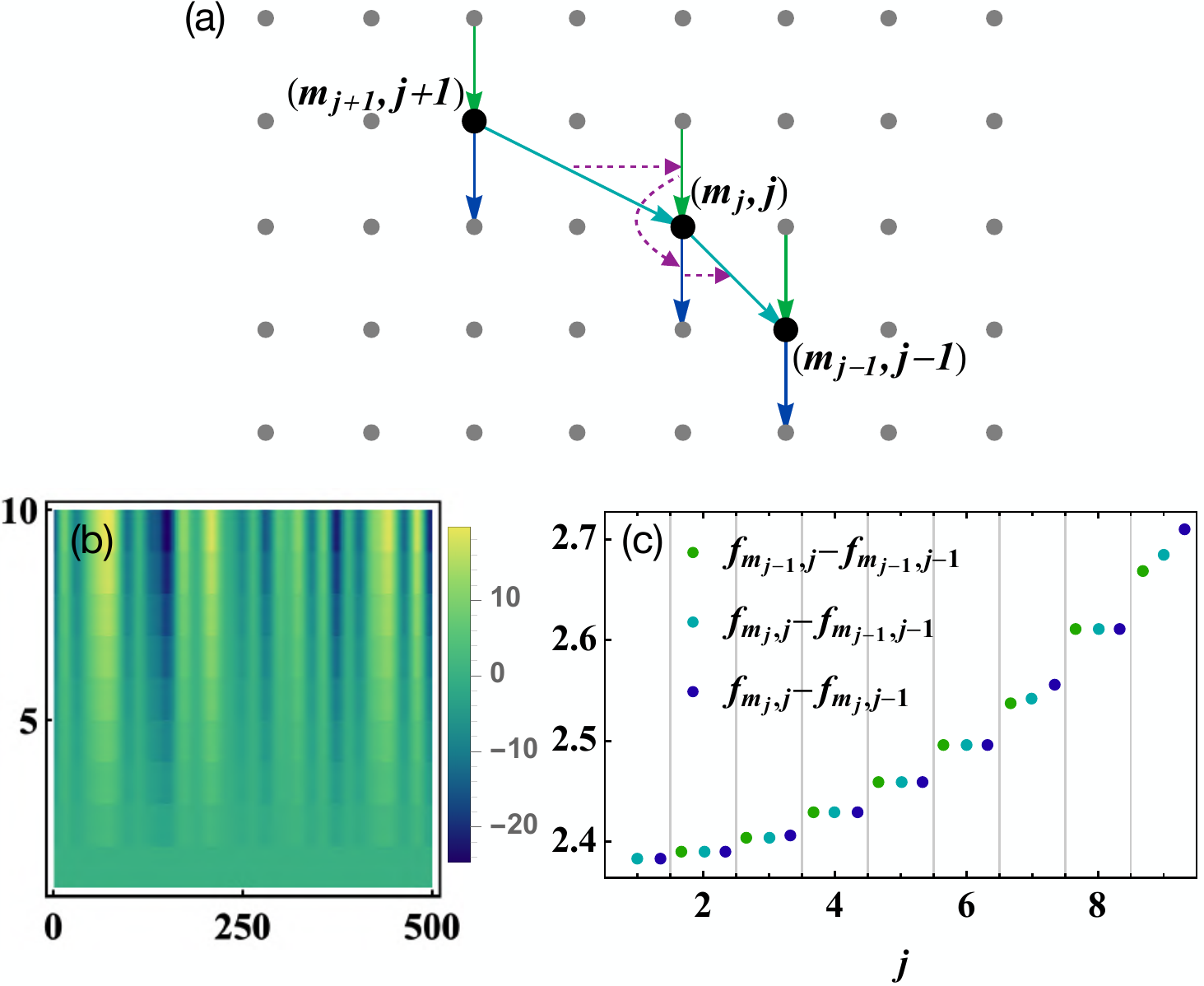}
       \end{center}
          \caption{(a) 2D array with the maximum of each row marked as $(m_j,j)$. The solid arrows denote the differences between the starting point to the end point, which are all positive. The dashed arrows highlight the order relations (descent along the arrow) between these differences. (b) A randomly generated array with $L=500$ and (c) the differences in (a) for the $j$th rows with $j=1,2,...,9$.}
          \label{fig13}
\end{figure}

In fact, a simple picture is available from a discrete version of the problem. We consider a 2D array $f:\mathbb{Z}\times\mathbb{Z}\to \mathbb{R}$, which satisfies that for $\forall j,j'\in\mathbb{Z}$, $f_{j,j'}=-f_{j,-j'}$ (so that $f_{j,0}=0$), $f_{j+L,j'}=f_{j,j'}$ ($L\in\mathbb{Z}^+$) and the discrete Laplace equation
\begin{equation} 
4f_{j,j'}=f_{j+1,j'}+f_{j-1,j'}+f_{j,j'+1}+f_{j,j'-1}.
\end{equation}
By imposing the boundary condition 
\begin{equation}
f_{j,1}=-f_{j,-1}=h\Delta'\left(\frac{2\pi j}{L}\right),
\end{equation}
we have $f(k,\kappa)\equiv{\rm Im}\Delta(k+i\kappa)=\lim_{L\to\infty,h\to0}f_{\lfloor \frac{kL}{2\pi}\rfloor,\lfloor \frac{\kappa}{h}\rfloor}$ since $f(k,\kappa)$ is a harmonic function satisfying 
\begin{equation}
\begin{split}
f(k,\kappa)=f(k+2\pi,\kappa)&=-f(k,-\kappa),\\
\partial_\kappa f(k,\kappa)|_{\kappa=0}&=\Delta'(k). 
\end{split}
\end{equation}
The discrete counterpart of the monotonicity of $\kappa^{-1}\max_{k\in[-\pi,\pi]}f(k,\kappa)$ reads
\begin{equation}
\frac{1}{j'+1}\max_{j\in\mathbb{Z}_L}f_{j,j'+1}\ge\frac{1}{j'}\max_{j\in\mathbb{Z}_L}f_{j,j'}
\label{dismono}
\end{equation} 
for $\forall j'\ge1$. In fact, we can prove a stronger result
\begin{equation}
\max_{j\in\mathbb{Z}_L}f_{j,j'+1}-\max_{j\in\mathbb{Z}_L}f_{j,j'}\ge\max_{j\in\mathbb{Z}_L}f_{j,j'}-\max_{j\in\mathbb{Z}_L}f_{j,j'-1},
\end{equation}
which implies Eq.~(\ref{dismono}). Denoting $m_{j'}$ as the horizontal label where $f_{m_{j'},j'}$ reaches the maximum for a given $j'$, %$\max_{j\in\mathbb{Z}_L}f_{j,j'}$, 
we have
\begin{equation}
\begin{split}
&\max_{j\in\mathbb{Z}_L}f_{j,j'+1}-\max_{j\in\mathbb{Z}_L}f_{j,j'}\\
\ge&f_{m_{j'},j'+1}-f_{m_{j'},j'} \\
=&3f_{m_{j'},j'}-f_{m_{j'}-1,j'}-f_{m_{j'}+1,j'}-f_{m_{j'},j'-1} \\
\ge &f_{m_{j'},j'}-f_{m_{j'},j'-1}\\
\ge &\max_{j\in\mathbb{Z}_L}f_{j,j'}-\max_{j\in\mathbb{Z}_L}f_{j,j'-1},
\end{split}
\end{equation}
which completes the proof. We provide a schematic illustration and a numerical verification in Fig.~\ref{fig13}. Similarly, we can prove the monotonicity of $\frac{1}{j'}\min_{j\in\mathbb{Z}_L}f_{j,j'}$ and thus the monotonicity of $\frac{1}{j'}\max_{j\in\mathbb{Z}_L}|f_{j,j'}|$.

The above idea can also be implemented directly in a continuous manner, provided that $k^*$, which makes $f(k^*,\kappa)$ maximal or minimal for a given $\kappa$, forms a smooth curve of $\kappa$. Let us focus on the case of maximum since the minimum counterpart is quite similar. By definition, along this curve we have
\begin{equation}
\partial^2_kfdk+\partial_k\partial_\kappa fd\kappa=0,\;\;\;\;
\partial^2_kf=-\partial^2_\kappa f\le0,
\label{p2f}
\end{equation}
%where the right equation is a result of $(\partial^2_k+\partial^2_\kappa)f=0$. 
so the second-order differential along this curve satisfies
\begin{equation}
\begin{split}
d^2f&=\partial^2_kfdk^2+\partial^2_\kappa fd\kappa^2+2\partial_k\partial_\kappa f dk d\kappa \\
&=\partial^2_\kappa fd\kappa^2-\partial^2_kfdk^2+2(\partial^2_kfdk+\partial_k\partial_\kappa fd\kappa)dk \\ 
&=\partial^2_\kappa f\left.\left[1+\left(\frac{dk}{d\kappa}\right)^2\right]\right|_{k=k^*}d\kappa^2.
\end{split}
\label{d2f}
\end{equation}
Introducing $F(\kappa)\equiv f(k^*(\kappa),\kappa)$, we find from Eqs.~(\ref{p2f}) and (\ref{d2f}) that
\begin{equation}
F''(\kappa)\ge0.
\label{Fpp}
\end{equation}
This result is sufficient to show
\begin{equation}
\kappa^{-1}_1F(\kappa_1)\ge\kappa^{-1}_2F(\kappa_2),\;\;\;\;\forall\kappa_1>\kappa_2\ge0.
\end{equation}
To see this, we consider the equivalent inequality
\begin{equation}
\frac{F(\kappa_1)-F(\kappa_2)}{\kappa_1-\kappa_2}\ge\frac{F(\kappa_2)-F(0)}{\kappa_2}.
\end{equation}
According to the mean-value theorem, there exists $\xi_1\in[\kappa_2,\kappa_1]$ and $\xi_2\in[0,\kappa_2]$ such that $F'(\xi_1)=\frac{F(\kappa_1)-F(\kappa_2)}{\kappa_1-\kappa_2}$ and $F'(\xi_2)=\frac{F(\kappa_2)-F(0)}{\kappa_2}$, so the above inequality becomes $F'(\xi_2)\ge F'(\xi_1)$, which is ensured by Eq.~(\ref{Fpp}).

\subsection{Example: SSH model}
We consider a general quench in the SSH model: 
\begin{equation}
\begin{split}
H_0(k)&=-(J_1+J_2\cos k)\sigma^x-J_2\sin k\sigma^y\\
\to\;\;H(k)&=-(J'_1+J'_2\cos k)\sigma^x-J'_2\sin k\sigma^y,
\end{split}
\label{SSHquench}
\end{equation}
where $\sigma^x$ and $\sigma^y$ are Pauli matrices. To evaluate $v_{\rm LR}$, we have only to numerically maximize 
\begin{widetext}
\begin{equation}
|{\rm Im}\;\epsilon_{k+i\kappa,\pm}|=\sqrt{\frac{1}{2}\left[\sqrt{(J'^2_1+J'^2_2+2J'_1J'_2\cos k\cosh\kappa)^2+(2J'_1J'_2\sin k\sinh\kappa)^2}-(J'^2_1+J'^2_2+2J'_1J'_2\cos k\cosh\kappa)\right]}
\end{equation}
\end{widetext}
over $k\in[-\pi,\pi]$. Note that $\kappa$ can be chosen freely, as long as $|\kappa|<|\ln\frac{J_1}{J_2}|$ (otherwise $P_<(k+i\kappa)$ may become nonanalytic). We plot $v_{\rm LR}=\kappa^{-1}\max_{k\in[-\pi,\pi]}|{\rm Im}\;\epsilon_{k+i\kappa,\pm}|$ in Fig.~\ref{fig7}(b) for the quench protocol used in the main text: $J_1=J'_2=0.5$ and $J_2=J'_1=1$. The $\kappa$ dependence of $v_{\rm LR}$ turns out to be rather weak, at least for $\kappa\in[0,\ln 2)$.

To evaluate $C$, we first write down the initial Bloch projector
\begin{equation}
P_<(k)=\frac{1}{2}\begin{bmatrix} 1 & \frac{J_1+J_2 e^{-ik}}{\sqrt{J^2_1+J^2_2+2J_1J_2\cos k}} \\  \frac{J_1+J_2 e^{ik}}{\sqrt{J^2_1+J^2_2+2J_1J_2\cos k}} & 1 \end{bmatrix}.
\label{SSHproj}
\end{equation}
By introducing
\begin{widetext}
\begin{equation}
F(k,\kappa,J_1,J_2)=\frac{J^2_1+J^2_2e^{2\kappa}+2J_1J_2e^{\kappa}
\cos k}{\sqrt{(J^2_1+J^2_2)^2+4J_1J_2(J^2_1+J^2_2)\cosh\kappa\cos k + 2J^2_1J^2_2(\cos 2k+\cosh2\kappa)}},
\end{equation}
\end{widetext}
we can express the norm of Eq.~(\ref{SSHproj}) with respect to a complex wave number as
\begin{equation}
\|P_<(k+i\kappa)\|=\frac{1}{2}[F(k,\kappa,J_1,J_2)+F(k,-\kappa,J_1,J_2)].
%=\frac{J^2_1+J^2_2\cosh2\kappa+2J_1J_2\cosh\kappa\cos k}{\sqrt{(J^2_1+J^2_2)^2+4J_1J_2(J^2_1+J^2_2)\cosh\kappa\cos k + 2J^2_1J^2_2(\cos 2k+\cosh2\kappa)}}
\label{Pless}
\end{equation}
Moreover, the left and right eigenvectors are found to be
\begin{equation}
\begin{split}
\langle k=0,\boldsymbol{a}|u^{\rm R}_{k+i\kappa,\pm}\rangle&=\frac{1}{\sqrt{2}}\begin{bmatrix} 1 \\ \mp\sqrt{\frac{J'_1+J'_2e^{ik-\kappa}}{J'_1+J'_2e^{-ik+\kappa}}} \end{bmatrix},\\
\langle k=0,\boldsymbol{a}|u^{\rm L}_{k+i\kappa,\pm}\rangle&=\frac{1}{\sqrt{2}}\begin{bmatrix} 1 \\ \mp\sqrt{\frac{J'_1+J'_2e^{ik+\kappa}}{J'_1+J'_2e^{-ik-\kappa}}} \end{bmatrix},
\end{split}
\end{equation}
where $\boldsymbol{a}$ labels the sublattice degrees of freedom, leading to
\begin{equation}
\begin{split}
\||u^{\rm R}_{k+i\kappa,\pm}\rangle\|&=\sqrt{\frac{1}{2}[1+F(k,-\kappa,J'_1,J'_2)]},\\
%=\sqrt{\frac{1}{2}+\frac{J'^2_1+J'^2_2e^{-2\kappa}+2J'_1J'_2e^{-\kappa}\cos k}{2\sqrt{(J'^2_1+J'^2_2)^2+4J'_1J'_2(J'^2_1+J'^2_2)\cosh\kappa\cos k + 2J'^2_1J'^2_2(\cos 2k+\cosh2\kappa)}}} \\ 
\||u^{\rm L}_{k+i\kappa,\pm}\rangle\|&=\sqrt{\frac{1}{2}[1+F(k,\kappa,J'_1,J'_2)]}.
%=\sqrt{\frac{1}{2}+\frac{J'^2_1+J'^2_2e^{2\kappa}+2J'_1J'_2e^{\kappa}\cos k}{2\sqrt{(J'^2_1+J'^2_2)^2+4J'_1J'_2(J'^2_1+J'^2_2)\cosh\kappa\cos k + 2J'^2_1J'^2_2(\cos 2k+\cosh2\kappa)}}} 
\end{split}
\label{uRuL}
\end{equation}
%To make a tighter estimation on the prefactor $C$, instead of Eq.~(\ref{newC}), we use
Substituting Eqs.~(\ref{Pless}) and (\ref{uRuL}) into Eq.~(\ref{newC}) yields
\begin{equation}
\begin{split}
C%&=\int^\pi_{-\pi}\frac{dk}{2\pi} \left(\sum_{\alpha=\pm}\||u^{\rm R}_{k+i\kappa,\alpha}\rangle\| \||u^{\rm L}_{k+i\kappa,\alpha}\rangle\|\right)^2 \|P_<(k+i\kappa)\| \\
&=\int^\pi_{-\pi}\frac{dk}{4\pi}[1+F(k,-\kappa,J'_1,J'_2)][1+F(k,\kappa,J'_1,J'_2)]\\
&\;\;\;\;\;\;\;\;\;\;\times[F(k,\kappa,J_1,J_2)+F(k,-\kappa,J_1,J_2)].
\end{split}
\end{equation}
%Combining with the previous results on $v_{\rm LR}$, we plot $Ce^{-\kappa(l-2v_{\rm LR}t)}$ in the right panel in Fig.~\ref{fig}, which confirms Thm.~\ref{fimp}.
In Fig.~\ref{fig6}(c), we choose $\kappa=0.6$, leading to %and the corresponding 
$C\simeq12.225$.

\section{Exact zero modes}
\label{EZM}
In Sec.~\ref{zeroEG}, we have argued that the emergent Kramers degeneracy and the nontrivial $\mathbb{Z}_2$ topology necessarily support zero modes in $R_{\rm d}$ in Eq.~(\ref{R}). Here, we prove this statement by showing that the invertibility of $R_{\rm d}$ implies a trivial $\mathbb{Z}_2$ topological index.  
\begin{theorem}
Given two skew-symmetric real matrices $R_{\rm d}$ and $R_{\rm o}$ such that $R\equiv\sigma^0\otimes R_{\rm d}+\sigma^x\otimes R_{\rm o}$ is unitary and $\Pf R=-1$, we must have $\det R_{\rm d}=0$.
\end{theorem}
\emph{Proof.---} If $\det R_{\rm d}\neq0$, we can apply the formula 
\begin{equation}
\Pf R=\Pf R_{\rm d}\Pf(R_{\rm d}-R_{\rm o}R^{-1}_{\rm d}R_{\rm o}).
\label{PfR}
\end{equation}
This formula can be derived from the identity
\begin{equation}
\Pf(B A B^{\rm T})=\det B\Pf A, 
\label{PfBAB}
\end{equation}
which is valid for arbitrary $A^{\rm T}=-A$ and $B$, and
\begin{equation}
\begin{split}
&\begin{pmatrix} R_{\rm d} & 0 \\ 0 & R_{\rm d}-R_{\rm o}R^{-1}_{\rm d}R_{\rm o} \end{pmatrix}\\
=&\begin{pmatrix} \mathbb{1} & 0 \\ -R_{\rm o}R^{-1}_{\rm d} & \mathbb{1} \end{pmatrix}
\begin{pmatrix} R_{\rm d} & R_{\rm o} \\ R_{\rm o} & R_{\rm d} \end{pmatrix}
\begin{pmatrix} \mathbb{1} & -R^{-1}_{\rm d}R_{\rm o} \\ 0 & \mathbb{1} \end{pmatrix}.
\end{split}
\end{equation}
Since $R$ is unitary, we have $\{R_{\rm d},R_{\rm o}\}=\mathbb{0}$ and $R^2_{\rm d}+R^2_{\rm o}=-\mathbb{1}$, leading to 
\begin{equation}
\begin{split}
R_{\rm d}-R_{\rm o}R^{-1}_{\rm d}R_{\rm o}
&=R_{\rm d}+R^{-1}_{\rm d}R^2_{\rm o}\\
&=R_{\rm d}+R^{-1}_{\rm d}(-\mathbb{1}-R^2_{\rm d})=-R^{-1}_{\rm d}.
\end{split}
\label{RdRo}
\end{equation}
Combining Eq.~(\ref{RdRo}) with Eq.~(\ref{PfR}), we have
\begin{equation}
\Pf R=\Pf R_{\rm d}\Pf(-R^{-1}_{\rm d})=\frac{(\Pf R_{\rm d})^2}{\det R_{\rm d}}=1,
\label{PfR1}
\end{equation}
where we have used Eq.~(\ref{PfBAB}) with $A=B=R^{-1}_{\rm d}$ and $\det A=(\Pf A)^2$ for $\forall A^{\rm T}=-A$. This result (\ref{PfR1}) contradicts the assumption $\Pf R=-1$, implying $\det R_{\rm d}=0$. $\square$

\section{Proof of Weyl's perturbation theorem}
\label{WPT}
In this appendix, we follow Ref.~\cite{Bhatia1997} to prove Weyl's perturbation theorem. To this end, we first introduce the \emph{min-max principle}.
\begin{lemma}[Min-max principle]
For any Hermitian operator $O$ on an $n$-dimensional Hilbert space, its $j$th largest eigenvalue $\lambda_j$ ($j=1,2,...,n$) is given by
\begin{equation}
\begin{split}
\lambda_j&=\max_{\dim V=j}\min_{|\psi\rangle\in V}\langle\psi|O|\psi\rangle \\
&=\min_{\dim W=n+1-j}\max_{|\psi\rangle\in W}\langle\psi|O|\psi\rangle,
\end{split}
\label{minmax}
\end{equation}
where $V$ and $W$ are Hilbert subspaces and $|\psi\rangle$ is a normalized state vector.
\end{lemma}
\emph{Proof.---} We denote the eigenstate with eigenvalue $\lambda_j$ as $|\psi_j\rangle$ ($j=1,2,...,n$) and define the following two special classes of Hilbert spaces with dimensions $j$ and $n+1-j$, respectively: 
\begin{equation}
\begin{split}
V_j&\equiv{\rm span}\{|\psi_1\rangle,|\psi_2\rangle,...,|\psi_j\rangle\},\\
W_j&\equiv{\rm span}\{|\psi_j\rangle,|\psi_{j+1}\rangle,...,|\psi_n\rangle\}.
\end{split}
\end{equation}
With $V$ chosen to be $V_j$, the rhs of the first line in Eq.~(\ref{minmax}) gives $\lambda_j$, implying
\begin{equation}
\lambda_j\le\max_{\dim V=j}\min_{|\psi\rangle\in V}\langle\psi|O|\psi\rangle.
\label{llm}
\end{equation}
Similarly, with a choice of $W=W_j$, the second line in Eq.~(\ref{minmax}) gives $\lambda_j$, implying
\begin{equation}
\lambda_j\ge\min_{\dim W=n+1-j}\max_{|\psi\rangle\in W}\langle\psi|O|\psi\rangle.
\label{gm}
\end{equation}
On the other hand, for an arbitrary Hilbert subspace $V$ with dimension $j$, we must have $\dim(V\bigcap W_j)\ge1$. Otherwise, if $V\bigcap W_j=\emptyset$, the full Hilbert space containing $V\bigcup W_j$ will be at least $n+1$ dimensional, contradicting the assumption. This implies that for $\forall V$ with $\dim V=j$, we have
\begin{equation}
\begin{split}
\min_{|\psi\rangle\in V}\langle\psi|O|\psi\rangle&\le\min_{|\psi\rangle\in V\bigcap W_j}\langle\psi|O|\psi\rangle \\
&\le \max_{|\psi\rangle\in W_j}\langle\psi|O|\psi\rangle=\lambda_j, 
\end{split}
\end{equation}
leading to
\begin{equation}
\lambda_j\ge\max_{\dim V=j}\min_{|\psi\rangle\in V}\langle\psi|O|\psi\rangle.
\label{lgm}
\end{equation}
Combining Eqs.~(\ref{llm}) and (\ref{lgm}), we obtain the first line in Eq.~(\ref{minmax}). By analogy, from $\dim(V_j\bigcap W)\ge1$ for $\forall W$ with $\dim W=n-j+1$, we can derive
\begin{equation}
\lambda_j\le\min_{\dim W=n+1-j}\max_{|\psi\rangle\in W}\langle\psi|O|\psi\rangle.
\label{lm}
\end{equation}
Combining Eqs.~(\ref{gm}) and (\ref{lm}), we obtain the second line in Eq.~(\ref{minmax}). $\square$

Now let us turn to the proof of Theorem~\ref{Weylper}. For $\forall V$ with $\dim V=j$, we can decompose $O$ as $O'+(O-O')$ to obtain
\begin{equation}
\begin{split}
&\min_{|\psi\rangle\in V}\langle\psi|O|\psi\rangle \\
=&\min_{|\psi\rangle\in V}(\langle\psi|O'|\psi\rangle+\langle\psi|O-O'|\psi\rangle) \\
\ge&\min_{|\psi\rangle\in V}\langle\psi|O'|\psi\rangle+\min_{|\psi\rangle\in V}\langle\psi|O-O'|\psi\rangle \\
\ge&\min_{|\psi\rangle\in V}\langle\psi|O'|\psi\rangle -\|O-O'\|.
\end{split}
\label{minV}
\end{equation}
After maximizing the rhs of Eq.~(\ref{minV}) with respect to $V$ and using the first line in Eq.~(\ref{minmax}), we obtain
\begin{equation}
\lambda_j\ge \lambda'_j-\|O-O'\|.
\label{lamg}
\end{equation}
Following a similar procedure, we can derive 
\begin{equation}
\max_{|\psi\rangle\in W}\langle\psi|O|\psi\rangle\le\max_{|\psi\rangle\in W}\langle\psi|O'|\psi\rangle+\|O-O'\|,
\end{equation}
which gives rise to (due to the second line in Eq.~(\ref{minmax}))
\begin{equation}
\lambda_j\le \lambda'_j+\|O-O'\|.
\label{laml}
\end{equation}
Theorem~\ref{Weylper} follows from the combination of Eqs.~(\ref{lamg}) and (\ref{laml}).

\section{Proof of Lemma~\ref{finitesize}}
\label{lem2}
We first show the following lemma which provides the relation between $|W^{(L)}_{j\alpha}\rangle$ and $|W^{(\infty)}_{j\alpha}\rangle$. %is exactly captured by
\begin{lemma}
Given $|W^{(\infty)}_{j\alpha}\rangle$ as a Wannier function of an infinite 1D lattice system, %which is localized at site $j$ and in band $\alpha$, 
the Wannier function $|W^{(L)}_{j\alpha}\rangle$ of the corresponding finite system with length $L$ reads
\begin{equation}
|W^{(L)}_{j\alpha}\rangle=\sum_{n\in\mathbb{Z}}P_{\mathbb{Z}_L}|W^{(\infty)}_{j+nL,\alpha}\rangle,
\label{WannierL}
\end{equation}
where $P_{\mathbb{Z}_L}\equiv\sum_{j\in\mathbb{Z}_L}\sum^d_{a=1}|ja\rangle\langle ja|$ is the projector onto the finite lattice, with $d$ and $|j,a\rangle$ being the number of internal states per site and the state localized at the $j$th site in an internal state $a$. %and $d$ being the number of internal states per site.
\label{fsWannier function}
\end{lemma}
\emph{Proof.---} Let us first write down the Wannier function defined on an infinite lattice:
\begin{equation}
|W^{(\infty)}_{j\alpha}\rangle=\int_{\rm B.Z.}\frac{dk}{2\pi}e^{ik(x-j)}|u_{k\alpha}\rangle,
\end{equation}
where $|u_{k\alpha}\rangle=\sum^d_{a=1} u_{\alpha a}(k)|k=0,a\rangle$ %with $u_{\alpha a}(k)$ 
is the Bloch wave function with $|k=0,a\rangle\equiv\sum_{j\in\mathbb{Z}}|ja\rangle$, and $x\equiv\sum_{j\in\mathbb{Z}}j|j\rangle\langle j|\otimes\mathbb{1}_{\rm I}$ is the position operator with $\mathbb{1}_{\rm I}$ being the identity acting on the internal-state Hilbert space. When the lattice is finite, say with length $L$, the Wannier function is defined in the form of a discrete Fourier transformation:
%degrees of freedom.
\begin{equation}
|W^{(L)}_{j\alpha}\rangle=\frac{1}{L}\sum_{k\in\frac{2\pi}{L}\mathbb{Z}_L}e^{ik(x_L-j)}|u_{k\alpha}\rangle,
\end{equation}
where $|u_{k\alpha}\rangle$ follows the previous definition for an infinite lattice except for $|k=0,a\rangle\equiv\sum_{j\in\mathbb{Z}_L}|ja\rangle$ and $e^{ik x_L}\equiv\sum_{j\in\mathbb{Z}_L}e^{ikj}|j\rangle\langle j|\otimes\mathbb{1}_{\rm I}$.

For $\forall j'\in\mathbb{Z}_L$ and $a=1,2,..,d$, the coefficient of $|j'a\rangle$ on the rhs in Eq.~(\ref{WannierL}) is given by
\begin{equation}
\begin{split}
\sum_{n\in\mathbb{Z}}\langle j'a|W^{(\infty)}_{j+nL,\alpha}\rangle=\sum_{n\in\mathbb{Z}}\int^{2\pi}_0\frac{dk}{2\pi}e^{ik(j'-j-nL)} u_{\alpha a}(k). %\langle l|u_{k,\alpha}\rangle
\end{split}
\label{ljnL}
\end{equation}
Using the identity
\begin{equation}
\sum_{n\in\mathbb{Z}}\delta(x-2n\pi)=\frac{1}{2\pi}\sum_{n\in\mathbb{Z}}e^{inx},
\end{equation}
we can further simplify the rhs of Eq.~(\ref{ljnL}) as
\begin{equation}
\sum_{n\in\mathbb{Z}}\int^{2\pi}_0\frac{dk}{2\pi}e^{ik(j'-j)}\delta(kL-2n\pi)u_{\alpha a}(k)=\langle j'a|W^{(L)}_{j\alpha}\rangle,
\end{equation}
which completes the proof of Lemma~\ref{fsWannier function}. $\square$

\begin{figure}
\begin{center}
       \includegraphics[width=8.5cm, clip]{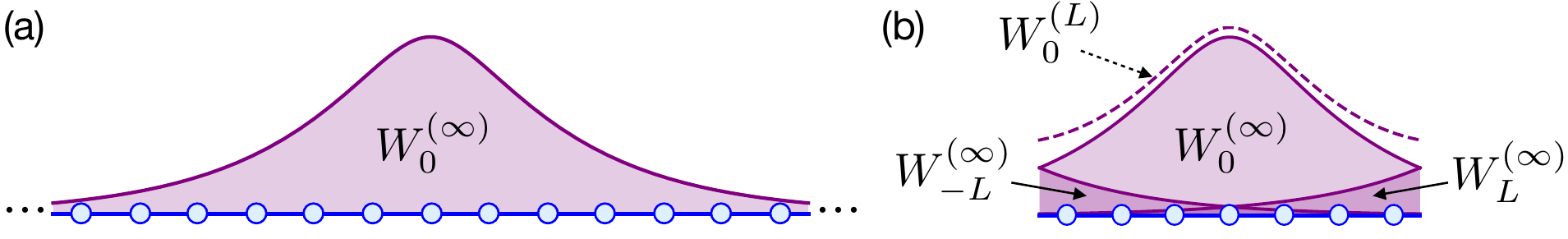}
       \end{center}
          \caption{(a) Wannier function $W^{(\infty)}_0$ centered at the origin of an infinite lattice. (b) Infinite-system Wannier functions $W^{(\infty)}_0$ and $W^{(\infty)}_{\pm L}$ projected onto a finite lattice $\mathbb{Z}_L$ under the periodic boundary condition. The finite-system Wannier function $W^{(L)}_0$ (dashed curve) is related to the infinite-system ones by Eq.~(\ref{WannierL}).}
          \label{fig14}
\end{figure}

A schematic illustration of Lemma~\ref{fsWannier function} is shown in Fig.~\ref{fig14}. Just like $|W^{(\infty)}_{j\alpha}\rangle$'s, we note that a finite-size Wannier function in Eq.~(\ref{WannierL}) satisfies the orthonormal relation:
\begin{equation}
\begin{split}
&\;\;\;\;\;\;\langle W^{(L)}_{j'\alpha'}|W^{(L)}_{j\alpha}\rangle\\
&=\sum_{n,n'\in\mathbb{Z}}\sum_{j''\in\mathbb{Z}_L,a}\langle W^{(\infty)}_{j'+n'L,\alpha'}|j''a\rangle\langle j''a|W^{(\infty)}_{j+nL,\alpha}\rangle \\
&=\sum_{n,n'\in\mathbb{Z}}\sum_{j''\in\mathbb{Z}_L,a}\langle W^{(\infty)}_{j'\alpha'}|j''-n'L,a\rangle \\
&\;\;\;\;\;\;\;\;\;\;\;\;\;\;\;\;\;\;\;\;\;\times\langle j''-n'L,a|W^{(\infty)}_{j+(n-n')L,\alpha}\rangle \\
&=\sum_{m\in\mathbb{Z}}\langle W^{(\infty)}_{j'\alpha'}|W^{(\infty)}_{j+mL,\alpha}\rangle=\delta_{j'j}\delta_{\alpha'\alpha},
\end{split}
\end{equation}
where we have used $\langle j'|W^{(\infty)}_{j\alpha}\rangle=\langle j'+p|W^{(\infty)}_{j+p,\alpha}\rangle$ for $\forall p\in\mathbb{Z}$, which is a property also inherited by $|W^{(L)}_{j\alpha}\rangle$ with $j',j\in\mathbb{Z}_L$.

Now we turn to prove Lemma~\ref{finitesize}. 
%\emph{Proof.---} 
We first express $\langle j| P^{(L)}_<| j'\rangle$, which is an operator acting on the internal-state Hilbert space, in terms of Wannier function projectors:
\begin{equation}
\langle j| P^{(L)}_<| j'\rangle =\sum_{j''\in\mathbb{Z}_L,\alpha\in\mathcal{O}}\langle j| W^{(L)}_{j''\alpha} \rangle\langle W^{(L)}_{j''\alpha} | j'\rangle.
\end{equation}
%where $\mathcal{O}$ consists of all the occupied bands. 
According to Eq.~(\ref{WannierL}), the above expression can be rewritten as
\begin{equation}
\langle j| P^{(L)}_<| j'\rangle = \sum_{\substack{j''\in\mathbb{Z}_L,\\\alpha\in\mathcal{O}}}\sum_{n,n'\in\mathbb{Z}}\langle j| W^{(\infty)}_{j''+nL,\alpha} \rangle\langle W^{(\infty)}_{j''+n'L,\alpha} | j'\rangle.
\label{PLele}
\end{equation}
On the other hand, we have
\begin{equation}
\langle j| P^{(\infty)}_< | j'\rangle = \sum_{\substack{j''\in\mathbb{Z}_L,\\\alpha\in\mathcal{O}}}\sum_{n\in\mathbb{Z}}\langle j| W^{(\infty)}_{j''+nL,\alpha} \rangle\langle W^{(\infty)}_{j''+nL,\alpha} | j'\rangle.
\label{Pinfele}
\end{equation}
Combining Eqs.~(\ref{PLele}) and (\ref{Pinfele}), we obtain
\begin{equation}
\begin{split}
&\;\;\;\;\;\;\langle j| P^{(L)}_<|j'\rangle - \langle j|P^{(\infty)}_< |j'\rangle \\
&= \sum_{j''\in\mathbb{Z}_L,\alpha\in\mathcal{O}}\sum_{n\neq n'\in\mathbb{Z}} \langle j| W^{(\infty)}_{j''+nL,\alpha} \rangle\langle W^{(\infty)}_{j''+n'L,\alpha} | j'\rangle \\
&= \sum_{\substack{j''\in\mathbb{Z}_L,\\\alpha\in\mathcal{O}}}\sum_{n\in\mathbb{Z}} \sum_{m\in\mathbb{Z}\backslash\{0\}} \langle j| W^{(\infty)}_{j''+nL,\alpha} \rangle\langle W^{(\infty)}_{j''+n L,\alpha} | j'+mL\rangle \\
&= \sum_{m\in\mathbb{Z}\backslash\{0\}}\langle j| P^{(\infty)}_< |j'+mL\rangle,
\end{split}
\end{equation}
where we again use $\langle j'|W^{(\infty)}_{j\alpha}\rangle=\langle j'+p|W^{(\infty)}_{j+p,\alpha}\rangle$ for $\forall p\in\mathbb{Z}$. %$\square$

\section{Basic properties of matrix-product unitaries}
\label{BPMPU}
We briefly review Ref.~\cite{Cirac2017} and introduce the several basic properties of MPUs, which are crucial for proving the main result. By definition, for $\forall L\in\mathbb{Z}^+$, $U$ in Eq.~(\ref{MPU}) in the main text obeys 
\begin{equation}
U^\dag U=\mathbb{I}\equiv\mathbb{1}^{\otimes L}\;\;\Rightarrow\;\;d^{-L}\Tr[U^\dag U]=1,
\end{equation}
implying that the spectrum of the quantum channel $\mathcal{E}_U(\cdot)=d^{-1}\sum^d_{j,j'=1}\mathcal{U}_{jj'}\cdot\mathcal{U}^\dag_{jj'}$ consists of a single $1$ and all the others being zero. This property enforces $d^{-\frac{1}{2}}\mathcal{U}_{jj'}$ to be a normal tensor \cite{Cirac2017b}, in the sense that $\mathcal{E}_U$ has a unique fixed point, which is Hermitian and positivie-definite. By choosing the gauge properly, i.e., performing a similarity transformation $\mathcal{U}_{jj'}\to X^{-1}\mathcal{U}_{jj'}X$ on the virtual level, $\mathcal{U}_{jj'}$ can always be made to satisfy 
\begin{equation}
\begin{tikzpicture}[scale=1.1]
\Text[x=-0.4]{$\frac{1}{d}$}
%ll
\draw[thick] (-0.15,0.375) -- (0.75,0.375) -- (0.75,-0.375) -- (-0.15,-0.375);
\draw[thick] (0.225,-0.75) -- (0.225,0.75);
\draw[thick] (0.225,-0.75) .. controls (0.25,-0.85) and (0.3,-0.85) .. (0.325,-0.7);
\draw[thick] (0.225,0.75) .. controls (0.25,0.85) and (0.3,0.85) .. (0.325,0.7);
\draw[thick,fill=blue!30] (0,0.15) rectangle (0.45,0.6);
\draw[thick,fill=blue!10]  (0,-0.15) rectangle (0.45,-0.6);
\draw[thick,fill=brown!10] (0.75,0) circle (0.15);
\Text[x=0.225,y=0.375,fontsize=\footnotesize]{$\overline{\mathcal{U}}$}
\Text[x=0.225,y=-0.375,fontsize=\footnotesize]{$\mathcal{U}$}
\Text[x=0.75,fontsize=\scriptsize]{$\rho$}
%ll
\Text[x=1.25]{$=$}
%lr
\draw[thick] (1.75,0.375) -- (2,0.375) -- (2,-0.375) -- (1.75,-0.375);
\draw[thick,fill=brown!10] (2,0) circle (0.15);
\Text[x=2,fontsize=\scriptsize]{$\rho$}
%lr
\Text[x=2.35,y=-0.2]{,}
\Text[x=3.45]{$\frac{1}{d}$}
%rl
\draw[thick] (4.6,0.375) -- (3.7,0.375) -- (3.7,-0.375) -- (4.6,-0.375);
\draw[thick] (4.225,-0.75) -- (4.225,0.75);
\draw[thick] (4.225,-0.75) .. controls (4.25,-0.85) and (4.3,-0.85) .. (4.325,-0.7);
\draw[thick] (4.225,0.75) .. controls (4.25,0.85) and (4.3,0.85) .. (4.325,0.7);
\draw[thick,fill=blue!30] (4,0.15) rectangle (4.45,0.6); 
\draw[thick,fill=blue!10] (4,-0.15) rectangle (4.45,-0.6);
\Text[x=4.225,y=0.375,fontsize=\footnotesize]{$\overline{\mathcal{U}}$}
\Text[x=4.225,y=-0.375,fontsize=\footnotesize]{$\mathcal{U}$}
%rl
\Text[x=5]{$=$}
%rr
\draw[thick] (5.75,0.375) -- (5.5,0.375) -- (5.5,-0.375) -- (5.75,-0.375);
%rr
\Text[x=6,y=-0.2]{,}
\end{tikzpicture}
\label{EUrho}
\end{equation}
where $\rho^\dag=\rho$, $\rho>0$ and $\Tr\rho=1$. It can be proved \cite{Cirac2017} that for $\forall k\in\mathbb{Z}^+$
\begin{equation}
\begin{tikzpicture}[scale=1.1]
%left
\draw[thick] (-0.3,0.375) -- (0.75,0.375) -- (0.75,-0.375) -- (-0.3,-0.375) -- cycle;
\draw[thick] (0.225,-0.75) -- (0.225,0.75);
\draw[thick,fill=blue!30] (0,0.15) rectangle (0.45,0.6);
\draw[thick,fill=blue!10]  (0,-0.15) rectangle (0.45,-0.6);
\draw[thick,fill=brown!10] (0.75,0) circle (0.15);
\Text[x=0.25,y=0.375,fontsize=\footnotesize]{$\overline{\mathcal{U}}_k$}
\Text[x=0.25,y=-0.375,fontsize=\footnotesize]{$\mathcal{U}_k$}
\Text[x=0.75,fontsize=\scriptsize]{$\rho$}
%left
\Text[x=1.25]{$=$}
%right
\draw[thick] (1.75,-0.7) -- (1.75,0.7);
\Text[x=1.85,y=-0.9,fontsize=\footnotesize]{$\mathbb{1}^{\otimes k}$}
%right
\Text[x=2.1,y=-0.2]{,}
\end{tikzpicture}
\label{Uid}
\end{equation}
where $\mathcal{U}_k$ is related to $\mathcal{U}$ by putting together $k$ sites into one, as given in Eq.~(\ref{block}).

An MPU is locality-preserving, in the sense that a local operator acting nontrivially on a finite segment stays local after being evolved by an MPU. This property can be understood from the following theorem \cite{Cirac2017}.
\begin{theorem}
It is always possible to make an MPU generated by $\mathcal{U}$ simple by putting together $k\le D^4_U$ ($D_U$: bond dimension of the MPU) sites into one (see Eq.~(\ref{block})). By simple, we mean that the building block $\mathcal{U}$ satisfies
\begin{equation}
\begin{tikzpicture}[scale=1.1]
%left
\draw[thick] (-0.15,0.375) -- (1.35,0.375) (-0.15,-0.375) -- (1.35,-0.375);
\draw[thick] (0.225,-0.75) -- (0.225,0.75) (0.975,-0.75) -- (0.975,0.75);
\draw[thick,fill=blue!30] (0,0.15) rectangle (0.45,0.6) (0.75,0.15) rectangle (1.2,0.6);
\draw[thick,fill=blue!10] (0,-0.15) rectangle (0.45,-0.6) (0.75,-0.15) rectangle (1.2,-0.6);
\Text[x=0.225,y=0.375,fontsize=\footnotesize]{$\overline{\mathcal{U}}$}
\Text[x=0.225,y=-0.375,fontsize=\footnotesize]{$\mathcal{U}$}
\Text[x=0.975,y=0.375,fontsize=\footnotesize]{$\overline{\mathcal{U}}$}
\Text[x=0.975,y=-0.375,fontsize=\footnotesize]{$\mathcal{U}$}
%left
\Text[x=1.85]{$=$}
%right
\draw[thick] (2.35,0.375) -- (3.25,0.375) -- (3.25,-0.375) -- (2.35,-0.375);
\draw[thick] (4.5,0.375) -- (3.6,0.375) -- (3.6,-0.375) -- (4.5,-0.375);
\draw[thick] (2.725,-0.75) -- (2.725,0.75) (4.125,-0.75) -- (4.125,0.75);
\draw[thick,fill=blue!30] (2.5,0.15) rectangle (2.95,0.6) (3.9,0.15) rectangle (4.35,0.6);
\draw[thick,fill=blue!10] (2.5,-0.15) rectangle (2.95,-0.6) (3.9,-0.15) rectangle (4.35,-0.6);
\draw[thick,fill=brown!10] (3.25,0) circle (0.15);
\Text[x=2.725,y=0.375,fontsize=\footnotesize]{$\overline{\mathcal{U}}$}
\Text[x=2.725,y=-0.375,fontsize=\footnotesize]{$\mathcal{U}$}
\Text[x=4.125,y=0.375,fontsize=\footnotesize]{$\overline{\mathcal{U}}$}
\Text[x=4.125,y=-0.375,fontsize=\footnotesize]{$\mathcal{U}$}
\Text[x=3.25,fontsize=\scriptsize]{$\rho$}
%right
\Text[x=4.8,y=-0.2]{,}
\end{tikzpicture}
\label{simple}
\end{equation}
where $\rho$ is given in Eq.~(\ref{EUrho}).
\label{MPUsimple}
\end{theorem}
Denoting $k_0$ as the smallest integer such that $\mathcal{U}_{k_0}$ is simple, the above theorem (\ref{MPUsimple}) implies that a local operator spreads by no more than $2k_0$ sites upon being evolved by the MPU. To see this, consider a general local operator $O_{\rm L}$ acting nontrivially on a segment with length $l_0$. Applying Eqs.~(\ref{Uid}) and (\ref{simple}) to $O'_{\rm L}=UO_{\rm L}U^\dag$ yields
\begin{equation*}
\begin{tikzpicture}[scale=1.1]
%left
\draw[thick] (-1.65,0.675) -- (1.35,0.675) (1.9,0.675) -- (4.15,0.675) (-1.65,-0.675) -- (1.35,-0.675) (1.9,-0.675) -- (4.15,-0.675);
\draw[thick] (-1.275,-1.05) -- (-1.275,1.05) (-0.525,-1.05) -- (-0.525,1.05) (0.225,-1.05) -- (0.225,1.05) (0.975,-1.05) -- (0.975,1.05) (2.275,-1.05) -- (2.275,1.05) (3.025,-1.05) -- (3.025,1.05) (3.775,-1.05) -- (3.775,1.05);
\draw[thick,fill=blue!30] (-1.5,0.45) rectangle (-1.05,0.9) (-0.75,0.45) rectangle (-0.3,0.9) (0,0.45) rectangle (0.45,0.9) (0.75,0.45) rectangle (1.2,0.9) (2.05,0.45) rectangle (2.5,0.9) (2.8,0.45) rectangle (3.25,0.9) (3.55,0.45) rectangle (4,0.9); 
\draw[thick,fill=blue!10] (-1.5,-0.45) rectangle (-1.05,-0.9) (-0.75,-0.45) rectangle (-0.3,-0.9) (0,-0.45) rectangle (0.45,-0.9) (0.75,-0.45) rectangle (1.2,-0.9) (2.05,-0.45) rectangle (2.5,-0.9) (2.8,-0.45) rectangle (3.25,-0.9) (3.55,-0.45) rectangle (4,-0.9);
\draw[thick,fill=yellow!10] (0,-0.225) rectangle (2.5,0.225);
\Text[x=1.625,y=0.675]{...}
\Text[x=1.625,y=-0.675]{...}
\Text[x=-1.95]{...}
\Text[x=4.45]{...}
\Text[x=-1.275,y=0.675,fontsize=\footnotesize]{$\overline{\mathcal{U}}$}
\Text[x=-0.525,y=0.675,fontsize=\footnotesize]{$\overline{\mathcal{U}}$}
\Text[x=0.225,y=0.675,fontsize=\footnotesize]{$\overline{\mathcal{U}}$}
\Text[x=0.975,y=0.675,fontsize=\footnotesize]{$\overline{\mathcal{U}}$}
\Text[x=2.275,y=0.675,fontsize=\footnotesize]{$\overline{\mathcal{U}}$}
\Text[x=3.025,y=0.675,fontsize=\footnotesize]{$\overline{\mathcal{U}}$}
\Text[x=3.775,y=0.675,fontsize=\footnotesize]{$\overline{\mathcal{U}}$}
\Text[x=-1.275,y=-0.675,fontsize=\footnotesize]{$\mathcal{U}$}
\Text[x=-0.525,y=-0.675,fontsize=\footnotesize]{$\mathcal{U}$}
\Text[x=0.225,y=-0.675,fontsize=\footnotesize]{$\mathcal{U}$}
\Text[x=0.975,y=-0.675,fontsize=\footnotesize]{$\mathcal{U}$}
\Text[x=2.275,y=-0.675,fontsize=\footnotesize]{$\mathcal{U}$}
\Text[x=3.025,y=-0.675,fontsize=\footnotesize]{$\mathcal{U}$}
\Text[x=3.775,y=-0.675,fontsize=\footnotesize]{$\mathcal{U}$}
\Text[x=1.25,fontsize=\footnotesize]{$O_{\rm L}$}
\draw[thick] (0.225,-1.1) .. controls (0.225,-1.15) and (1.25,-1.05) .. (1.25,-1.25);
\draw[thick] (2.275,-1.1) .. controls (2.275,-1.15) and (1.25,-1.05) .. (1.25,-1.25);
\Text[x=1.25,y=-1.4,fontsize=\footnotesize]{$l_0$}
%left
\end{tikzpicture}
\end{equation*}
\begin{equation}
\begin{tikzpicture}[scale=1.1]
\Text[x=5]{$=$}
%middle
\draw[thick] (6.025,-1.05) -- (6.025,1.05) (6.775,-1.05) -- (6.775,1.05) (7.525,-1.05) -- (7.525,1.05);
\draw[thick] (5.5,-0.675) -- (5.5,0.675) -- (8.05,0.675) -- (8.05,-0.675) -- cycle;
\draw[thick,fill=blue!30] (5.8,0.45) rectangle (6.25,0.9) (6.55,0.45) rectangle (7,0.9) (7.3,0.45) rectangle (7.75,0.9);
\draw[thick,fill=blue!10] (5.8,-0.45) rectangle (6.25,-0.9) (6.55,-0.45) rectangle (7,-0.9) (7.3,-0.45) rectangle (7.75,-0.9);
\draw[thick,fill=yellow!10] (6.55,-0.225) rectangle (7,0.225);
\draw[thick,fill=brown!10] (8.05,0) circle (0.15);
\Text[x=6.05,y=0.675,fontsize=\scriptsize]{$\overline{\mathcal{U}}_{k_0}$}
\Text[x=6.8,y=0.675,fontsize=\scriptsize]{$\overline{\mathcal{U}}_{l_0}$}
\Text[x=7.55,y=0.675,fontsize=\scriptsize]{$\overline{\mathcal{U}}_{k_0}$}
\Text[x=6.05,y=-0.675,fontsize=\scriptsize]{$\mathcal{U}_{k_0}$}
\Text[x=6.8,y=-0.675,fontsize=\scriptsize]{$\mathcal{U}_{l_0}$}
\Text[x=7.55,y=-0.675,fontsize=\scriptsize]{$\mathcal{U}_{k_0}$}
\Text[x=6.775,fontsize=\footnotesize]{$O_{\rm L}$}
\Text[x=8.05,fontsize=\scriptsize]{$\rho$}
%middle
\end{tikzpicture}
\label{lplus2k}
\end{equation}
\begin{equation*}
\begin{tikzpicture}[scale=1.1]
\Text[x=8.55]{$=$}
%right
\draw[thick] (9.225,-0.5) -- (9.225,0.5) (9.975,-0.5) -- (9.975,0.5) (11.275,-0.5) -- (11.275,0.5);
\draw[thick,fill=yellow!10] (9,-0.225) rectangle (11.5,0.225);
\Text[x=10.625,y=0.4]{...}
\Text[x=10.625,y=-0.4]{...}
\Text[x=10.25,fontsize=\footnotesize]{$O'_{\rm L}$}
\draw[thick] (9.225,-0.55) .. controls (9.225,-0.6) and (10.25,-0.5) .. (10.25,-0.7);
\draw[thick] (11.275,-0.55) .. controls (11.275,-0.6) and (10.25,-0.5) .. (10.25,-0.7);
\Text[x=10.25,y=-0.85,fontsize=\footnotesize]{$l_0+2k_0$}
%right
\Text[x=11.7,y=-0.2]{,}
\end{tikzpicture}
\end{equation*}
where local identities are omitted. Note that Fig.~2(c) in the main text is the special case of $l_0=1$, which is nevertheless sufficient for obtaining Eq.~(\ref{lplus2k}) since a local operator can generally be expressed as $O_{\rm L}=\sum_{\{j_s\}^{l_0}_{s=1}}c_{j_1j_2...j_{l_0}}O_{j_1}\otimes O_{j_2}\otimes ... \otimes O_{j_{l_0}}$ with $O_{j_s}$ being an on-site operator \cite{Gong2018c}.

Besides the locality-preserving property, an MPU is by definition unitary, implying the following lemma.
\begin{lemma}
Given an MPS $|\Psi\rangle$ generated by $\{A_j\}^d_{j=1}$ and an MPU $U$ generated by $\{\mathcal{U}_{jj'}\}^d_{j,j'=1}$, with the associated unital channel of the evolved MPS $|\Psi'\rangle=U|\Psi\rangle$ denoted as $\mathcal{E}'(\cdot)\equiv\sum^d_{j=1}A'_j\cdot A'^\dag_j$, the spectrum of $\mathcal{E}'$ is the same as that of $\mathcal{E}(\cdot)\equiv\sum^d_{j=1}A_j\cdot A^\dag_j$, the unital channel associated with $|\Psi\rangle$.
\label{invspe}
\end{lemma}
\emph{Proof.---} The main idea is already mentioned in Ref.~\cite{Gong2018c} --- we only have to combine Lemma~9 in Ref.~\cite{Perez2016} with $\Tr[(\sum_jA'_j\otimes\bar A'_j)^L]=\Tr[(\sum_jA_j\otimes\bar A_j)^L]$, $\forall L\in\mathbb{Z}^+$, which results from the unitary nature of the time evolution:
\begin{widetext}
\begin{equation}
\begin{tikzpicture}[scale=1.1]
%left
\draw[thick] (-3.775,-0.4) -- (-3.775,0.4) (-3.025,-0.4) -- (-3.025,0.4) (-1.975,-0.4) -- (-1.975,0.4);
\draw[thick] (-4.15,0.375) -- (-2.65,0.375)  (-2.35,0.375) -- (-1.6,0.375) (-4.15,-0.375) -- (-2.65,-0.375)  (-2.35,-0.375) -- (-1.6,-0.375);
\draw[thick] (-4.15,0.375) .. controls (-4.25,0.325) and (-4.25,0.3) .. (-4.125,0.25);
\draw[thick] (-1.6,0.375) .. controls (-1.5,0.325) and (-1.5,0.3) .. (-1.625,0.25);
\draw[thick] (-4.15,-0.375) .. controls (-4.25,-0.425) and (-4.25,-0.45) .. (-4.125,-0.5);
\draw[thick] (-1.6,-0.375) .. controls (-1.5,-0.425) and (-1.5,-0.45) .. (-1.625,-0.5);
\draw[thick,fill=purple!30] (-4,0.15) rectangle (-3.55,0.6) (-3.25,0.15) rectangle (-2.8,0.6) (-2.2,0.15) rectangle (-1.75,0.6); 
\draw[thick,fill=purple!10] (-4,-0.15) rectangle (-3.55,-0.6) (-3.25,-0.15) rectangle (-2.8,-0.6) (-2.2,-0.15) rectangle (-1.75,-0.6); 
\Text[x=-2.5]{...}
\Text[x=-3.775,y=0.375,fontsize=\footnotesize]{$\bar A'$}
\Text[x=-3.775,y=-0.375,fontsize=\footnotesize]{$A'$}
\Text[x=-3.025,y=0.375,fontsize=\footnotesize]{$\bar A'$}
\Text[x=-3.025,y=-0.375,fontsize=\footnotesize]{$A'$}
\Text[x=-1.975,y=0.375,fontsize=\footnotesize]{$\bar A'$}
\Text[x=-1.975,y=-0.375,fontsize=\footnotesize]{$A'$}
\draw[thick] (-3.775,-0.7) .. controls (-3.775,-0.75) and (-2.875,-0.65) .. (-2.875,-0.85);
\draw[thick] (-1.975,-0.7) .. controls (-1.975,-0.75) and (-2.875,-0.65) .. (-2.875,-0.85);
\Text[x=-2.875,y=-1,fontsize=\footnotesize]{$L$}
%left
\Text[x=-0.9]{$=$}
%middle
\draw[thick] (-0.15,1.125) -- (1.35,1.125) (1.65,1.125) -- (2.4,1.125) (-0.15,0.375) -- (1.35,0.375) (1.65,0.375) -- (2.4,0.375) (-0.15,-0.375) -- (1.35,-0.375) (1.65,-0.375) -- (2.4,-0.375) (-0.15,-1.125) -- (1.35,-1.125) (1.65,-1.125) -- (2.4,-1.125);
\draw[thick] (-0.15,1.125) .. controls (-0.25,1.075) and (-0.25,1.05) .. (-0.125,1.0);
\draw[thick] (2.4,1.125) .. controls (2.5,1.075) and (2.5,1.05) .. (2.375,1.0);
\draw[thick] (-0.15,0.375) .. controls (-0.25,0.325) and (-0.25,0.3) .. (-0.125,0.25);
\draw[thick] (2.4,0.375) .. controls (2.5,0.325) and (2.5,0.3) .. (2.375,0.25);
\draw[thick] (-0.15,-0.375) .. controls (-0.25,-0.425) and (-0.25,-0.45) .. (-0.125,-0.5);
\draw[thick] (2.4,-0.375) .. controls (2.5,-0.425) and (2.5,-0.45) .. (2.375,-0.5);
\draw[thick] (-0.15,-1.125) .. controls (-0.25,-1.175) and (-0.25,-1.2) .. (-0.125,-1.25);
\draw[thick] (2.4,-1.125) .. controls (2.5,-1.175) and (2.5,-1.2) .. (2.375,-1.25);
\draw[thick] (0.225,-1.3) -- (0.225,1.3) (0.975,-1.3) -- (0.975,1.3) (2.025,-1.3) -- (2.025,1.3);
\draw[thick,fill=blue!30] (0,0.15) rectangle (0.45,0.6) (0.75,0.15) rectangle (1.2,0.6) (1.8,0.15) rectangle (2.25,0.6); 
\draw[thick,fill=red!30] (0,0.9) rectangle (0.45,1.35) (0.75,0.9) rectangle (1.2,1.35) (1.8,0.9) rectangle (2.25,1.35); 
\draw[thick,fill=blue!10] (0,-0.15) rectangle (0.45,-0.6) (0.75,-0.15) rectangle (1.2,-0.6) (1.8,-0.15) rectangle (2.25,-0.6);
\draw[thick,fill=red!10] (0,-0.9) rectangle (0.45,-1.35) (0.75,-0.9) rectangle (1.2,-1.35) (1.8,-0.9) rectangle (2.25,-1.35); 
\Text[x=1.5]{...}
\Text[x=0.225,y=1.125,fontsize=\footnotesize]{$\bar A$}
\Text[x=0.225,y=0.375,fontsize=\footnotesize]{$\overline{\mathcal{U}}$}
\Text[x=0.225,y=-0.375,fontsize=\footnotesize]{$\mathcal{U}$}
\Text[x=0.225,y=-1.125,fontsize=\footnotesize]{$A$}
\Text[x=0.975,y=1.125,fontsize=\footnotesize]{$\bar A$}
\Text[x=0.975,y=0.375,fontsize=\footnotesize]{$\overline{\mathcal{U}}$}
\Text[x=0.975,y=-0.375,fontsize=\footnotesize]{$\mathcal{U}$}
\Text[x=0.975,y=-1.125,fontsize=\footnotesize]{$A$}
\Text[x=2.025,y=1.125,fontsize=\footnotesize]{$\bar A$}
\Text[x=2.025,y=0.375,fontsize=\footnotesize]{$\overline{\mathcal{U}}$}
\Text[x=2.025,y=-0.375,fontsize=\footnotesize]{$\mathcal{U}$}
\Text[x=2.025,y=-1.125,fontsize=\footnotesize]{$A$}
\draw[thick] (0.225,-1.45) .. controls (0.225,-1.5) and (1.125,-1.4) .. (1.125,-1.6);
\draw[thick] (2.025,-1.45) .. controls (2.025,-1.5) and (1.125,-1.4) .. (1.125,-1.6);
\Text[x=1.125,y=-1.75,fontsize=\footnotesize]{$L$}
%middle
\Text[x=3.1]{$=$}
%right
\draw[thick] (4.225,-0.4) -- (4.225,0.4) (4.975,-0.4) -- (4.975,0.4) (6.025,-0.4) -- (6.025,0.4);
\draw[thick] (3.85,0.375) -- (5.35,0.375)  (5.65,0.375) -- (6.4,0.375) (3.85,-0.375) -- (5.35,-0.375)  (5.65,-0.375) -- (6.4,-0.375);
\draw[thick] (3.85,0.375) .. controls (3.75,0.325) and (3.75,0.3) .. (3.875,0.25);
\draw[thick] (6.4,0.375) .. controls (6.5,0.325) and (6.5,0.3) .. (6.375,0.25);
\draw[thick] (3.85,-0.375) .. controls (3.75,-0.425) and (3.75,-0.45) .. (3.875,-0.5);
\draw[thick] (6.4,-0.375) .. controls (6.5,-0.425) and (6.5,-0.45) .. (6.375,-0.5);
\draw[thick,fill=red!30] (4,0.15) rectangle (4.45,0.6) (4.75,0.15) rectangle (5.2,0.6) (5.8,0.15) rectangle (6.25,0.6); 
\draw[thick,fill=red!10] (4,-0.15) rectangle (4.45,-0.6) (4.75,-0.15) rectangle (5.2,-0.6) (5.8,-0.15) rectangle (6.25,-0.6); 
\Text[x=5.5]{...}
\Text[x=4.225,y=0.375,fontsize=\footnotesize]{$\bar A$}
\Text[x=4.225,y=-0.375,fontsize=\footnotesize]{$A$}
\Text[x=4.975,y=0.375,fontsize=\footnotesize]{$\bar A$}
\Text[x=4.975,y=-0.375,fontsize=\footnotesize]{$A$}
\Text[x=6.025,y=0.375,fontsize=\footnotesize]{$\bar A$}
\Text[x=6.025,y=-0.375,fontsize=\footnotesize]{$A$}
\draw[thick] (4.225,-0.7) .. controls (4.225,-0.75) and (5.125,-0.65) .. (5.125,-0.85);
\draw[thick] (6.025,-0.7) .. controls (6.025,-0.75) and (5.125,-0.65) .. (5.125,-0.85);
\Text[x=5.125,y=-1,fontsize=\footnotesize]{$L$}
%right
\Text[x=7,y=-0.2]{.\;\;\;\;$\square$}
\end{tikzpicture}
\end{equation}
\end{widetext}
A direct corollary of Lemma~\ref{invspe} is that the spectrum of $\mathcal{E}$, i.e., the transfer matrix of an MPS, stays invariant during the stroboscopic time evolution governed by an MPU. In particular, $\mu$ as the spectral radius of $\mathcal{E}-\mathcal{E}^\infty$ is conserved.

\section{Interacting systems undergoing continuous evolution}
\label{ISCE}
We argue that the Lieb-Robinson bound on the many-body entanglement gap for MPUs implies qualitatively the same result for continuous time evolutions generated by local Hamiltonians. To this end, we first prove that the difference in time-evolution operators, which can be highly nonlocal, rigorously bounds the difference in any reduced density operator after the time evolution. 
\begin{lemma}
Given an arbitrary wave function $|\Psi_0\rangle$ defined on a bipartite system $S\bigcup\bar S$ and two unitaries $U$ and $U'$ satisfying $\|U-U'\|\le\epsilon$, denoting $\rho_S\equiv\Tr_{\bar S}|\Psi\rangle\langle\Psi|$ and $\rho'_S\equiv\Tr_{\bar S}|\Psi'\rangle\langle\Psi'|$ as the density operators of the evolved wave functions $|\Psi\rangle\equiv U|\Psi_0\rangle$ and $|\Psi'\rangle\equiv U'|\Psi_0\rangle$, we have $\|\rho_S-\rho'_S\|\le\epsilon$.
\label{UrhoS}
\end{lemma}
\emph{Proof.---} We first point out a useful norm inequality for the commutator of a positive-semidefinite Hermitian operator $A$ ($A^\dag=A$ and $A\ge0$) and an arbitrary operator $B$ \cite{Kittaneh2008}:
\begin{equation}
\|[A,B]\|\le\|A\|\|B\|.
\label{ABcom}
\end{equation}
Note that there is an improvement of factor $\frac{1}{2}$ compared with $\|[A,B]\|\le2\|A\|\|B\|$, which holds for arbitrary $A$ and $B$. 

Let us move on to prove Lemma~\ref{UrhoS}. Noting that $\rho_S-\rho'_S$ is Hermitian, according to the definition of the operator norm, we have
\begin{equation}
\begin{split}
&\|\rho_S-\rho'_S\|=\max_{\||\psi\rangle\|=1}\langle\psi|\rho_S-\rho'_S|\psi\rangle \\
=&\max_{P_\psi}\Tr[P_\psi\otimes\mathbb{1}_{\bar S}(|\Psi\rangle\langle\Psi|-|\Psi'\rangle\langle\Psi'|)]\\
%&\max_{\||\psi\rangle\|=1}\Tr[(|\psi\rangle\langle\psi|\otimes\mathbb{1}_{\bar S})(|\Psi\rangle\langle\Psi|-|\Psi'\rangle\langle\Psi'|)]\\
=&\max_{P_\psi} \langle\Psi_0|U(P_\psi\otimes\mathbb{1}_{\bar S})U^\dag-U'(P_\psi\otimes\mathbb{1}_{\bar S})U'^\dag|\Psi_0\rangle \\
\le&\max_{P_\psi}\|U(P_\psi\otimes\mathbb{1}_{\bar S})U^\dag-U'(P_\psi\otimes\mathbb{1}_{\bar S})U'^\dag\| \\
\le&\max_{P_\psi}\|[P_\psi\otimes\mathbb{1}_{\bar S},U^\dag U']\|,
\end{split}
\label{rhoSSp}
\end{equation}
where $P_\psi\equiv|\psi\rangle\langle\psi|$ is a rank-one projector, i.e., $P^2_\psi=P_\psi$ and $\Tr P_\psi=1$. Since $P_\psi\otimes\mathbb{1}_{\bar S}\ge0$ and $\|P_\psi\otimes\mathbb{1}_{\bar S}\|=1$, %for $\forall P_\psi$, 
we find from Eq.~(\ref{ABcom}) that for $\forall P_\psi$
\begin{equation}
\begin{split}
&\|[P_\psi\otimes\mathbb{1}_{\bar S},U^\dag U']\|\\
=&\|[P_\psi\otimes\mathbb{1}_{\bar S},U^\dag U'-\mathbb{1}]\|\\
\le&\|U^\dag U'-\mathbb{1}\|=\|U-U'\|.
\end{split}
\label{PpsiU}
\end{equation}
Combining Eq.~(\ref{PpsiU}) with Eq.~(\ref{rhoSSp}), we obtain
\begin{equation}
\|\rho_S-\rho'_S\|\le\|U-U'\|\le\epsilon.\;\square
\end{equation}

\begin{figure}
\begin{center}
       \includegraphics[width=8cm, clip]{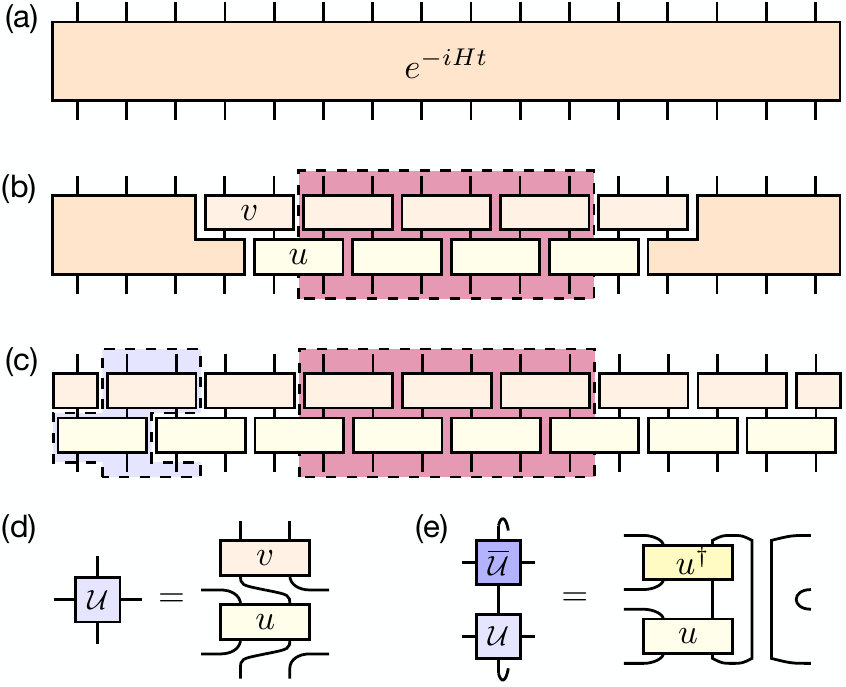}
       \end{center}
          \caption{(a) Continuous time evolution $e^{-iHt}$ generated by a local Hamiltonian $H$ and its (b) partial and (c) complete approximations by local unitaries, which are denoted as $U'_{\rm c}$ and $U_{\rm c}$.  If we focus on a subsystem marked in the red rectangle, (b) and (c) make no difference --- starting from the same initial state, the reduced density operators of the states evolved by (b) and (c) coincide with each other. (d) Building block of (c), which can be regarded as an MPU. (e) Quantum channel associated with (d) takes the form $\rho\Tr[...]$.}
          \label{fig15}
\end{figure}

Now let us discuss how to combine Lemma~\ref{UrhoS} with Theorem~\ref{MBEG} and the main result of Ref.~\cite{Osborne2006} to bound the many-body entanglement gap in a continuous quench dynamics starting from an SPT MPS. According to Ref.~\cite{Osborne2006}, given a local Hamiltonian $H=\sum_j h_j$, we can approximate it with a bilayer unitary circuit or quantum cellular automaton $U_{\rm c}$ such that   
\begin{equation}
\|e^{-iHt}- U_{\rm c}\|\le O\left(\frac{L}{|\Omega|}\right) e^{-\kappa_{\rm c}(|\Omega|-v_{\rm c}t)},
\end{equation}
where $L$ is the total system size, $|\Omega|$ is the number of sites that a single unitary acts on, and $\kappa_{\rm c}$ and $v_{\rm c}$ are constants independent of $L$. While the bound diverges in the thermodynamic limit, we can make a cut off of the circuit approximation at the length scale of the subsystem without changing the reduced density operator. For such a truncated circuit $U'_{\rm c}$, we have 
\begin{equation}
\|e^{-iHt}- U'_{\rm c}\|\le O\left(\frac{l}{|\Omega|}\right) e^{-\kappa_{\rm c}(|\Omega|-v_{\rm c}t)},
\label{Ucp}
\end{equation}
where the rhs is finite in the thermodynamic limit. See Figs.~(\ref{fig15})(a), (b) and (c) for the relations and distinctions in $e^{-iHt}$, $U'_{\rm c}$ and $U_{\rm c}$.

On the other hand, to apply Theorem~\ref{MBEG}, we should still regard the approximated reduced density operator as resulting from the time evolution by $U_{\rm c}$. After putting $|\Omega|$ sites into one, we can regard $U_{\rm c}$ as an MPU generated by the block given in Fig.~\ref{fig15}(d), which is simple (see Fig.~\ref{fig15}(e)). Under such a rescaling, the parameters in Theorem~\ref{MBEG} reads $k_0=t=1$, $D_U=d^{|\Omega|}$ and
\begin{equation}
\mu\to\mu^{|\Omega|},\;\;\;\;l\to\frac{l}{|\Omega|},
\end{equation}
while $D$ stays invariant. Therefore, the many-body entanglement gap of the approximated density operator $\rho'_{[1,l]}$ is bounded by  
\begin{equation}
\Delta'^{\rm mb}_{\rm E}\le {\rm poly}\left(\frac{l}{|\Omega|}\right)e^{-\kappa l+\kappa' |\Omega|},
\label{mbEGp}
\end{equation}
where $\kappa=-\ln\mu$ and
\begin{equation}
\kappa'=\ln d+(D^2+1)\kappa,
\label{kappap}
\end{equation}
provided that
\begin{equation}
\frac{l}{|\Omega|}-2\ge\coth\left(\frac{\kappa}{2}|\Omega|\right).
%\frac{1+\mu^{|\Omega|}}{1-\mu^{|\Omega|}}
\label{lOmega}
\end{equation}

As for the many-body entanglement gap of the exact density operator $\rho_{[1,l]}$, we have
\begin{equation}
\begin{split}
\Delta^{\rm mb}_{\rm E}&\equiv|\zeta_{r^2}-\zeta_1| \\
&\le |\zeta_{r^2}-\zeta'_{r^2}|+|\zeta_{r^2}-\zeta'_1|+|\zeta'_{r^2}-\zeta'_1| \\
&\le 2\|\rho_{[1,l]}-\rho'_{[1,l]}\|+\Delta'^{\rm mb}_{\rm E} \\
&\le 2\|e^{-iHt}-U'_{\rm c}\|+\Delta'^{\rm mb}_{\rm E},
\end{split}
\label{mbEGr}
\end{equation}
where we have used Lemma~\ref{UrhoS} in deriving the last inequality. Combining Eqs.~(\ref{Ucp}) and (\ref{mbEGp}) with Eq.~(\ref{mbEGr}) and taking (we use $\simeq$ since $|\Omega|$ should be an even integer dividing $l$)
\begin{equation}
|\Omega|\simeq \frac{\kappa_{\rm c}v_{\rm c}t+\kappa' l}{\kappa_{\rm c}+\kappa'},
\end{equation}
we obtain
\begin{equation}
\Delta^{\rm mb}_{\rm E}\le O(1) e^{-\frac{\kappa_{\rm c}\kappa}{\kappa_{\rm c}+\kappa'}(l-\frac{\kappa'}{\kappa}v_{\rm c}t)},
\end{equation}
which indeed takes the form of Lieb-Robinson bound. We note that, at large length scales such that $\coth(\frac{\kappa}{2}|\Omega|)=1+o(1)$, Eq.~(\ref{lOmega}) can be satisfied by
\begin{equation}
t<\frac{\kappa_{\rm c}+\kappa'}{\kappa_{\rm c}}\left(\frac{1}{3}-\frac{\kappa}{\kappa_{\rm c}+\kappa'}-o(1)\right)\frac{l}{v_{\rm c}},
\end{equation}
where the rhs is positive due to $\kappa'>5\kappa$, which arises from Eq.~(\ref{kappap}) and $D\ge2$ for an arbitrary SPT MPS.

%approximate MPS, future work
On the other hand, it seems that we cannot derive a Lieb-Robinson bound by simply combining Theorem~\ref{MBEG} with the errors in approximating ground states with MPSs \cite{Verstraete2006}. Even if we only require the MPS approximation to be locally good \cite{Huang2019,Brandao2019}, the needed bond dimension scales like a polynomial of the inverse of error, implying an exponentially large bond dimension and a doubly exponentially large (due to the prefactor) bound predicted by Theorem~\ref{MBEG} for an exponentially small error. We leave this problem for future work, which probably requires new ideas to directly estimate the entanglement gap in the exact ground state.

\section{Convergence bounds for unital channels}
\label{uniconv}
We briefly review Ref.~\cite{Wolf2015} and discuss how to bound $\|\mathcal{E}^l-\mathcal{E}^\infty\|$ by using function algebra. Applying Lemma~\ref{fnorm} to $\mathcal{M}=\mathcal{E}-\mathcal{E}^\infty$ leads to the following theorem.
\begin{theorem}[Main result of Ref.~\cite{Wolf2015} for unital channels]
Let $\mathcal{E}$ be a unital channel acting on a $D\times D$-dimensional operator space such that $\|\mathcal{E}^n\|\le C$ for $\forall n\in\mathbb{N}$. Let $\mu\equiv\lim_{n\to\infty}\|\mathcal{E}^n-\mathcal{E}^\infty\|^{\frac{1}{n}}$ be the spectral radius of $\mathcal{E}-\mathcal{E}^\infty$. We denote the minimal polynomial of $\mathcal{E}-\mathcal{E}^\infty$ as $m(z)=m_{\mathcal{E}-\mathcal{E}^\infty}(z)=\sum^J_{j=1}(z-\mu_j)^{s_j}$, where $\mu_j$'s are different eigenvalues of $\mathcal{E}-\mathcal{E}^\infty$ and $s_j$ is the size of the largest Jordan block with eigenvalue $\mu_j$, and define the corresponding Blaschke product as
\begin{equation}
B(z)\equiv\prod^J_{j=1}\left(\frac{z-\mu_j}{1-\bar\mu_j z}\right)^{s_j}.
\label{Bz}
\end{equation} 
Then, for $l>\frac{\mu}{1-\mu}$, we have
\begin{equation}
\begin{split}
\|\mathcal{E}^l-\mathcal{E}^\infty\| &\le 2C\|z^l\|_{W/mW} \\
&\le \mu^{l+1}\frac{4e^2C\sqrt{|m|}(|m|+1)}{l[1-(1+l^{-1})\mu]^{\frac{3}{2}}} \\
&\times\sup_{|z|=(1+l^{-1})\mu}\left|\frac{1}{B(z)}\right|,
\end{split}
\label{Wolfbound}
\end{equation}
where $|m|=\sum^J_{j=1}s_j$ is the degree of $m$ and $C$ is upper bounded by $\sqrt{\frac{D}{2}}$ \cite{Perez2006}.
\label{mainwolf}
\end{theorem}
Since the detailed proof in Ref.~\cite{Wolf2015} is rather technical, it is worthwhile to sketch the outline here. First, we note that ``$\sup$" in Eq.~(\ref{Wolfbound}) arises from the following Cauchy-Schwarz inequality:
\begin{equation}
\begin{split}
\|f_r\|_W&=\sum_{p\in\mathbb{N}} r^p|f_p| \\
&\le \sqrt{\left(\sum_{p\in\mathbb{N}}r^{2p}\right)\left(\sum_{p\in\mathbb{N}}|f_p|^2\right)} \\
&=\sqrt{\frac{1}{1-r^2}\sup_{0\le\rho<1}\int^{2\pi}_0\frac{d\phi}{2\pi}|f(\rho e^{i\phi})|^2} \\
&\le \frac{\|f\|_{H^\infty}}{\sqrt{1-r^2}},\;\;\;\;\forall f\in W,
\end{split}
\label{fWfinf}
\end{equation}
where $r\in(0,1)$ can arbitrarily be chosen, $f_r(z)\equiv f(rz)$ and $\|f\|_{H^\infty}\equiv\sup_{z\in\mathbb{D}}|f(z)|$. In the typical case with $s_j=1$, i.e., if $\mathcal{M}=\mathcal{E}-\mathcal{E}^\infty$ is diagonalizable \footnote{If this is not true, we can add an arbitrarily small perturbation $\mathcal{P}$ such that $\mathcal{M}+\mathcal{P}$ is diagonalizable, and the tiny spectrum shift is again bounded by Weyl's perturbation theorem. See also Ref.~\cite{Wolf2015}.}, $\|z^l\|_{W/mW}$ is, by definition, upper bounded by $\|g_r\|_W$ as long as $g_r(\mu_j)=g(r\mu_j)=\mu^l_j$ for $\forall j=1,2,...,J$. An example is 
\begin{equation}
g(z)=\sum^J_{j=1}\mu^l_j\frac{\tilde B_j(z)}{\tilde B_j(r\mu_j)},
\label{gz}
\end{equation}
where $\tilde B_j(z)\equiv\frac{\tilde B(z)}{z-r\mu_j}$ and $\tilde B(z)\equiv\prod^J_{j=1}\frac{z-r\mu_j}{1-r\bar\mu_j z}$ is the modified Blaschke product. Combining Eqs.~(\ref{fWfinf}) and (\ref{gz}), we obtain
\begin{equation}
\begin{split}
\|z^l\|_{W/mW}&\le\|g_r\|_W \le \frac{\|g\|_{H^\infty}}{\sqrt{1-r^2}} =\frac{1}{\sqrt{1-r^2}}\sup_{|z|=1}|g(z)| \\
&=\frac{1}{\sqrt{1-r^2}}\sup_{|z|=1}\left| \sum^J_{j=1}\frac{\mu^l_j}{(z-r\mu_j)\tilde B_j(r\mu_j)} \right|,
\end{split}
\label{mitten}
\end{equation}
where $|\tilde B(z)|=1$ for $\forall |z|=1$ is used. We will eventually arrive at Eq.~(\ref{Wolfbound}) by further bounding the rightmost expression in Eq.~(\ref{mitten}), which can be rewritten in terms of a contour integral as 
\begin{equation}
\oint_{|w|=(1+l^{-1})\mu} \frac{dw}{2\pi i}\frac{w^l}{\tilde B_r(w)(z-rw)}
\end{equation}
for $l>\frac{\mu}{1-\mu}$, and set 
\begin{equation}
r=\sqrt{1-\frac{1-(1+l^{-1})\mu}{|m|}}.
\end{equation}

\section{Generalization to finite interacting systems}%lattices}
We generalize Theorem~\ref{MBEG} to the case of finite $L$. While we still have the decomposition given in Eq.~(\ref{PsipsiPhi}), $|\Phi_{\alpha\beta}\rangle$'s are no longer orthogonal to each other. To compute the ES, we use the following generalization of Lemma~\ref{rhoma}:
\begin{lemma}
For a bipartite state $|\Psi\rangle=\sum^J_{j=1}|\phi_j\rangle|\psi_j\rangle$, where $|\phi_j\rangle$'s and $|\psi_j\rangle$'s are generally neither normalized nor orthogonal to each other, the entanglement spectrum under such a bipartition coincides with the spectrum %the nonzero part of the spectrum 
of $\bar M^{\frac{1}{2}}_\psi M_\phi \bar M^{\frac{1}{2}}_\psi$ ($\bar M_\psi$: complex conjugation of $M_\psi$) or $\bar M^{\frac{1}{2}}_\phi M_\psi \bar M^{\frac{1}{2}}_\phi$ (except for zeros), where $[M_\phi]_{jj'}=\langle\phi_j|\phi_{j'}\rangle$ and $[M_\psi]_{jj'}=\langle\psi_j|\psi_{j'}\rangle$.
\label{ABhalf}
\end{lemma}
\emph{Proof.---} It is equivalent to consider the spectrum of $\rho_\phi=\Tr_\psi|\Psi\rangle\langle\Psi|$ and that of $\rho_\psi=\Tr_\phi|\Psi\rangle\langle\Psi|$. To be specific, we focus on the former, which can be explicitly written as
\begin{equation}
\begin{split}
\rho_\phi&=\sum_{j,j'}|\phi_j\rangle\langle\phi_{j'}|\Tr[|\psi_j\rangle\langle\psi_{j'}|]\\
&=\sum_{j,j'}[M_\psi]_{jj'}|\phi_j\rangle\langle\phi_{j'}|.
\end{split}
\end{equation}
Since $M_\psi$ is Hermitian, it can be expressed as $U^\dag\Lambda U$ with $\Lambda={\rm diag}\{\Lambda_k\}^J_{k=1}$ and $U$ being unitary. To be concrete, we have $[M_\psi]_{jj'}=\sum_{j''}\bar U_{j''j'}\Lambda_{j''} U_{j''j}$. Introducing $|\tilde\phi_j\rangle=\sum_{j'}\sqrt{\Lambda_j}U_{jj'}|\phi_{j'}\rangle$, we can rewrite $\rho_\phi$ as
\begin{equation}
\rho_\phi=\sum_j|\tilde\phi_j\rangle\langle\tilde\phi_j|
\end{equation}
to which we can apply Lemma~\ref{rhoma} -- the spectrum of $\rho_\phi$ coincides with that of 
\begin{equation}
\begin{split}
\langle\tilde\phi_j|\tilde\phi_{j'}\rangle&=\sum_{j'',j'''}\sqrt{\Lambda_j\Lambda_{j'}}U_{j'j'''}\bar U_{jj''}\langle\phi_{j''}|\phi_{j'''}\rangle \\
&=\sum_{j'',j'''}\sqrt{\Lambda_j}\bar U_{jj''}[M_\phi]_{j''j'''}U^{\rm T}_{j'''j'}\sqrt{\Lambda_{j'}} \\
&=[\sqrt{\Lambda}\bar UM_\phi U^{\rm T}\sqrt{\Lambda}]_{jj'}.
\end{split}
\end{equation}
From the fact that unitary conjugation preserves the spectrum, we know that the spectrum of $\rho_\phi$ should be given by that of
\begin{equation}
U^{\rm T}\sqrt{\Lambda}\bar UAU^{\rm T}\sqrt{\Lambda}\bar U=\bar M^{\frac{1}{2}}_\psi M_\phi \bar M^{\frac{1}{2}}_\psi.
\end{equation}
Note that the spectrum of $\bar M_\psi^{\frac{1}{2}}M_\phi \bar M_\psi^{\frac{1}{2}}$ is nothing but the squared absolute values of the singular values of $M_\phi^{\frac{1}{2}}\bar M_\psi^{\frac{1}{2}}$, which are the same as those of $(M_\phi^{\frac{1}{2}}\bar M_\psi^{\frac{1}{2}})^{\rm T}=M_\psi^{\frac{1}{2}}\bar M_\phi^{\frac{1}{2}}$ and thus give the same spectrum as that of $\bar M_\phi^{\frac{1}{2}}M_\psi\bar M_\phi^{\frac{1}{2}}$ (this can directly be obtained by considering $\rho_\psi$ following a similar analysis as above). $\square$ \newline
In fact, this result has already been obtained in Ref.~\cite{Cirac2011b}, where it is used to calculate the ES of a projected entangled-pair state. 

To bound the many-body entanglement gap, we need the following lemma.
\begin{lemma}
Let $M$, $M'$, $M_0$, $M'_0$ be non-negative definite Hermitian matrices and let the $j$th largest eigenvalue of $\bar M'^{\frac{1}{2}}M\bar M'^{\frac{1}{2}}$ and that of $\bar M'^{\frac{1}{2}}_0M_0\bar M'^{\frac{1}{2}}_0$ be denoted as  $\lambda_j$ and $\lambda_{0j}$, respectively. Then for $\forall j$, we have
\begin{equation}
\begin{split}
%\|B^{\frac{1}{2}}AB^{\frac{1}{2}}-B^{\frac{1}{2}}_0A_0B^{\frac{1}{2}}_0\|
|\lambda_j-\lambda_{0j}|\le\min\{\|\bar M'_0\delta\|+\|\bar M^{\frac{1}{2}}_0\delta'\bar M^{\frac{1}{2}}_0\|,\\
\|\bar M_0\delta'\|+\|\bar M'^{\frac{1}{2}}_0\delta\bar M'^{\frac{1}{2}}_0\|\}+\|\bar\delta'\delta\|,
%\|\bar M'_0(M-M_0)\|+\|\bar M^{\frac{1}{2}}_0(M'-M'_0)\bar M^{\frac{1}{2}}_0\|+\|(\bar M'-\bar M'_0)(M-M_0)\|,
\end{split}
\label{rootbound}
\end{equation}
where $\delta\equiv M-M_0$ and $\delta'\equiv M'-M'_0$.
\label{ABAB0}
\end{lemma}
\emph{Proof.---} We first note the following useful norm inequality: for any two Hermitian matrices $A$ and $B$ with $B\ge0$ (so that $B^\frac{1}{2}$ is well-defined), we have
\begin{equation}
\|B^\frac{1}{2}AB^\frac{1}{2}\|\le\|AB\|.
\label{ABnormal}
\end{equation}
This is because the spectrum of $B^\frac{1}{2}AB^\frac{1}{2}$ coincides with that of $AB$ due to $\Tr[(B^\frac{1}{2}AB^\frac{1}{2})^n]=\Tr[(AB)^n]$ for $\forall n\in\mathbb{N}$ \cite{Perez2016}. Moreover, $B^\frac{1}{2}AB^\frac{1}{2}$ is Hermitian so its norm is nothing but the spectral radius, which is no more than the norm of $AB$. This result is a special case of Proposition IX.1.1 in Ref.~\cite{Bhatia1997}.

Let us turn to the proof of the lemma. Denoting the $j$th largest eigenvalue of $\bar M'^{\frac{1}{2}}M_0\bar M'^{\frac{1}{2}}$ as $\lambda'_j$, which is also the $j$th largest eigenvalue of $\bar M^{\frac{1}{2}}_0M'\bar M^{\frac{1}{2}}_0$, we use Weyl's perturbation theorem to obtain
\begin{equation}
\begin{split}
|\lambda_j-\lambda_{0j}|&\le|\lambda_j-\lambda'_j|+|\lambda'_j-\lambda_{0j}|\\
&\le\|\bar M'^{\frac{1}{2}}M\bar M'^{\frac{1}{2}}-\bar M'^{\frac{1}{2}}M_0\bar M'^{\frac{1}{2}}\|\\
&+\|\bar M^{\frac{1}{2}}_0M'\bar M^{\frac{1}{2}}_0-\bar M^\frac{1}{2}_0M'_0\bar M^{\frac{1}{2}}_0\|\\
&=\|\bar M'^{\frac{1}{2}}\delta\bar M'^{\frac{1}{2}}\|+\|\bar M^{\frac{1}{2}}_0\delta'\bar M^{\frac{1}{2}}_0\|\\
&\le\|\bar M'\delta\|+\|\bar M^{\frac{1}{2}}_0\delta'\bar M^{\frac{1}{2}}_0\| \\
&\le\|\bar M'_0\delta\|+\|\bar M^{\frac{1}{2}}_0\delta'\bar M^{\frac{1}{2}}_0\|+\|\bar\delta'\delta\|,%\\
%&\le\|\bar M'_0(M-M_0)\|+\|\bar M_0(M'-M'_0)\|+\|(\bar M'-\bar M'_0)(M-M_0)\|,%\;\;\;\;\square
\end{split}
\label{lamj0j}
\end{equation}
%which completes the proof. 
where we have used $\|A+B\|\le \|A\|+\|B\|$ in the last step. Replacing $M$ and $M_0$ with $M'$ and $M'_0$, respectively, and following a similar procedure, we obtain
\begin{equation}
|\lambda_j-\lambda_{0j}|\le\|\bar M_0\delta'\|+\|\bar M'^{\frac{1}{2}}_0\delta\bar M'^{\frac{1}{2}}_0\|+\|\bar\delta\delta'\|.
\label{lamj0jp}
\end{equation}
Combining Eqs.~(\ref{lamj0j}) and (\ref{lamj0jp}), we obtain Eq.~(\ref{rootbound}). $\square$

To apply Lemma~\ref{ABAB0} to the many-body entanglement gap, we have only to choose 
\begin{equation}
\begin{split}
&M_{\alpha\beta,\alpha'\beta'}=a_L\langle\alpha'|\mathcal{E}^l(|\beta'\rangle\langle\beta|)|\alpha\rangle, \\
&M'_{\alpha\beta,\alpha'\beta'}=a_L\langle\beta'|\mathcal{E}^{L-l}(|\alpha'\rangle\langle\alpha|)|\beta\rangle,\\
&M_0=a_L\mathbb{1}_{\rm v}\otimes\Lambda,\;\;\;M'_0=a_L\Lambda\otimes\mathbb{1}_{\rm v}, 
\end{split}
\end{equation}
where $a_L=(\Tr\mathcal{E}^L)^{-\frac{1}{2}}$ is a finite-size normalization factor. Following the derivation of Eq.~(\ref{ElEinf}) in the main text, we can upper bound each term on the rhs of Eq.~(\ref{rootbound}) as 
\begin{equation}
\begin{split}
&\|\bar M'_0\delta\|\le a^2_L\left(\frac{D}{2}\right)^{\frac{3}{4}}\|\mathcal{E}^l-\mathcal{E}^\infty\|,\\
&\|\bar M_0\delta'\|\le a^2_L\left(\frac{D}{2}\right)^{\frac{3}{4}}\|\mathcal{E}^{L-l}-\mathcal{E}^\infty\|,\\
&\|\bar M'^{\frac{1}{2}}_0\delta\bar M'^{\frac{1}{2}}_0\|\le a^2_L\sqrt{\frac{D}{2}}\|\mathcal{E}^l-\mathcal{E}^\infty\|,\\
&\|\bar M^{\frac{1}{2}}_0\delta'\bar M^{\frac{1}{2}}_0\|\le a^2_L\sqrt{\frac{D}{2}}\|\mathcal{E}^{L-l}-\mathcal{E}^\infty\|,\\
&\|\bar\delta'\delta\|\le a^2_LD^2\|\mathcal{E}^l-\mathcal{E}^\infty\|\|\mathcal{E}^{L-l}-\mathcal{E}^\infty\|.
\end{split}
\end{equation}
Using the techniques in Sec.~\ref{MBMR} to bound $\|\mathcal{E}^l-\mathcal{E}^\infty\|$ associated with a time evolved MPS, we obtain the following theorem.
\begin{theorem}[Finite interacting systems]
Starting from an SPT MPS with length $L$ and bond dimension $D$ subject to the periodic boundary condition, the many-body entanglement gap of a length-$l$ subsystem after $t$ time-evolution steps by a trivial symmetric MPU with bond dimension $D_U$ is bounded from above by
\begin{equation}
\begin{split}
%\|\mathcal{E}^l_t-\mathcal{E}_\infty\|\le
%\max\{{\rm Cor}(X:Y)_t,\Delta_{{\rm S},t}\}
\Delta^{\rm mb}_{\rm E}\le(\Tr\mathcal{E}^L)^{-1}[\min\{b_{1/2}(l,t)+b_{3/4}(L-l,t),\\
b_{1/2}(L-l,t)+b_{3/4}(l,t)\}+4b_1(l,t)b_1(L-l,t)]
%+\frac{1}{2}|1-(\Tr\mathcal{E}^L)^{-1}|
%C_0(l-2k_0t)^{D^2_0-1}e^{-\frac{l-vt}{\xi}}
\end{split}
\label{LRSG}
\end{equation}
for any $\min\{l,L-l\}-2k_0t\ge\frac{1+\mu}{1-\mu}$, where 
\begin{equation}
b_\alpha(l,t)=C_\alpha(l-2k_0t)^{D^2-1}e^{-\frac{l-v_\alpha t}{\xi}},
\end{equation}
with $k_0$, $\mu$ and $\xi$ being the same as those in Theorem~\ref{MBEG}, $v_\alpha=2k_0-(\alpha+\frac{1}{2})\frac{\ln D_U}{\ln\mu}$, and the coefficient
\begin{equation}
C_\alpha=e^22^{\frac{5}{2}-\alpha}D^{\frac{3}{2}+\alpha}(D^2+1)\mu^{1-D^2}(1+\mu)^{D^2+\frac{1}{2}}(1-\mu)^{D^2-\frac{5}{2}} 
\end{equation}
depends only on the initial state. 
\label{TLRB}
\end{theorem} 

Two remarks are in order here. First, we note that in the thermodynamic limit of the entire system, we have $\Tr\mathcal{E}^L=1$ and $b_\alpha(\infty,t)=0$ so that Theorem~\ref{MBEG} is reproduced. Even if $L$ is finite, we still find that $\Delta^{\rm mb}_{\rm E}$ is exponentially small up to 
\begin{equation}
t\sim\min\left\{\frac{\min\{l,L-l\}}{v_{1/2}},\frac{\max\{l,L-l\}}{v_{3/4}},\frac{L}{v_1}\right\},
\end{equation}
provided that $\min\{l,L-l\}>\frac{v_{1/2}(1+\mu)}{(v_{1/2}-2k_0)(1-\mu)}$. %but $\Delta_S(t)$ is of the order of root square of that in the $L\to\infty$ limit. 
Second, in the special case of the zero correlation length, i.e., $\mu=0$, which corresponds to fixed points of entanglement renormalization \cite{Verstraete2005b}, we can infer from Theorem~\ref{TLRB} that the degeneracy is \emph{exact} up to $t\sim \frac{l}{2k_0}$. This is intuitively rather clear since the entanglement edge modes are absolutely localized (without an exponential tail) for fixed-point states and it takes a finite time for a nonzero overlap to develop between the edge modes by a locality-preserving MPU.

\section{Details on ES dynamics upon partial symmetry breaking}
\subsection{Flat-band model for class BDI $\to$ class D}
\label{DESDF}
Thanks to the flat-band nature of the Hamiltonians given in Eqs.~(\ref{H0N}) and (\ref{HN}), it suffices to consider the $2N$ sites closest to the entanglement cut (purple dashed line in Fig.~\ref{fig10}(a)). Restricted to the single-particle Hilbert subspace of these sites, the projector onto the Fermi sea of $H_0$ reads $P_0=\frac{1}{2}\bigoplus^N_{j=1}(\sigma^0+\sigma^x)$. The ES dynamics is then determined by the spectrum of $E_S(t)=P_SP(t)P_S$, where $P_S=\frac{1}{2}\bigoplus^N_{j=1}(\sigma^0+\sigma_z)$ is the projector onto the left half $N$ sites and
\begin{equation}
P(t)=e^{-iH^{\rm sp}t}P_0e^{iH^{\rm sp}t},\;\;
H^{\rm sp}=1\oplus\bigoplus^{N-1}_{j=1}e^{it\sigma_x}\oplus1,
\end{equation}
where we have already set $J=1$. After straightforward calculations, we obtain the following matrix form of $E_S(t)$:
\begin{widetext}
\begin{equation}
E_S(t)=\frac{1}{2}\begin{bmatrix} 1 & -i\sin t & 0 & 0 & \cdots & 0 & 0 \\ i\sin t & 1 & -i\sin t \cos t & 0 & \cdots & 0 & 0 \\ 0 & i\sin t\cos t & 1 & -i\sin t\cos t & \cdots & 0 & 0 \\ 0 & 0 & i\sin t\cos t & 1 & \cdots & 0 & 0 \\ \vdots & \vdots & \vdots & \vdots & \ddots & \vdots & \vdots \\ 0 & 0 & 0 & 0 & \cdots & 1 & -i\sin t\cos t \\ 0 & 0 & 0 & 0 & \cdots & i\sin t\cos t & 1 \end{bmatrix}_{N\times N},
\end{equation}
\end{widetext}
whose characteristic polynomial reads
\begin{equation}
\begin{split}
f_N(\xi;t)&\equiv\det[\xi\mathbb{I}_{N\times N}-E_S(t)] \\
&=%\prod^N_{j=1}\left[\xi-\frac{1}{2}(1+\eta_j(t)\sin t)\right]
\left(\xi-\frac{1}{2}\right)F_{N-1}\left(\xi-\frac{1}{2};\frac{\sin2t}{4}\right) \\
&-\frac{\sin^2 t}{4}F_{N-2}\left(\xi-\frac{1}{2};\frac{\sin2t}{4}\right).
\end{split}
\end{equation}
Here, $F_N(x;a)$ is defined recursively as
\begin{equation}
F_N(x;a)=xF_{N-1}(x;a)-a^2F_{N-2}(x;a),
\label{FNxa}
\end{equation}
with initial conditions $F_1(x;a)=x$ and $F_2(x;a)=x^2-a^2$ (or $F_{-1}(x;a)=0$ and $F_0(x;a)=1$). In fact, we have an analytic expression $F_N(x;a)\equiv\prod^N_{j=1}(x-2a\cos\frac{j\pi}{N+1})$, which enjoys the properties $F_N(bx;ab)=b^NF(x;a)$, $F_N(x;a)=F_N(x;-a)$ and, in particular, $F_N(-x;a)=(-)^NF_N(x;a)$. Using the recursive relation (\ref{FNxa}), we can rewrite $f_N(\xi;t)$ into
\begin{equation}
\begin{split}
f_N(\xi;t)&=%\frac{\sin^N2t}{2^N}\left[F_N\left(\frac{2\xi-1}{\sin 2t};\frac{1}{2}\right)-\frac{\tan^2 t}{4}F_{N-2}\left(\frac{2\xi-1}{\sin 2t};\frac{1}{2}\right)\right]
F_N\left(\xi-\frac{1}{2};\frac{\sin2t}{4}\right) \\
&-\frac{\sin^4t}{4}F_{N-2}\left(\xi-\frac{1}{2};\frac{\sin2t}{4}\right).
\end{split}
\end{equation}
The roots of $f_N(\xi;t)$ give the ES dynamics. Since $F_N(x;a)$ is an odd function of $x$ for odd $N$, we have $F_{2n+1}(0;a)=0$ so that $\xi=\frac{1}{2}$ is always a root of $f_N(\xi;t)$ if $N$ is odd. Moreover, defining $g_{2n+1}(\xi;t)\equiv f_{2n+1}(\xi;t)/(\xi-\frac{1}{2})$, we find
\begin{equation}
\begin{split}
g_{2n+1}\left(\frac{1}{2};t\right)&=\left(-\frac{\sin^22t}{16}\right)^n\left(1+\frac{n}{\cos^2t}\right),\\
f_{2n}\left(\frac{1}{2};t\right)&=\left(-\frac{\sin^22t}{16}\right)^n\frac{1}{\cos^2t},
\end{split}
\end{equation}
which are generally nonzero except for some special time points. Therefore, for an even $N$, all the initial topological entanglement modes at $\xi=\frac{1}{2}$ split, while one and only one $\xi=\frac{1}{2}$ mode survives for an odd $N$.

Let us calculate the full ES dynamics for $N=1,2,3,4,5$. When $N=1$, we have $f_1(\xi;t)=\xi-\frac{1}{2}$, implying the persistence of the topological entanglement mode at $\xi=\frac{1}{2}$. This is a trivial result since $H=0$ and there is no dynamics. When $N=2$, we find $f_2(\xi;t)=\left(\xi-\frac{1}{2}\right)^2-\frac{1}{4}\sin^2t$, so the two entanglement modes oscillate as 
\begin{equation}
\xi=\frac{1}{2}(1\pm\sin t). 
\end{equation}
When $N=3$, we find $f_3(\xi;t)=\xi-\frac{1}{2})[(\xi-\frac{1}{2})^2-\frac{1}{4}\sin^2t(1+\cos^2 t)]$, implying a persistent topological mode at $\xi=\frac{1}{2}$ and two oscillating modes 
\begin{equation}
\xi=\frac{1}{2}(1\pm\sin t\sqrt{1+\cos^2t}). 
\end{equation}
When $N=4$, we find $f_4(\xi;t)=(\xi-\frac{1}{2})^4-\frac{1}{4}\sin^2t(1+2\cos^2t)(\xi-\frac{1}{2})^2+\frac{1}{16}\sin^4t\cos^2t$, so all of the four modes oscillate in time as 
\begin{equation}
\xi=\frac{1}{2}\pm\frac{\sin t}{4}\sqrt{2+4\cos^2t\pm2\sqrt{1+4\cos^4t}}.
\end{equation} 
Finally, we find the characteristic polynomial for $N=5$ to be $f_5(\xi;t)=(\xi-\frac{1}{2})[(\xi-\frac{1}{2})^4-\frac{1}{4}\sin^2t(1+3\cos^2t)(\xi-\frac{1}{2})^2+\frac{1}{16}\sin^4t\cos^2t(2+\cos^2t)]$. Except for a constant solution $\xi=\frac{1}{2}$, the other four modes oscillate as 
\begin{equation}
\xi=\frac{1}{2}\pm\frac{\sin t}{4}\sqrt{2+6\cos^2t\pm2\sqrt{5\cos^4t-2\cos^2t+1}}. 
\end{equation}
These exact ES dynamics are plotted in Fig.~\ref{fig10}(b).

\subsection{Mathematical formulation}
\label{KSPT}
Having the above simple example in mind, we are ready to introduce a general mathematical formalism for dealing with partial symmetry breaking. We first recall that the classification of gapped free-fermion systems at equilibrium is given by the homotopy group $\pi_{d_{\rm s}}(\mathcal{S})$, where $d_{\rm s}$ is the spatial dimension and $\mathcal{S}$ is the classifying space satisfying symmetry constraints. If we are interested in the stable topology, we can take the limit of infinite bands and obtain the well-known $K$-theory classification \cite{Kitaev2009}. For example, $\mathcal{S}=\mathcal{R}_1\equiv\lim_{n\to\infty}{\rm O}(n)$ for class BDI and $\mathcal{S}=\mathcal{R}_2\equiv\lim_{n\to\infty}{\rm O}(2n)/{\rm U}(n)$ for class D. However, we emphasize that the general formalism applies equally to the finite-band case, although the practical calculations could be intractable. Denoting $\mathcal{S}$ and $\tilde{\mathcal{S}}$ as the classifying spaces subject to symmetries $G$ and $\tilde G$ with $\tilde G<G$, we have $\mathcal{S}\subset\tilde{\mathcal{S}}$ since a $G$-symmetric system is always $\tilde G$-symmetric but the converse is generally not true. Therefore, we have a natural inclusion $\iota:\mathcal{S}\to\tilde{\mathcal{S}}$, which is a morphism (continuous map) in $\bold{Top}_*$ (category of pointed topological spaces). %with the point being a trivial flat-band system in the flat band limit. 
Such an inclusion induces a group homomorphism $\iota_*:\pi_{d_{\rm s}}(\mathcal{S})\to\pi_{d_{\rm s}}(\tilde{\mathcal{S}})$ through $[f]\to[\iota\circ f]$, where $f:S^{d_{\rm s}}\to\mathcal{S}$ is a continuous map from $d_{\rm s}$D sphere to $\mathcal{S}$ and $[f]$ is its homotopy equivalence class. In fact, $\pi_{d_{\rm s}}$ can be regarded as a \emph{functor} from $\bold{Top}_*$ to $\bold{Grp}$ (category of groups) or $\bold{Ab}$ (category of Abelian groups) if $d_{\rm s}\ge2$, which maps not only pointed topological spaces into homotopy groups but also continuous maps between topological spaces into group homomorphisms between the corresponding homotopy groups \cite{Hatcher2002}. In particular, $\iota_*$ is the image of $\iota$ that makes the following diagram commute:
\begin{equation}
\begin{tikzpicture}
%objects
\Text[x=-1,y=0.5]{$\mathcal{S}$}
\Text[x=1,y=0.5]{$\tilde{\mathcal{S}}$}
\Text[x=-1.25,y=-1,fontsize=\small]{$\pi_{d_{\rm s}}(\mathcal{S})$}
\Text[x=1.25,y=-1,fontsize=\small]{$\pi_{d_{\rm s}}(\tilde{\mathcal{S}})$}
%morphisms
\Text[x=0,y=0.7,fontsize=\small]{$\iota$}
\Text[x=0.125,y=-0.75,fontsize=\small]{$\iota_*$}
\Text[x=-1.3,y=-0.125,fontsize=\small]{$\pi_{d_{\rm s}}$}
\Text[x=1.3,y=-0.125,fontsize=\small]{$\pi_{d_{\rm s}}$}
\begin{scope}[>=latex] %axis
\draw[->] (-0.75,0.5) -- (0.75,0.5);
\draw[->] (-1,0.25) -- (-1,-0.75);
\draw[->] (1,0.25) -- (1,-0.75);
\draw[->] (-0.7,-1) -- (0.7,-1);
\end{scope}
\end{tikzpicture}
\end{equation}
For class BDI $\to$ class D in 1D, we have $d_{\rm s}=1$, $\pi_1(\mathcal{R}_1)=\mathbb{Z}$, $\pi_1(\mathcal{R}_2)=\mathbb{Z}_2$ and $\iota_*(N)=N\mod2$ for $\forall N\in\mathbb{Z}$, as illustrated in the previous subsection.

Let us move on to discuss interacting SPT systems classified by group cohomology \cite{Chen2013}. We note that the group-cohomology classification is actually \emph{complete} in 1D \cite{Chen2011,Chen2011b,Schuch2011}, although not in higher dimensions \cite{Kapustin2014}. Instead of the classifying spaces, we focus directly on the symmetry groups. The inclusion $\iota$ of $\tilde G$ into $G$, which is a natural group homomorphism, induces another group homomorphism from $H^{d_{\rm st}}(G,{\rm U}(1))$ to $H^{d_{\rm st}}(\tilde G,{\rm U}(1))$, where $d_{\rm st}\equiv d_{\rm s}+1\in\mathbb{Z}^+$ is the \emph{spacetime} dimension. To see this, we only have to note that an $d_{\rm st}$-cocycle $\omega:G^{\times d_{\rm st}}\to {\rm U}(1)$ can naturally be restricted to $\tilde\omega:\tilde G^{\times d_{\rm st}}\to {\rm U}(1)$ through $\tilde\omega(\tilde g_1,\tilde g_2,...,\tilde g_{d_{\rm st}})=\omega(\tilde g_1,\tilde g_2,...,\tilde g_{d_{\rm st}})$, which obviously satisfies the cocycle property $d\tilde\omega=1$. In fact, %the well-definedness of such a induced group homomorphism is rooted in the \emph{functoriality} of group cohomology. That is, 
any group cohomology $h:G\to H^{d_{\rm st}}(G,{\rm U}(1))$ (${\rm U}(1)$ can actually be replaced by other Abelian groups) is a \emph{contravariant} functor from $\bold{Grp}$ to $\bold{Ab}$, which map not only groups into Abelian cohomology groups but also group homomorphisms into those between cohomology groups \cite{Brown1982}. By contravariant, we mean that the directions of morphisms are reversed by the functor. In particular, the reduction of SPT phases is determined by the induced map $\iota^*$ that makes the following diagram commute:
\begin{equation}
\begin{tikzpicture}
%objects
\Text[x=-1,y=0.5]{$\tilde G$}
\Text[x=1,y=0.5]{$G$}
\Text[x=-1.3,y=-1,fontsize=\small]{$H^{d_{\rm st}}(\tilde G,{\rm U}(1))$}
\Text[x=1.3,y=-1,fontsize=\small]{$H^{d_{\rm st}}(G,{\rm U}(1))$}
%morphisms
\Text[x=0,y=0.7,fontsize=\small]{$\iota$}
\Text[x=0.125,y=-0.75,fontsize=\small]{$\iota^*$}
\Text[x=-1.25,y=-0.125,fontsize=\small]{$h$}
\Text[x=1.25,y=-0.125,fontsize=\small]{$h$}
\begin{scope}[>=latex] %axis
\draw[->] (-0.75,0.5) -- (0.75,0.5);
\draw[->] (-1,0.25) -- (-1,-0.75);
\draw[->] (1,0.25) -- (1,-0.75);
\draw[->] (0.25,-1) -- (-0.25,-1);
\end{scope}
\end{tikzpicture}
\end{equation}
In the main text, we have given the simplest nontrivial example with $d_{\rm st}=2$, $H^2(G,{\rm U}(1))=\mathbb{Z}_N$, $H^2(\tilde G,{\rm U}(1))=\mathbb{Z}_n$ and $\iota^*(\nu)=\frac{N}{n}\nu\mod n$.

Finally, let us comment on the impact of SPT-order reduction on the ES dynamics. For free-fermion systems, the bulk-edge correspondence usually has a very simple form --- the number of edge states, or the degeneracy in ES, is simply given by the bulk topological number \cite{Hatsugai1993,Ryu2002,Prodan2016}. As a prototypical example (e.g., class BDI $\to$ class D in 1D), a single (no) $\xi=\frac{1}{2}$ mode survives a surjective $\mathbb{Z}\to\mathbb{Z}_2$ reduction if the original topological number is odd (even). For interacting SPT systems in 1D, the open boundary ES degeneracy $r$ is determined by the minimal dimension of the \emph{irreducible projective representations} of the symmetry group. Such a minimal dimension is $1$ if the projective representation can be lifted to a linear representation, but is otherwise no less than $2$ according to Lemma~\ref{minESdeg}. In fact, there is a character theory for projective representations \cite{Cheng2015}, which shares many similarities with the conventional character theory for linear representations. For example, denoting the dimension of the $\alpha$th irreducible projective representation with respect to a $2$-cocycle $\omega$ as $d_\alpha$, we have $d_\alpha||G|$ and $\sum^{R_\omega}_{\alpha=1} d^2_\alpha=|G|$, where $R_\omega$ is the number of \emph{$\omega$-regular conjugacy classes}, i.e., those conjugacy classes with a representative element $g$ satisfying $\omega(g,h)=\omega(h,g)$ for $\forall h\in N_g\equiv\{h\in G:gh=hg\}<G$. An immediate collorary is that, for $G=\mathbb{Z}_N\times\mathbb{Z}_N$ with $N$ being a prime, we have $r=N$ for any $G$-symmetric SPT state. By simple analysis, we can also infer that possible $r>1$ for $G=\mathbb{Z}_6\times\mathbb{Z}_6$ is $2$, $3$ and $6$. These conclusions are consistent with the results in Table~\ref{table2}.

\subsection{Minimal models for quenched interacting SPT systems}
\label{MMSPT}
\begin{figure}
\begin{center}
       \includegraphics[width=8.5cm, clip]{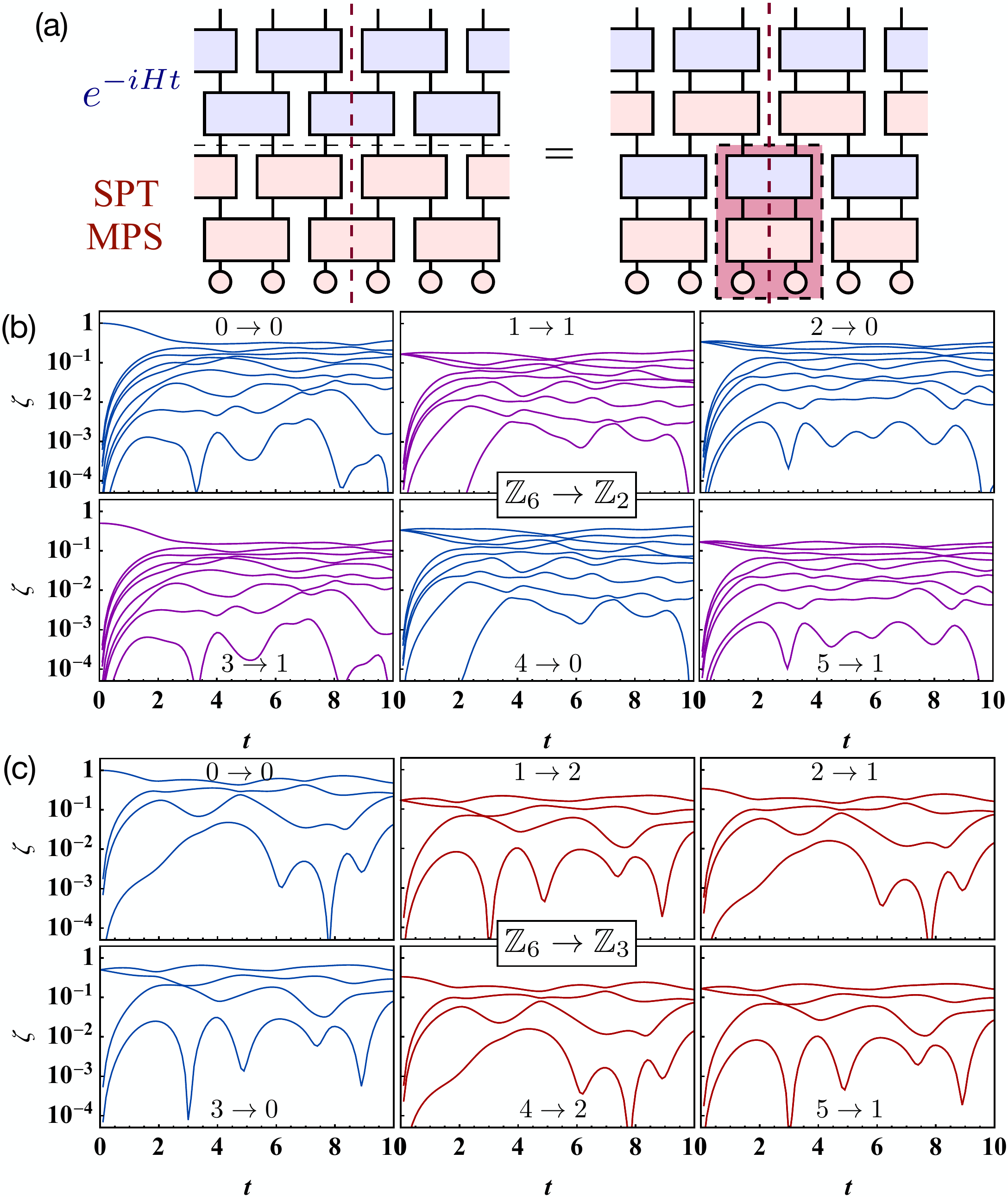}
       \end{center}
   \caption{(a) Reduction of the many-body ES into a two-site ES from the commutativity of all the two-site unitaries. The array of circles is the product state $|G\rangle^{\otimes L}$ with $|G\rangle\equiv|G|^{-\frac{1}{2}}\sum_g|g\rangle$. The red and blue two-site unitaries are given by $\sum_{g,g'}\omega(g^{-1}g',g'^{-1})^{-1}|gg'\rangle\langle gg'|$ and $\sum_{g,g'}e^{-ih(g,g')t}|gg'\rangle\langle gg'|$, respectively. The two-site state marked in the red rectangle reads $\sum_{g,g'}[M(t)]_{gg'}|gg'\rangle$, where $[M(t)]_{gg'}$ is given in Eq.~(\ref{Mt}). (b) Many-body ES dynamics for a topological number reduction $\mathbb{Z}_6\to\mathbb{Z}_2$ from a partial symmetry breaking quench $G=\mathbb{Z}_6\times\mathbb{Z}_6\to\tilde G=\mathbb{Z}_2\times\mathbb{Z}_2$. (c) Same as (b) but for $\mathbb{Z}_6\to\mathbb{Z}_3$ from $G=\mathbb{Z}_6\times\mathbb{Z}_6\to\tilde G=\mathbb{Z}_3\times\mathbb{Z}_3$. In (b) and (c), blue, purple and red curves are of degeneracy $1$, $2$ and $3$, respectively. The expressions $j\to j'$ ($j\in\mathbb{Z}_6$ and $j'\in\mathbb{Z}_{2,3}$) specify the group homomorphism from $H^2(G,{\rm U}(1))$ to $H^2(\tilde G,{\rm U}(1))$. See also Table~\ref{table2}.}
   \label{fig16}
\end{figure}

Let us introduce the interacting counterparts of flat-band free-fermion models, in the sense that these minimal models have \emph{zero} correlation lengths. In $(1+1)$D spacetime, a $G$-symmetric SPT state with zero correlation length can be built from a $2$-cocyle %$\omega:G^{\times2}\to{\rm U}(1)$ 
(which satisfies $\omega(gh,k)\omega(g,h)=\omega(g,hk)\omega(h,k)$ for $\forall g,h,k\in G$) as \cite{Chen2013}
\begin{equation}
|\Psi\rangle=\frac{1}{|G|^{\frac{L}{2}}}\sum_{\{g_j\}^L_{j=1}}\prod^L_{j=1}\omega(g_j^{-1}g_{j+1},g_{j+1}^{-1})^{-1}|g_1g_2...g_L\rangle.
\label{fpSPT}
\end{equation}
Here the local Hilbert space $\mathbb{C}^{|G|}$ is spanned by $\{|g\rangle:g\in G\}$ and the on-site symmetry representation is regular: $\rho_g=\sum_{h\in G}|gh\rangle\langle h|$ for $\forall g\in G$. Note that such a construction (\ref{fpSPT}) applies equally to continuous symmetries if we replace $\sum_g$ by $\int dg$ \cite{Chen2013}. To demonstrate the impact of partial symmetry breaking on the ES dynamics, we consider the simplest case in which the Hamiltonian is a sum of commutative two-site operators:
\begin{equation}
H=\sum^L_{j=1}h_j,\;\; h_j=\sum_{g_j,g_{j+1}}h(g_j,g_{j+1})|g_jg_{j+1}\rangle\langle g_jg_{j+1}|,
\end{equation}
whose eigenstates are simply Fock states. To partially break the symmetry from $G$ to $\tilde G$, we require
\begin{equation}
h(g,g')=h(\tilde gg,\tilde gg')\in\mathbb{R},\;\;\;\;\forall g,g'\in G\;{\rm and}\;\tilde g\in\tilde G,
\label{tdG}
\end{equation}
and otherwise 
\begin{equation}
h(g,g')\neq h(g''g,g''g'),\;\;\;\;\forall g''\in G\backslash \tilde G. 
\end{equation}
As illustrated in Fig.~\ref{fig16}(a), due to the fact that $h_j$'s commute with each other, the open boundary ES at time $t$ are simply the squared singular values of
\begin{equation}
[M(t)]_{gg'}=e^{-ih(g,g')t}\omega(g^{-1}g',g'^{-1})^{-1}.
\label{Mt}
\end{equation}
We can then numerically determine the degeneracies $r$ and $\tilde r$ in the initial ES and that after the quench.

In Figs.~\ref{fig16}(b) and (c), we plot the ES dynamics for two different partial symmetry breaking quenches $G=\mathbb{Z}_6\times\mathbb{Z}_6\to\tilde G=\mathbb{Z}_2\times\mathbb{Z}_2$ and $G=\mathbb{Z}_6\times\mathbb{Z}_6\to\tilde G=\mathbb{Z}_3\times\mathbb{Z}_3$. We randomly sample $h(g,g')$ among $[-1,1]$ while keeping the $\tilde G$-symmetry requirement (\ref{tdG}). The behaviors of ES splitting agree perfectly with the results in Table~\ref{table2}. While not shown here, we have also checked that the ES always becomes non-degenerate upon the partial symmetry breaking quench $G=\mathbb{Z}_4\times\mathbb{Z}_4\to\tilde G=\mathbb{Z}_2\times\mathbb{Z}_2$. This is fully consistent with the fact that the induced group homomorphism from $\mathbb{Z}_4$ to $\mathbb{Z}_2$ is trivial, as mentioned in Sec.~\ref{subsubIS}.

\section{Numerical simulations for disordered systems}
\label{NSDS}
In this appendix, we provide some numerical pieces of evidence to support the qualitative discussions in Sec.~\ref{Stoerung}.

\begin{figure}
\begin{center}
       \includegraphics[width=6.5cm, clip]{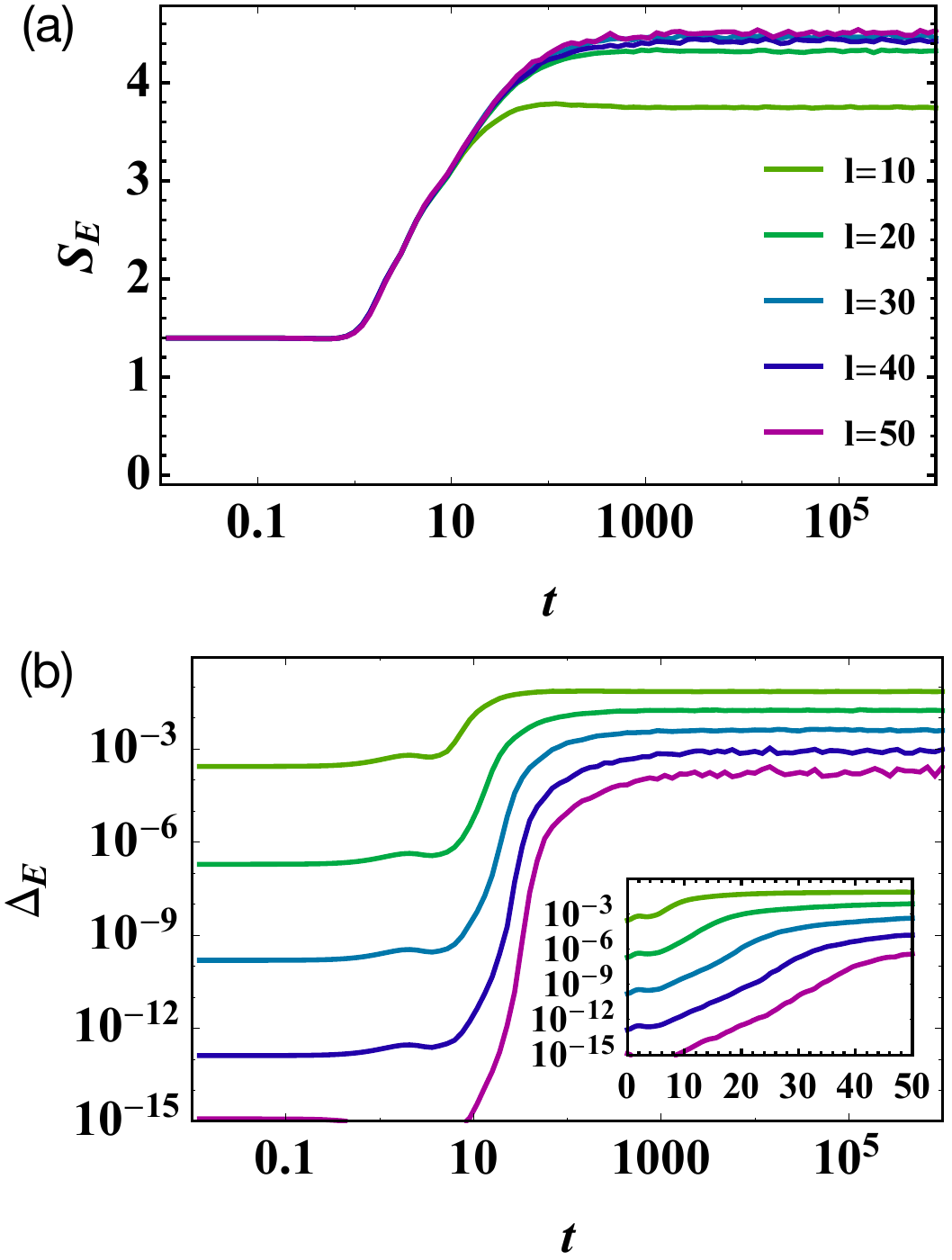}
       \end{center}
          \caption{Dynamics of (a) the entanglement entropy and (b) the single-particle entanglement gap after a quench to the disordered SSH model (\ref{disSSH}) from a clean state. Inset in (b): the same as the main panel but in the log-linear scale. We choose $L=2l+1$ so that the entanglement bipartition is asymmetric and the initial entanglement gap is finite. The numbers of disorder realizations for $l=10,20,30,40$ and $50$ are $10^4$, $5\times10^3$, $2\times10^3$, $10^3$ and $5\times10^2$, respectively.  The parameters are quenched as $(\bar J,\bar J',f)=(0.5,1,0)\to(1,0.5,0.6)$. Note that the early-time data of $\Delta^{\rm sp}_{\rm E}$ for $l=50$ (only a part of which is visible) are not reliable due to a finite numerical resolution.}
          \label{fig17}
\end{figure}

\subsection{Disordered SSH model}
\label{DSSH}
We consider the entanglement dynamics in a disordered SSH model described by the Hamiltonian
\begin{equation}
H=-\sum_j(J_j b^\dag_ja_j+ J'_j a^\dag_{j+1}b_j+{\rm H.c.}),
\label{disSSH}
\end{equation}
where $a_j$ and $b_j$ denote the sublattice fermionic modes in the $j$th unit cell, and the hopping amplitudes
\begin{equation}
J_j\in[(1-f)\bar J,(1+f)\bar J],\;\;\;\;
J'_j\in[(1-f)\bar J',(1+f)\bar J']
\end{equation}
are uniformly and independently sampled. We start from a topological state with $(\bar J,\bar J',f)=(0.5,1,0)$ (no disorder) and quench the parameters to $(\bar J,\bar J',f)=(1,0.5,0.6)$. The length of the subsystem is chosen to be $l=\frac{1}{2}(L-1)$ so that the entanglement gap in the initial state becomes nonzero due to the asymmetric entanglement bipartition.

As shown in Fig.~\ref{fig17}(a), we find that the entanglement entropy grows logarithmically --- a feature usually associated with many-body localized systems. Since there is no interaction in our model, one may na\"ively expect that the entanglement grows extremely slowly like $\ln\ln t$ \cite{Igloi2012}, as might be inferred from a dynamical version of the strong-disorder renormalization group \cite{Altman2013}. However, there is a crucial difference in our setup --- the intial state $|\Psi_0\rangle$ is not a product state (in real space) and has a nonzero correlation length. We denote $U_0$ as a local unitary such that $U_0|\Psi_0\rangle$ becomes a product state; then the quench dynamics in this new frame is governed by $U_0HU^\dag_0$ ($H$ is given by Eq.~(\ref{disSSH}) with the postquench parameters), which now involves small but finite long-range hoppings. We expect this effect to dramatically change the common paradigm of entanglement growth in Anderson insulators. On the other hand, the entanglement entropy eventually becomes saturated at a constant independent of (sufficiently large) $l$, as a manifestation of a finite localization length. 

In stark contrast to the slow growth of entanglement entropy, the numerical results (see Fig.~\ref{fig17}(b)) suggests an exponentially fast growth of the single-particle entanglement gap before saturation, which is similar to the clean case. Nevertheless, due again to the finiteness of the localization length, the saturation value decreases exponentially with respect to $l$, implying that the topological entanglement edge modes are stable even in the long-time limit.

\begin{figure}
\begin{center}
       \includegraphics[width=6.5cm, clip]{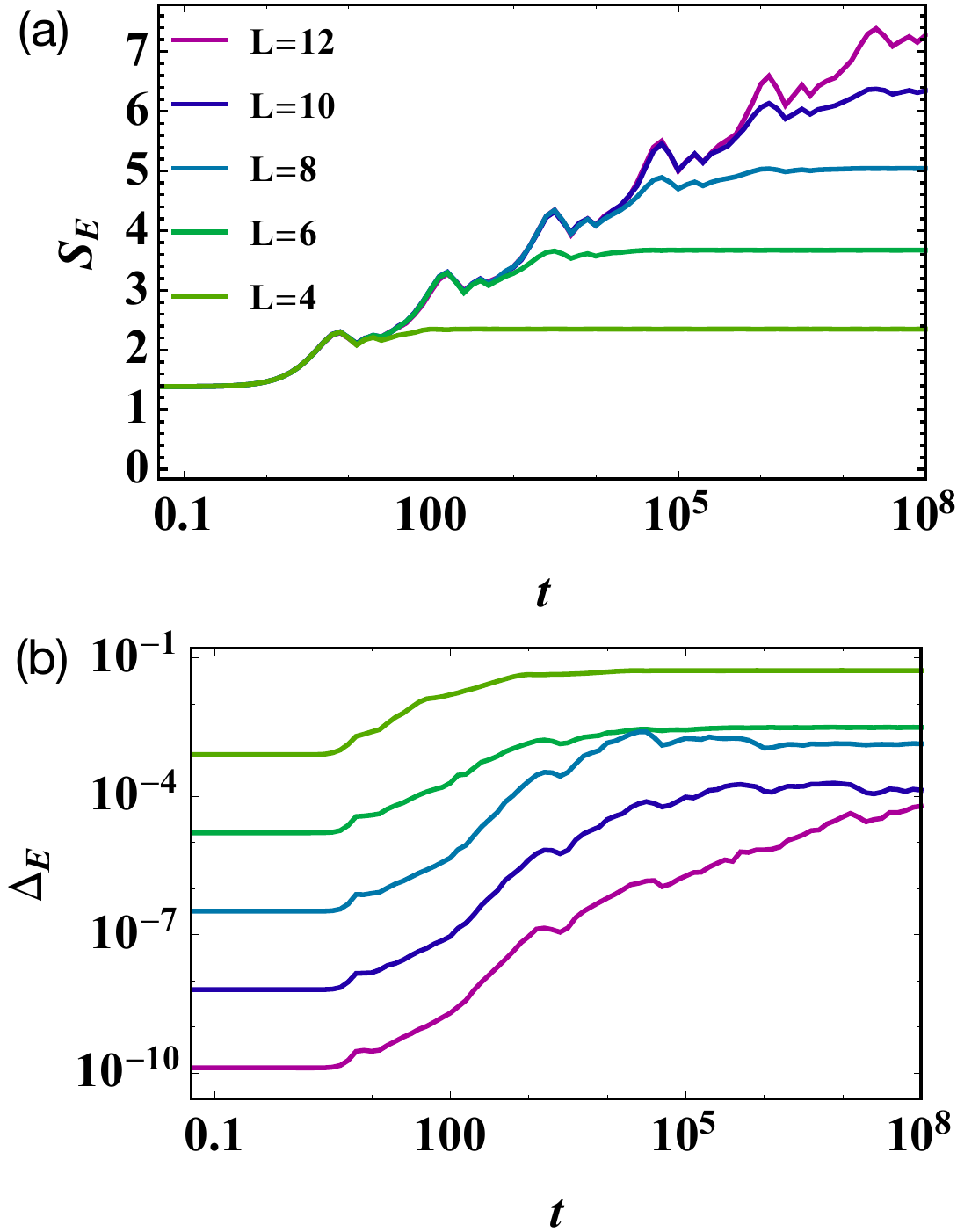}
       \end{center}
          \caption{Dynamics of (a) the half-chain entanglement entropy and (b) the many-body entanglement gap starting from the $\mathbb{Z}_2\times\mathbb{Z}_2$ SPT MPS (\ref{Z2Z2MPS}) and evolving in time according to the disordered Hamiltonian given in Eq.~(\ref{phMBL}). The numbers of disorder realizations for $L=4,6,8,10$ and $12$ are $5\times10^4$, $10^4$, $5\times10^3$, $5\times10^2$ and $3\times10^2$, respectively. The parameters in Eqs.~(\ref{A1234}) and (\ref{phMBL}) are set to be $p=q=0.49$, $J_0=3$ and $\kappa=3$.}
          \label{fig18}
\end{figure}

\subsection{Phenomenological model for many-body localization}
\label{PMMBL}
The minimal group that supports an interacting SPT phase protected by unitary symmetries is $G=\mathbb{Z}_2\times\mathbb{Z}_2=\{(m,n):m,n\in\mathbb{Z}_2\}$, whose second cohomology group reads $H^2(\mathbb{Z}_2\times\mathbb{Z}_2,{\rm U}(1))=\mathbb{Z}_2$. By specifying the on-site symmetry as the regular representation $\rho_{(m,n)}=Z^m\otimes Z^n$ with $Z=|0\rangle\langle 0|-|1\rangle\langle 1|$ acting on a local qubit, we can construct a nontrivial MPS as
\begin{equation}
|\Psi_0\rangle=\sum_{\{\boldsymbol{j}_s=00,01,10,11\}}\Tr[A_{\boldsymbol{j}_1}A_{\boldsymbol{j}_2}...A_{\boldsymbol{j}_L}]|\boldsymbol{j}_1\boldsymbol{j}_2...\boldsymbol{j}_L\rangle,
\label{Z2Z2MPS}
\end{equation}
where a local state $|\boldsymbol{j}\rangle$ consists of two qubits and the injective tensor $A_{\boldsymbol{j}}$ is given by
\begin{equation}
\begin{split}
A_{00}&=\sqrt{(1-p)(1-q)}\sigma^0,%\mathbb{1}_{\rm v},
\;\;\;\;
A_{01}=\sqrt{q(1-p)}\sigma^x,\\%\;\;\;\;
A_{10}&=i\sqrt{p(1-q)}\sigma^y,\;\;\;\;\;\;\;\;\;\;\;\;
A_{11}=\sqrt{pq}\sigma^z,
\end{split}
\label{A1234}
\end{equation}
with $p,q\in(0,1)$. We can check that the projective representation on the virtual level is nontrivial:
\begin{equation}
\begin{split}
V_{(0,0)}&=\sigma^0,%\mathbb{1}_{\rm v},
\;\;\;\;
V_{(1,0)}=\sigma^x,\\%\;\;\;\;
V_{(0,1)}&=\sigma^y,\;\;\;\;
V_{(1,1)}=\sigma^z.
\end{split}
\end{equation}
To slightly lift the exact four-fold degeneracy in the ES of a finite segment, we choose $p=q=0.49$ in Eq.~(\ref{A1234}).

To simulate the effect of many-body localization, we recall that a fully many-body localized Hamiltonian can generally be written as 
\begin{equation}
H_{\rm MBL}=\sum_j h_j\tau^z_j+\sum_{j,j'}J_{jj'}\tau^z_j\tau^z_{j'}+\cdots, 
\end{equation}
where $\tau^z_j=U_{\rm loc}Z_jU^\dag_{\rm loc}$ is the spin operator of the logic qubit which is related to the physical one by a local unitary transformation \cite{Nandkishore2015}. Inspired by this phenomenology, we expect that an Ising Hamiltonian with random and exponentially decay long-range interations,
\begin{equation}
H'=\sum_{j< j'}J_{jj'}Z_jZ_{j'},\;\;
J_{jj'}\in [-J_0e^{-\kappa(j'-j)},J_0e^{-\kappa(j'-j)}],
\label{phMBL}
\end{equation}
where $J_{jj'}$'s are sampled uniformly and independently, would be sufficient to capture the essential physics of the dynamical interplay between many-body localization and SPT order. Note that this Ising Hamiltonian (\ref{phMBL}) respects the $\mathbb{Z}_2\times\mathbb{Z}_2$ symmetry.

Figure~\ref{fig18}(a) shows that the half-chain entanglement entropy of the state $|\Psi_t\rangle=e^{-iH't}|\Psi_0\rangle$ essentially follows a logarithmic growth. The separation between two successive local peaks is very close to $e^\kappa$ (as a multiple, since we take the logarithmic scale), consistent with the argument that the logarithmic growth of entanglement entropy results basically from the decoherence of remote spins \cite{Abanin2013}. In stark contrast with the previous noninteracting case, the entanglement entropy eventually becomes saturated at an \emph{extensive} quantity (i.e., proportional to $L$). Regarding the growth of the many-body entanglement gap, the numerical results in Fig.~\ref{fig18}(b) seem to suggest a power law. This result is to some extent expected once we accept that the entanglement gap in an SPT state is bounded essentially by the correlation at the subsystem length scale, which we have proved for clean systems.

\bibliography{GZP_references}

\begin{thebibliography}{196}
\bibitem [{ant()}]{antisym}%
  \BibitemOpen
  \href@noop {} {}\bibinfo {note} {A unitary or anti-unitary operator is a
  symmetry (anti-symmetry) if it commutes (anti-commutes) with the
  Hamiltonian.}\BibitemShut {Stop}%
\end{thebibliography}

\end{document}